\documentclass[12pt]{article}

\pdfoutput=1

\usepackage[T1]{fontenc}
\usepackage[utf8]{inputenc}
\usepackage[english]{babel}
\usepackage{amsmath,amssymb,amsfonts,mathtools}
\usepackage{amsthm}
\usepackage{latexsym}
\usepackage[mathcal]{euscript}
\usepackage{graphicx}
\usepackage{microtype}
\graphicspath{{./}{figures/}}
\usepackage{epsfig}
\usepackage{subfig}
\usepackage{psfrag}
\usepackage{array}
\usepackage{float}
\usepackage{placeins}
\usepackage{cancel}
\usepackage{mathrsfs}
\usepackage{bm}
\usepackage[normalem]{ulem}
\usepackage{xcolor}
\usepackage{slashed}
\usepackage{appendix}
\usepackage{jheppub}

\newcommand{\be}{\begin{equation}}
\newcommand{\ee}{\end{equation}}
\newcommand{\bea}{\begin{eqnarray}}
\newcommand{\eea}{\end{eqnarray}}

\newcommand{\bi}{\begin{itemize}}
\newcommand{\ei}{\end{itemize}}

\def\Tr{\mathop{\rm Tr}}

\newcommand{\cO}{{\cal O}}

\newcommand{\cF}{{\cal F}}
\newcommand{\cE}{{\cal E}}
\newcommand{\cW}{{\cal W}}

\begin{document}

\title{Energy plateau cascades from complex-time local quenches in thermal two-dimensional CFTs}

\author[a,b]{Dmitry S. Ageev}
\author[b,c]{and Vladimir A. Bykov}

\affiliation[a]{Department of Mathematical Methods for Quantum Technologies, Steklov Mathematical Institute of Russian Academy of Sciences, Gubkina str. 8, Moscow 119991, Russia}
\affiliation[b]{Institute for Theoretical and Mathematical Physics, Lomonosov Moscow State University, 119991 Moscow, Russia}
\affiliation[c]{National University of Science and Technology MISIS, Moscow, 119049 Russia}

\emailAdd{ageev@mi-ras.ru}
\emailAdd{vbykov@itmp.msu.ru}

\abstract{
In this work, we study local operator quenches in thermal two-dimensional conformal field theories with complex-time evolution, which can be viewed as an effective description of conditional dynamics in a system interacting with an environment. In contrast to quenches of the vacuum, the thermal initial state allows several conformal families to participate in the evolution. We show that their competition can produce cascades of approximately homogeneous energy plateaux, with successive plateaux approaching the energies of different primary states. Using conformal Ward identities on the cylinder and torus, we determine the local energy density and illustrate this behavior in free-field theories, the Ising model and Wess--Zumino--Witten models. We develop a cold-limit spectral description in which the dominant primary channel determines the height of each plateau, while conformal dimensions and operator matrix elements govern the transitions between them. For a fixed compact unitary CFT, we show that only finitely many primary channels can contribute to the leading plateau sequence. Finally, we discuss the large-central-charge regime, relating light-operator plateaux to few-particle sectors in thermal AdS and examining the limitations of this picture for heavy quenches.
}

\maketitle

\clearpage
\section{Introduction}
\label{sec:intro}

Understanding the nonequilibrium dynamics of quantum many-body systems remains a central problem in condensed matter physics, statistical mechanics and quantum field theory. One of the fundamental questions is how isolated quantum systems can exhibit relaxation and approach thermal equilibrium despite evolving unitarily. While generic interacting systems may display thermalization, integrability and disorder can lead to qualitatively different behavior and prevent conventional thermal equilibration~\cite{Sachdev:2011,Polkovnikov:2011,DAlessio:2016,GogolinEisert:2016,EisertFriesdorfGogolin:2015,NandkishoreHuse:2015,AbaninAltmanBlochSerbyn:2019,AmicoFazioOsterlohVedral:2008}. These questions are closely connected to the growth of entanglement and the spreading of quantum information, which provide complementary ways to characterize the evolution of correlations in many-body systems~\cite{NielsenChuang:2010,Swingle:2018,XuSwingle:2024}. Related problems arise in high-energy physics, where thermalization, quantum chaos and information scrambling play important roles in black-hole dynamics and the black-hole information problem~\cite{HaydenPreskill:2007,SekinoSusskind:2008,ShenkerStanford:2014,MaldacenaShenkerStanford:2016,Harlow:2016,AlmheiriHartmanMaldacenaShaghoulianTajdini:2021}. A common objective in these settings is to understand how initially localized perturbations spread, how local observables relax, and what information about the initial state remains accessible during the evolution.

Quantum quenches provide a useful framework for investigating these processes. By preparing a system away from equilibrium, either through a change of its Hamiltonian or by introducing a localized excitation, one can study the subsequent propagation of energy, correlations and entanglement. Conformal field theories are particularly well suited to this approach because conformal symmetry strongly constrains the relevant correlation functions and their time evolution. The theoretical framework for global and local quenches in CFT was established in pioneering works by Calabrese and Cardy~\cite{CalabreseCardy:2006,CalabreseCardy:2007,CalabreseCardyLocal:2007,CalabreseCardy:2016}. In two dimensions, these protocols have revealed characteristic features of nonequilibrium dynamics, including light-cone propagation of excitations, entanglement growth and saturation, and finite-size revivals~\cite{Cardy:2014}.

An especially useful class of protocols consists of local operator quenches, in which an excitation is created by the insertion of a local operator into the initial state~\cite{Nozaki:2014hna}. Such quenches provide a direct connection between the operator content of a CFT and its real-time dynamics. In rational conformal field theories, for example, the entanglement generated by local primary excitations is related to their quantum dimensions, whereas more general theories can exhibit behavior beyond a simple quasiparticle description~\cite{HeNumasawaTakayanagiWatanabe:2014,KudlerFlamKusukiRyu:2021}. Related joining and splitting protocols have also been investigated in holographic CFTs~\cite{ShimajiTakayanagiWei:2019}. Extensions to thermal initial states and partitioning protocols allow one to study the response of mixed states, as well as nonequilibrium energy and charge transport~\cite{SotiriadisCalabreseCardy:2009,BernardDoyon:2016}. More recent applications have explored the role of initial correlations, inhomogeneous state preparation, information recovery and complexity-related observables~\cite{MikhaylenkoSemenov:2026,JiangMezei:2025,NozakiTamaokaTan:2024,GotoNozakiRyuTamaokaTan:2024,CaputaDiGiulio:2025,ZhaoHeOuyangZengZhang:2026}. Beyond conformal field theory, quenches of masses, couplings and localized degrees of freedom have been used to investigate universal short-time behavior and nonequilibrium dynamics in massive and interacting quantum field theories~\cite{SotiriadisCardy:2009,DasGalanteMyers:2014,DasGalanteMyers:2015,AgeevBelokonPushkarev:2023,RadovskayaSemenov:2023,AgeevPushkarev:2025}.

The holographic correspondence provides another perspective on these phenomena by relating nonequilibrium dynamics in strongly coupled conformal field theories to gravitational evolution in asymptotically anti-de Sitter spacetime. Global quenches can be modeled by infalling matter shells, allowing the relaxation of correlation functions and the growth of entanglement to be studied geometrically~\cite{Balasubramanian:2011,LiuSuh:2014,LiuSonner:2020}. A gravitational description of a local operator quench in terms of an infalling particle in AdS was proposed in~\cite{NozakiNumasawaTakayanagi:2013}. Subsequent work has examined the associated gravitational backreaction, entanglement dynamics, complexity and the response of confining backgrounds~\cite{AsplundBernamontiGalliHartman:2015,AgeevArefevaBagrovKatsnelson:2018,AgeevBykov:2026}. Related methods have also been extended to quantum fields in de Sitter space and time-dependent cosmological geometries~\cite{AlbrechtHolmanRichard:2015,CarrilhoRibeiro:2017,AlvesCamilo:2019,BanerjeeChoudhury:2023,Belkovich:2025izl}.

An alternative approach to nonequilibrium dynamics arises when a quantum system interacts with auxiliary degrees of freedom and its evolution is conditioned on selected measurement outcomes. In such protocols, repeated interactions with ancillary systems followed by postselection can generate effective nonunitary evolution~\cite{BreuerPetruccione:2002,DalibardCastinMolmer:1992,PlenioKnight:1998}. In particular, suitable sequences of weak measurements can realize an effective Euclidean evolution, which modifies the relative weights of energy eigenstates and progressively suppresses higher-energy contributions to the conditioned state~\cite{LapierrePelliconiRyuSonner:2025}. Local operator quenches combined with projections have also been investigated as probes of information transfer in conformal field theory~\cite{GyongyosiHollowoodKumar:2025}. When Euclidean filtering is combined with ordinary Lorentzian evolution, the resulting dynamics can be described by evolution along a complex-time contour. Importantly, this construction describes a conditioned quantum trajectory associated with a selected measurement record, rather than the unconditional evolution obtained by averaging over measurement outcomes.

Complex-time evolution also has a natural geometric interpretation in quantum field theory and gravity. The role of complex metrics and integration contours has been studied in connection with Wick rotation, the convergence of path integrals and the admissibility of complex spacetime geometries~\cite{LoukoSorkin:1997,KontsevichSegal:2021,WittenComplex:2021}. Related constructions in conformal field theory combine Euclidean and Lorentzian evolution, sometimes generated by different Hamiltonians, and demonstrate that the structure and ordering of these evolutions can affect energy profiles, correlations and entanglement~\cite{MaoNozakiTamaokaTan:2024,MaoNozaki:2026}. Nonunitary local-operator quenches have also been explored holographically, revealing different late-time entanglement behavior and the role of gravitational backreaction for heavy excitations~\cite{MaoMiyataNozakiOmidi:2026}.

A concrete realization of complex-time local-quench dynamics was recently studied by Bai et al.~\cite{Bai:2026}. They considered local operator excitations of the vacuum in two-dimensional CFTs on a spatial circle and supplemented Lorentzian evolution with a time-dependent Euclidean filter. The filtering protocol was also related to repeated interactions with ancillary spins followed by postselection. An important result of their analysis is that the filtered evolution can drive subsystem observables toward those of the primary state created by the quench operator.

In the present work, we extend this framework to local quenches of an initial thermal state and investigate the resulting evolution of the energy density. The presence of thermal excitations introduces additional input states that can be connected by the quench operator to different conformal families. Under Euclidean filtering, these contributions compete because their weights depend differently on the initial inverse temperature and the filtering interval. As a result, the energy density can develop a sequence of approximately homogeneous plateaux, each associated with a different dominant primary output state. We refer to this behavior as a cascade of energy plateaux.

We characterize this mechanism through a spectral description of the filtered state. In the low-temperature regime, the competition between primary channels is governed by their conformal dimensions and the matrix elements of the quench operator. The dominant contributions can be identified from the lower envelope of the corresponding spectral costs, which determines the leading plateau heights and crossover positions. For a fixed compact unitary CFT, we establish that only finitely many primary pairs can contribute to this leading envelope, despite the potentially infinite number of conformal families in the full spectrum. The number of observable plateaux nevertheless depends on the allowed operator couplings and the separation between competing spectral channels.

We compute the local energy density using conformal Ward identities on the cylinder and torus and illustrate the resulting dynamics in free-field theories, the Ising model and Wess--Zumino--Witten models. We further apply the spectral description to compact-boson vertex quenches, unitary minimal models and affine-primary quenches in WZW theories, where increasingly long plateau sequences can arise. Finally, we examine the large-central-charge regime and discuss the conditions under which these plateaux admit a holographic interpretation in terms of few-particle states in thermal AdS or distinct backreacted gravitational configurations.

In parallel with the present study, Junlong Chen has been pursuing an independent investigation of closely related questions, with partially overlapping results~\cite{Chen:2026:inprep}. His manuscript is currently in preparation.

The paper is organized as follows. In Sec.~\ref{sec:cylinder}, we introduce the complex-time local-quench protocol for thermal states and use the conformal Ward identity to derive the energy density on the cylinder and its late-time asymptotics. In Sec.~\ref{sec:torus}, we extend the calculation to the torus and investigate the energy evolution in several conformal field theories, illustrating both the emergence of intermediate energy plateaux and their subsequent relaxation. Section~\ref{sec:effective} develops a general description of the plateau hierarchy in the low-temperature regime. We identify the dominant primary channels, establish a bound restricting the number of contributing channels, and determine the plateau heights and crossover times from conformal data. We illustrate this framework in the Ising model, the compact boson, unitary minimal models, and WZW theories. In Sec.~\ref{sec:holographic}, we examine the large-$c$ regime, relating light-operator plateaux to few-particle sectors in thermal AdS$_3$ and discussing the conditions under which heavy quenches may admit an interpretation in terms of distinct gravitational geometries. We conclude with a summary and discussion of possible extensions. The appendices provide the proofs of the cold-limit selection results and local-stress bounds, together with the tricritical-Ising sewing construction and numerical checks.

\section{Complex-time local quenches in thermal CFTs}
\label{sec:cylinder}

\subsection{Setup}

We consider a two-dimensional conformal field theory initially prepared in a thermal state at inverse temperature $\beta$ and investigate the evolution of its energy density following a local operator quench. Our setup extends the conventional local-quench protocol by introducing a complex-time evolution, following the prescription of~\cite{Bai:2026}. In this approach, the usual unitary evolution in Lorentzian time is supplemented by Euclidean evolution, which acts as an energy filter and gradually suppresses higher-energy contributions to the evolving state.

The quench is generated by inserting a primary operator $\cO(x_0)$, with conformal weights $(h,\bar h)$, at the spatial position $x_0$. The resulting time-dependent density matrix is
\be
\rho(t)=\frac{e^{-itH}e^{-sH}\cO(x_0)e^{-\beta H}
\cO^{\dagger}(x_0)e^{-sH}e^{itH}}{Q(\beta,s)}.
\label{eq:rho}
\ee
Here $H$ is the Hamiltonian of the theory, and $t$ denotes Lorentzian time. The factor $e^{-\beta H}$ describes the initial thermal ensemble, while the operator insertions create a localized excitation above this background. The Euclidean evolution factors $e^{-sH}$ act on both sides of the operator-excited density matrix, modifying the relative contributions of different energy eigenstates. In contrast to purely unitary evolution, this filtering changes the energy distribution of the state as time progresses.

We choose the Euclidean evolution parameter to depend linearly on Lorentzian time,
\be
s(t)=\epsilon+\delta t,\qquad
\epsilon>0,\qquad \delta>0.
\label{eq:filter}
\ee
The parameter $\epsilon$ provides the initial Euclidean regularization of the local operator insertion, ensuring that the excited state is well defined. The parameter $\delta$ determines the strength of the additional Euclidean evolution per unit Lorentzian time. Consequently, as $t$ increases, the filtering becomes progressively stronger, favoring lower-energy components of the operator-excited state.

Since Euclidean evolution is nonunitary, the density matrix must be normalized at each time. The corresponding normalization factor is
\be
Q(\beta,s)=\Tr\left(e^{-2sH}\cO(x_0)e^{-\beta H}
\cO^\dagger(x_0)\right),
\label{eq:Q}
\ee
which follows from the cyclicity of the trace and ensures that $\Tr\rho(t)=1$.

The normalization factor also has a natural interpretation in the Euclidean path-integral formulation. Taking the trace identifies the initial and final Euclidean time slices, producing a periodic Euclidean geometry with temporal circumference $\beta+2s$. The two operator insertions divide the circle into a thermal arc of length $\beta$ and a filtering arc of length $2s$. Thus, $Q(\beta,s)$ can be viewed as an unnormalized two-point function on a Euclidean cylinder whose temporal circumference depends on the filtering parameter. This geometric representation provides the starting point for computing the energy density using conformal Ward identities.

Finally, in the zero-temperature limit $\beta\to\infty$, the initial thermal ensemble reduces to the vacuum. The density matrix then describes a pure state prepared by applying $\cO(x_0)$ to the vacuum and evolving it with $e^{-itH}e^{-sH}$, followed by normalization. At finite temperature, however, the initial ensemble contains contributions from excited states, allowing the local operator to connect different energy eigenstates and conformal families. In particular, the competition between these contributions under Euclidean filtering plays an important role in the resulting energy dynamics.

Our main observable is the local energy density in the time-dependent state~(\ref{eq:rho}). In a two-dimensional conformal field theory, the energy density can be decomposed into left- and right-moving contributions, whose expectation values we denote by
\be
\begin{aligned}
\cE_L(X,t)&=\Tr\bigl[\rho(t)\widehat{\cE}_L(X)\bigr],\\
\cE_R(X,t)&=\Tr\bigl[\rho(t)\widehat{\cE}_R(X)\bigr],
\end{aligned}
\label{eq:energytrace}
\ee
where $X$ is the spatial observation point. We adopt the normalization in which the physical energy density is related to these chiral contributions by
\be
\langle T_{tt}(X)\rangle_{\rho(t)}
=\frac{\cE_L(X,t)+\cE_R(X,t)}{2\pi},
\ee
with the Hamiltonian given by $H=\int dX\,T_{tt}(X)$.

\begin{figure}[t]
\centering
\includegraphics[width=\textwidth]{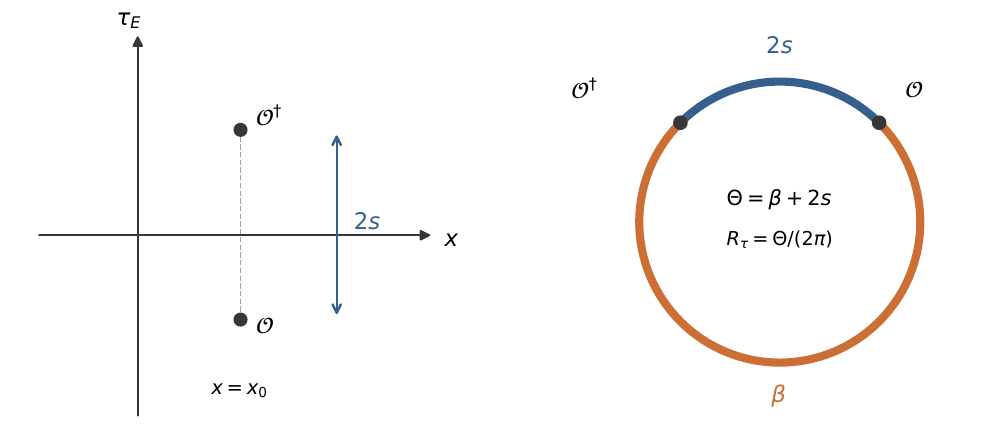}
\caption{Euclidean quench contour. The operator insertions at $\tau_E=\pm s$ (left) lie on the time circle obtained by taking the trace (right), with arcs of lengths $2s$ and $\beta$ and circumference $\Theta=\beta+2s$.}
\label{fig:geometry}
\end{figure}

To evaluate these expectation values, it is convenient to formulate the problem in Euclidean signature, where the energy density can be obtained from correlation functions involving the stress tensor. As discussed above, the trace defining the normalization factor~(\ref{eq:Q}) identifies the endpoints of the Euclidean evolution, producing a periodic Euclidean time direction with circumference
\be
\Theta(t)=\beta+2s(t).
\label{eq:theta}
\ee
The trace geometry is illustrated in Fig.~\ref{fig:geometry}.

We introduce the complex Euclidean coordinates
\be
w=x+i\tau_E,\qquad \bar w=x-i\tau_E,
\label{eq:euclcoord}
\ee
so that the periodicity of Euclidean time is expressed as $w\sim w+i\Theta$. Choosing the origin of Euclidean time midway between the two insertions, their positions are
\be
w_1=x_0-is,\qquad w_2=x_0+is,
\label{eq:insertions}
\ee
together with $\bar w_1=x_0+is$ and $\bar w_2=x_0-is$.

In this representation, the effect of the local quench on the energy density is encoded in correlation functions of the stress tensor with the two quench operators. Let $\Sigma$ denote the relevant Euclidean geometry. We introduce the two-point function
\be
G_\Sigma(w_1,w_2)=\langle\cO(w_1,\bar w_1)
\cO^\dagger(w_2,\bar w_2)\rangle_{\Sigma},
\label{eq:euclG}
\ee
and the corresponding stress-tensor three-point function
\be
\mathcal{T}_\Sigma(w)=\langle T_E(w)\cO(w_1,\bar w_1)
\cO^\dagger(w_2,\bar w_2)\rangle_{\Sigma}.
\label{eq:euclT}
\ee
Here the Euclidean correlation functions are normalized by the partition function of the underlying geometry. Their ratio,
\be
\cW_E(w)=\frac{\mathcal{T}_\Sigma(w)}{G_\Sigma(w_1,w_2)},
\label{eq:euclideanratio}
\ee
gives the expectation value of the holomorphic stress tensor in the presence of the operator insertions, normalized by the corresponding two-point function. This ratio is the Euclidean quantity from which the left-moving energy density will be obtained.

The relation to the original density-matrix formulation follows directly from the path-integral representation of the trace. In particular, the two-point function satisfies
\be
G_\Sigma(w_1,w_2)=\frac{Q(\beta,s)}{Z_\Sigma(\Theta)},
\label{eq:euclnorm}
\ee
where $Z_\Sigma(\Theta)$ is the partition function on the Euclidean surface. Thus, the normalization by $G_\Sigma$ in~(\ref{eq:euclideanratio}) implements the same normalization as the factor $Q(\beta,s)$ in the density matrix. The antiholomorphic contribution is obtained analogously from the correlator involving $\bar T_E(\bar w)$.

The precise form of these correlation functions depends on the spatial geometry. We consider two cases. When the theory is defined on the infinite line, $x\in\mathbb R$, the periodic Euclidean time direction gives a cylinder of circumference $\Theta$. If instead the spatial direction is compactified to a circle of length $L$, the additional identification $x\sim x+L$ produces a torus with periods $L$ and $i\Theta$. The same general construction applies in both cases, with the appropriate correlation functions evaluated on the corresponding Euclidean surface. For the infinite line, the normalized local correlators can be understood as the thermodynamic limits of their finite-volume counterparts.

Having expressed the energy density in terms of Euclidean correlation functions, we obtain the real-time evolution by analytic continuation. We first perform the continuation of the observation time, $\tau_E\to it$, while keeping the Euclidean insertion points fixed. The holomorphic and antiholomorphic observation coordinates then become
\be
w\longrightarrow X-t,\qquad
\bar w\longrightarrow X+t.
\label{eq:wickcoordinates}
\ee
With our conventions for the Euclidean stress tensor, the Lorentzian chiral energy densities are related to the analytically continued correlators by
\be
\begin{aligned}
\cE_L(X,t)&=-\left.\cW_E(w)\right|_{w=X-t},\\
\cE_R(X,t)&=-\left.\bar{\cW}_E(\bar w)
\right|_{\bar w=X+t}.
\end{aligned}
\label{eq:wickenergy}
\ee
It is important that the Euclidean separation between the operator insertions is held fixed during this continuation. Only after the Euclidean correlators have been analytically continued do we impose the time-dependent filtering protocol,
$s=\epsilon+\delta t$, together with $\Theta=\beta+2s(t)$. In this way, the Lorentzian time dependence enters both through the propagation of the chiral excitations and through the changing Euclidean geometry induced by the filtering.

\subsection{Energy evolution on the thermal cylinder}

We first consider the case of an infinite spatial line, where the periodicity in Euclidean time defines a cylinder of circumference $\Theta$. The calculation can be reduced to the plane by the conformal map
\be
z=e^{\frac{2\pi}{\Theta}w}.
\label{eq:cylmap}
\ee
The normalized stress-tensor three-point function on the plane is determined by the conformal Ward identity
\be
\frac{\langle T_E(z)\cO(z_1,\bar z_1)
\cO^\dagger(z_2,\bar z_2)\rangle_{\mathbb C}}
{\langle\cO(z_1,\bar z_1)
\cO^\dagger(z_2,\bar z_2)\rangle_{\mathbb C}}
=
\sum_{j=1}^{2}
\left[
\frac{h}{(z-z_j)^2}
+\frac{1}{z-z_j}\partial_{z_j}\log G_{\mathbb C}
\right],
\label{eq:planeward}
\ee
where $G_{\mathbb C}$ is the two-point function of the primary operator $\mathcal{O}(x)$
\be
G_{\mathbb C}(z_1,z_2)
=\frac{1}{(z_1-z_2)^{2h}
(\bar z_1-\bar z_2)^{2\bar h}}.
\label{eq:planetwopoint}
\ee
The corresponding antiholomorphic expression follows from the same Ward identity with $h$ replaced by $\bar h$. To obtain the result on the cylinder, we use the transformation law of the stress tensor
\be
T_E(w)=\left(\frac{dz}{dw}\right)^2T_E(z)
+\frac{c}{12}\{z,w\},
\label{eq:stressmap}
\ee
where $\{z,w\}$ is the Schwarzian derivative, which contributes the term proportional to the central charge $c$.

\begin{figure}[t]
\centering
\includegraphics[width=\textwidth]{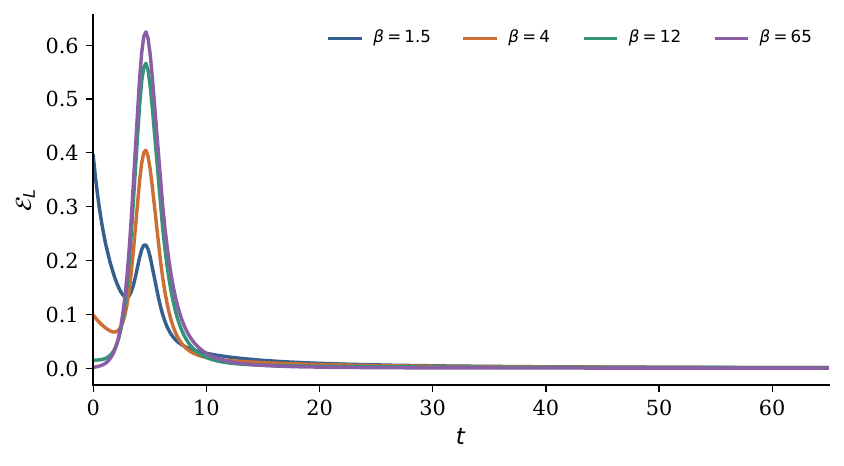}
\caption{Left-moving energy density $\cE_L$ on the thermal cylinder for $c=3/2$, $h=3/4$ and the indicated inverse temperatures. Parameters: $X=5$, $x_0=0$, $\epsilon=0.5$, $\delta=0.35$.}
\label{fig:cylinder}
\end{figure}

Substituting (\ref{eq:planetwopoint}) into the Ward identity (\ref{eq:planeward}), mapping back to the cylinder
and performing the analytic continuation
(\ref{eq:wickenergy}), we obtain
\be
\begin{aligned}
\cE_L(X,t)
&=\frac{\pi^2c}{6\Theta^2}
+\frac{h\pi^2}{\Theta^2}
\frac{\sin^2(2\pi s/\Theta)}
{\left[
\sinh^2\!\left(\frac{\pi(X-x_0-t)}{\Theta}\right)
+\sin^2\!\left(\frac{\pi s}{\Theta}\right)
\right]^2},\\
\cE_R(X,t)
&=\frac{\pi^2c}{6\Theta^2}
+\frac{\bar h\pi^2}{\Theta^2}
\frac{\sin^2(2\pi s/\Theta)}
{\left[
\sinh^2\!\left(\frac{\pi(X-x_0+t)}{\Theta}\right)
+\sin^2\!\left(\frac{\pi s}{\Theta}\right)
\right]^2}.
\end{aligned}
\label{eq:cylstress}
\ee
The first term gives the energy density of the thermal cylinder with circumference $\Theta$, while the remaining contribution describes the energy generated by the local operator insertion. The time evolution for different initial temperatures is shown in Fig.~\ref{fig:cylinder}.

We now consider the late-time behavior at fixed $X$, $\beta$ and $\delta>0$. Expanding (\ref{eq:cylstress}) for $t\to\infty$ gives
\be
\cE_L(X,t)
=\frac{\pi^2c}{6\Theta(t)^2}
+\frac{h\pi^4\beta^2}{16\delta^4}
\operatorname{sech}^4\!\left(\frac{\pi}{2\delta}\right)
t^{-4}
+o(t^{-4}).
\label{eq:cylasymp}
\ee
The right-moving expression is obtained by replacing $h$ with $\bar h$. Since $\Theta(t)=\beta+2\epsilon+2\delta t$, the thermal contribution decays as $t^{-2}$, while the contribution of the local operator decays as $t^{-4}$.

For comparison, in the zero-temperature limit $\beta\to\infty$, the initial state is prepared by acting with $\cO(x_0)$ on the vacuum. The corresponding left-moving energy density is
\be
\cE_L^{\beta\rightarrow \infty}(X,t)
=\frac{4hs(t)^2}
{\left[(X-x_0-t)^2+s(t)^2\right]^2}
=\frac{4h\delta^2}{(1+\delta^2)^2}\,t^{-2}
+O(t^{-3}).
\label{eq:operatorplane}
\ee
This reproduces the result for the operator-prepared state studied in~\cite{Bai:2026}. The different late-time powers in (\ref{eq:cylasymp}) and (\ref{eq:operatorplane}) reflect the noncommutativity of the limits $t\to\infty$ and $\beta\to\infty$.

\section{Energy plateaux and relaxation following local quenches on the torus}
\label{sec:torus}

We now extend the analysis to a conformal field theory defined on a spatial circle of circumference $L$. In contrast to the infinite-line geometry considered in the previous section, the spatial coordinate is periodically identified, $x\sim x+L$. Combined with the Euclidean time periodicity $\tau_E\sim\tau_E+\Theta$, this gives a torus
\be
\mathbb T^2=\mathbb C/(L\mathbb Z+i\Theta\mathbb Z),
\label{eq:torusgeom}
\ee
with fundamental periods $\omega_1=L$ and $\omega_2=i\Theta$. The corresponding modular parameter is
\be
\tau=\frac{\omega_2}{\omega_1}=\frac{i\Theta}{L}.
\label{eq:torusmodulus}
\ee
We introduce the fixed inverse spatial scale $\kappa\equiv 2\pi/L$, which sets the natural finite-volume energy scale. Throughout the torus analysis, we express the local chiral energy densities in units of $\kappa^2$; for a primary state of left conformal weight $h$, the corresponding homogeneous level is $\cE_L/\kappa^2=h-c/24$. This normalization concerns the finite-volume theory and is not used for the noncompact cylinder of the preceding section.
As in the cylinder case, the Euclidean time circumference depends on the filtering parameter through $\Theta=\beta+2s(t)$. Consequently, the complex structure of the torus evolves with time, providing an additional dependence of the correlation functions on the filtering protocol.

Unlike the cylinder, the torus cannot be mapped conformally to the complex plane. The stress-tensor correlation functions must therefore be evaluated directly on the genus-one geometry. The relevant tool is the torus conformal Ward identity~\cite{Eguchi:1987,Brehm:2022}, which relates a stress-tensor insertion to derivatives of the operator correlation function with respect to both the insertion coordinates and the torus modulus. For the two-point function of the quench operator, it takes the form
\be
\begin{aligned}
\langle T_E(w)\cO(w_1,\bar w_1)
\cO^\dagger(w_2,\bar w_2)\rangle_{\mathbb T^2}=
&\,\langle T_E\rangle_{\mathbb T^2}
G_{\mathbb T^2}
+\frac{2\pi i}{L}\partial_{\omega_2}G_{\mathbb T^2}\\
&+\sum_{j=1}^{2}K_1(w,w_j)\partial_{w_j}G_{\mathbb T^2}
+h\sum_{j=1}^{2}K_2(w,w_j)G_{\mathbb T^2}.
\end{aligned}
\label{eq:torusward}
\ee
Here $G_{\mathbb T^2}$ denotes the normalized torus two-point function introduced in~(\ref{eq:euclG}). The first term contains the stress-tensor expectation value in the absence of operator insertions, while the remaining terms describe the response of the two-point function to variations of the torus geometry and the insertion positions.

The stress-tensor one-point function is determined by the torus partition function,
\be
\langle T_E\rangle_{\mathbb T^2}=\frac{2\pi i}{L}\partial_{\omega_2}\log Z_{\mathbb T^2}(\tau,\bar\tau),
\label{eq:torusT1}
\ee
where $Z_{\mathbb T^2}$ is the partition function of the theory. The derivative with respect to $\omega_2$ accounts for the dependence on the complex structure of the torus and is taken at fixed operator insertion coordinates and fixed antiholomorphic modulus.

The dependence on the stress-tensor insertion point $w$ is encoded in the kernels $K_1$ and $K_2$, defined by
\be
K_1(w,w_j)=\zeta(w-w_j)+\frac{2\eta_1}{L}w_j,
\label{eq:K1}
\ee
and
\be
K_2(w,w_j)=\wp(w-w_j)+\frac{2\eta_1}{L}.
\label{eq:K2}
\ee
Here $\zeta$ and $\wp$ denote the Weierstrass zeta and elliptic functions associated with the periods $L$ and $i\Theta$, and $\eta_1=\zeta(L/2)$. These kernels play a role analogous to the simple- and double-pole terms in the conformal Ward identity on the complex plane, while incorporating the periodic structure of the torus.

To simplify the resulting expressions, it is convenient to introduce two functions of the coordinate difference $v$,
\be
J_\Theta(v)=
\partial_v\log\vartheta_1
\left(\frac{\pi v}{L}\Big|\tau\right)
=\zeta(v)-\frac{2\eta_1}{L}v,
\label{eq:Jtheta}
\ee
and
\be
P_\Theta(v)=-\partial_vJ_\Theta(v)
=\wp(v)+\frac{2\eta_1}{L},
\label{eq:Ptheta}
\ee
where $\vartheta_1$ is the odd Jacobi theta function. These functions provide a convenient representation of the torus Ward kernels in terms of the separation between the stress-tensor insertion and the quench operators.

Since the torus is translationally invariant, the two-point function depends on the insertion coordinates only through their difference. Consequently,
$\partial_{w_1}\log G_{\mathbb T^2}
=-\partial_{w_2}\log G_{\mathbb T^2}$.
Using this relation, dividing the Ward identity~(\ref{eq:torusward}) by $G_{\mathbb T^2}$, and expressing the kernels in terms of $J_\Theta$ and $P_\Theta$, we obtain
\be
\begin{aligned}
\cW_E(w)=
&\,\langle T_E\rangle_{\mathbb T^2}
+\frac{2\pi i}{L}\partial_{\omega_2}\log G_{\mathbb T^2}\\
&+h\sum_{j=1}^{2}P_\Theta(w-w_j)\\
&+\partial_{w_1}\log G_{\mathbb T^2}
\big(J_\Theta(w-w_1)-J_\Theta(w-w_2)\big).
\end{aligned}
\label{eq:masterward}
\ee
Equation~(\ref{eq:masterward}) separates the background stress tensor from the operator contributions. The theory dependence enters through the partition function and the quench two-point function, while $J_\Theta$ and $P_\Theta$ encode the common torus geometry.

To obtain the Lorentzian energy density, we follow the analytic continuation described in the previous section. The quench operators remain at the Euclidean positions $w_1=x_0-is$ and $w_2=x_0+is$, while the stress-tensor insertion is continued to $w=X-t$. The left-moving energy density then follows from~(\ref{eq:wickenergy}), and the right-moving contribution is obtained from the corresponding antiholomorphic Ward identity, with $\bar w=X+t$ and conformal weight $\bar h$. As before, the Euclidean insertion coordinates are held fixed during the continuation, and the time dependence $s=\epsilon+\delta t$ and $\Theta=\beta+2s(t)$ is imposed afterward.

We now evaluate the required partition functions and correlators in several CFTs, comparing their energy profiles, late-time relaxation and intermediate plateaux.

\subsection{Noncompact free boson}
\label{subsec:noncompact}

We begin with the noncompact free boson, which provides a simple example in which the effects of Euclidean filtering can be studied analytically. The theory has central charge $c=1$ and, in addition to its oscillator excitations, contains a continuous momentum zero mode. In particular, the presence of this continuous degree of freedom has important consequences for the late-time relaxation of the energy density.

The torus partition function of the noncompact boson is well known~\cite{ItzyksonZuber:1986,DiFrancescoSaleurZuber:1987,DijkgraafVerlindeVerlinde:1988}. Since the target space is noncompact, the integration over the constant scalar mode produces an infinite volume factor $V_\phi$. After factoring out this volume, the partition function takes the form
\be
\frac{Z(\tau,\bar\tau)}{V_\phi}
\propto\frac{1}{\sqrt{\Theta/L}\,|\eta(\tau)|^2}.
\label{eq:znc}
\ee
Here $\eta(\tau)$ denotes the Dedekind eta function. The factor $|\eta(\tau)|^{-2}$ accounts for the oscillator excitations, while the factor $(\Theta/L)^{-1/2}$ originates from the integration over the continuous momentum zero mode.

We consider quenches generated by the holomorphic current $J=i\partial\phi_L$, with conformal weights $(1,0)$, and by the vertex operator $V_a=:e^{ia\phi}:$, with weights $(a^2/2,a^2/2)$. These operators provide complementary examples of the filtering dynamics, since the current probes fluctuations of the scalar field, whereas the vertex operator carries a definite momentum charge.

To determine their two-point functions, we use the Green function of the free scalar on the torus. For the rectangular geometry with modular parameter $\tau=i\Theta/L$, it is given by
\be
G_\phi(w_2-w_1)=
-\log\left|
\frac{\vartheta_1\left(\frac{\pi (w_2-w_1)}{L}\big|\tau\right)}
{\vartheta_1'(0|\tau)}
\right|^2
+\frac{2\pi}{L\Theta}(\operatorname{Im}(w_2-w_1))^2.
\label{eq:ncpropagator}
\ee
Here the constant scalar mode has been removed from the Green function. The first term describes the fluctuations expressed through the Jacobi theta function, while the additional nonholomorphic term ensures the required periodicity under translations along the torus cycles.

We first consider the current quench. Since the current is linear in derivatives of the scalar field, its two-point function follows by differentiating the Green function. With the convention appropriate for the conjugate-current insertion, we obtain
\be
G_J(w_1,w_2)
=-P_\Theta(w_2-w_1)+\frac{\pi}{L\Theta}.
\label{eq:nccurrent}
\ee
The first term contains the contribution associated with the scalar fluctuations, while the second arises from the variance of the continuous momentum zero mode. Both contributions are relevant for the energy density generated by the quench.

\begin{figure}[t]
\centering
\includegraphics[width=\textwidth]{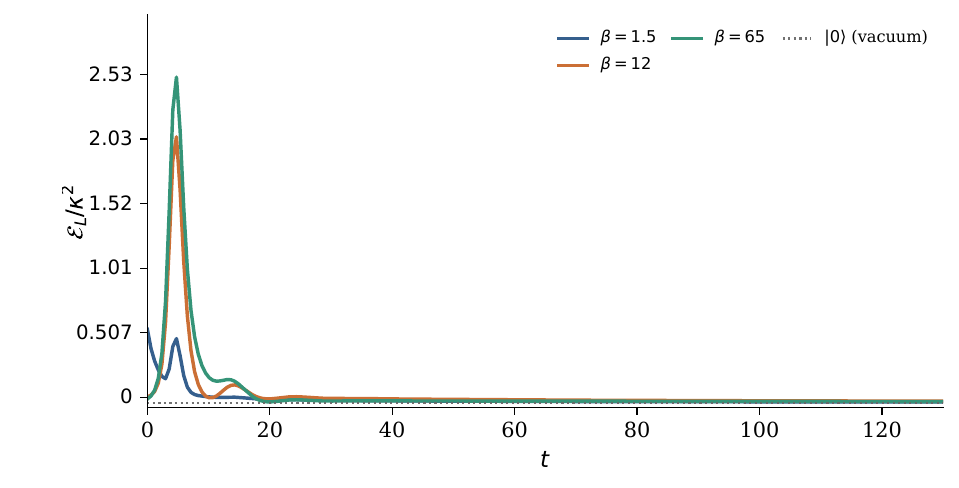}
\caption{Current-quench energy density $\cE_L/\kappa^2$ in the noncompact boson at $\beta=1.5,12,65$. The dotted line is the vacuum level $-1/24$. The calculation includes the continuous momentum zero mode and $220$ oscillator modes; $L=10$, $X=5$, $x_0=0$, $\epsilon=0.5$, $\delta=0.35$.}
\label{fig:nccurrent}
\end{figure}

Substituting this correlator, together with the partition function~(\ref{eq:znc}), into the torus Ward identity~(\ref{eq:masterward}), we obtain the left-moving energy density shown in Fig.~\ref{fig:nccurrent} for several initial temperatures.

We next turn to the vertex-operator quench. For a free scalar field, the two-point function of normal-ordered exponential operators can be evaluated directly using Gaussian contractions,
\be
\begin{aligned}
G_{V_a}(w_1,w_2)
&=\big\langle:e^{ia\phi(w_1)}::e^{-ia\phi(w_2)}:\big\rangle_{\mathbb T^2}\\
&\propto\exp\big[a^2 G_\phi(w_2-w_1)\big].
\end{aligned}
\label{eq:ncvertexwick}
\ee
Evaluating this expression at the Euclidean insertion points of the quench, for which $w_2-w_1=2is$, gives
\be
G_{V_a}(w_1,w_2)\propto
\left|
\frac{\vartheta_1'(0|\tau)}
{\vartheta_1\left(\frac{2\pi is}{L}\big|\tau\right)}
\right|^{2a^2}
\exp\left(\frac{8\pi a^2s^2}{L\Theta}\right).
\label{eq:ncvertex}
\ee
The theta-function factor describes the contribution of the scalar fluctuations, while the exponential factor reflects the nonholomorphic part of the torus Green function. Together, they determine the dependence of the vertex-operator correlator on the Euclidean separation and the torus modulus.

\begin{figure}[t]
\centering
\includegraphics[width=\textwidth]{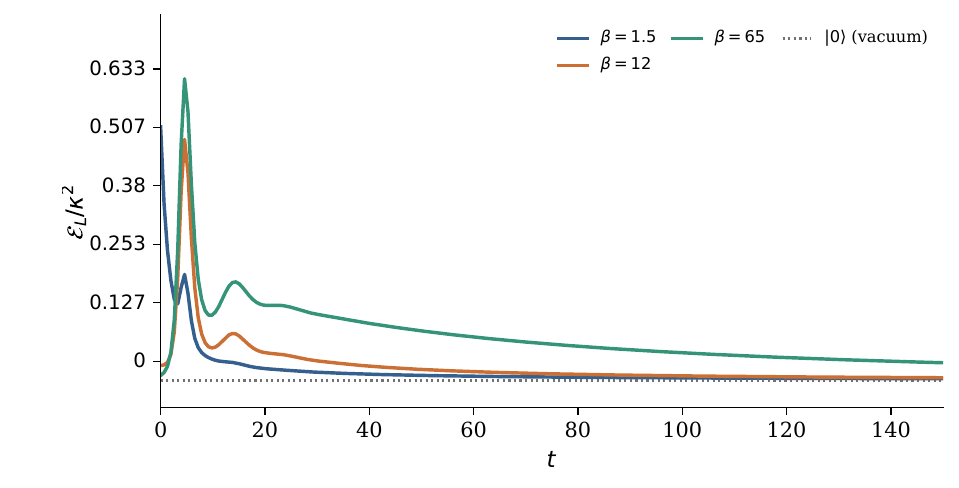}
\caption{Energy density $\cE_L/\kappa^2$ following a noncompact momentum-vertex quench with $a=1/\sqrt2$, $(h,\bar h)=(1/4,1/4)$ and $\beta=1.5,12,65$. The dotted line is the vacuum level $-1/24$. The oscillator sum retains $220$ modes; other parameters are as in Fig.~\ref{fig:nccurrent}.}
\label{fig:ncvertex}
\end{figure}

Using~(\ref{eq:ncvertex}) in the Ward identity~(\ref{eq:masterward}), we obtain the corresponding energy profile shown in Fig.~\ref{fig:ncvertex}.

An important feature of the noncompact theory is that its continuous momentum spectrum leads to algebraic late-time relaxation. To make this behavior explicit, we consider the limit $t\to\infty$ at fixed inverse temperature $\beta>0$ and observation point $X$. Under the linear filtering protocol, the Euclidean time circumference grows as
$\Theta(t)=\beta+2s(t)\sim2\delta t$.
Although both the circumference and the separation between the operator insertions increase, the complementary Euclidean interval remains fixed at length $\beta$.

For the current quench, the oscillator contraction at separation $2is$ consequently approaches the finite-temperature quantity
$g_\beta=\sum_{m=1}^{\infty}m e^{-\kappa m\beta}$. Combining this limiting oscillator contribution with the Gaussian moments of the continuous momentum zero mode, and using the torus Ward identity~(\ref{eq:masterward}), we find
\be
\begin{aligned}
\frac{\cE_{L,J}(X,t)}{\kappa^2}
&=-\frac{1}{24}+\frac{1}{4\kappa\Theta(t)}
+\frac{1}{4\kappa^2g_\beta\Theta(t)^2}
+O(\Theta^{-3})+O(e^{-\kappa s(t)}),\\
\frac{\cE_{L,V}(X,t)}{\kappa^2}
&=-\frac{1}{24}+\frac{1}{4\kappa\Theta(t)}
+\frac{a^2\beta^2}{2\Theta(t)^2}
+O(e^{-\kappa s(t)}/\Theta(t)),
\qquad t\to\infty.
\end{aligned}
\label{eq:ncasymp}
\ee
The leading constant term is the vacuum Casimir energy, while the remaining terms describe the relaxation toward this value. In particular, both quenches exhibit the same leading time-dependent correction, proportional to $1/\Theta(t)$. Since $\Theta(t)\sim2\delta t$, this correction takes the form $L/(16\pi\delta t)$ in units of $\kappa^2$.

The common leading behavior originates from the continuous Gaussian momentum zero mode and is therefore independent of the choice of quench operator. The distinction between the current and vertex quenches first appears at order $\Theta^{-2}$, where the coefficients depend on the oscillator contribution $g_\beta$ and the vertex charge $a$, respectively.

For the current quench, the asymptotic expansion requires $\kappa\Theta\gg g_\beta^{-1}$ and becomes nonuniform in the low-temperature limit $\beta\to\infty$, where $g_\beta$ is exponentially suppressed. More generally, the algebraic relaxation reflects the absence of a finite gap above the bottom of the continuous zero-mode spectrum. Unlike in a compact theory with a discrete spectrum, Euclidean filtering cannot exponentially isolate a normalizable lowest-energy state in this sector. This distinction will become particularly important when we turn to the compact boson, where momentum quantization leads to qualitatively different late-time behavior.

\subsection{Compact boson}
\label{subsec:compact}

We next consider the compact free boson, which provides a useful contrast to the noncompact theory discussed above. Although the two theories share the same local oscillator dynamics, compactification replaces the continuous momentum spectrum with a discrete set of momentum and winding sectors. This distinction has important consequences for the dynamics under Euclidean filtering. In particular, the compact theory exhibits exponential late-time relaxation and, at sufficiently low initial temperatures, can develop a sequence of energy plateaux associated with different primary states.

The compact boson is defined by imposing the identification
\be
\phi\sim\phi+2\pi R,
\label{eq:compactident}
\ee
so that the target space becomes a circle of radius $R$. This periodicity leads to two independent integer quantum numbers. First, the wavefunctions of the constant scalar mode must be single-valued on the target-space circle, which quantizes the corresponding momentum in units of $1/R$. Second, the field configuration may wind around the target-space circle as the spatial coordinate traverses the circle $x\sim x+L$. These sectors are labeled by the momentum number $n\in\mathbb Z$ and the winding number $m\in\mathbb Z$, respectively.

In the normalization adopted in Section~\ref{subsec:noncompact}, the corresponding left- and right-moving charges are~\cite{Ginsparg:1988,DiFrancesco:1997}
\be
p_L=\frac nR+\frac{mR}{2},\qquad
p_R=\frac nR-\frac{mR}{2},\qquad n,m\in\mathbb Z.
\label{eq:compactcharges}
\ee
The compactification leaves the oscillator spectrum unchanged, but replaces the continuous zero-mode integral by a sum over the allowed charge lattice. Consequently, the torus partition function factorizes into an oscillator contribution and a sum over momentum and winding sectors,
\be
\begin{aligned}
Z_R(\tau,\bar\tau)
&=\frac{1}{|\eta(\tau)|^2}
\sum_{n,m\in\mathbb Z}q^{p_L^2/2}\bar q^{p_R^2/2}\\
&=\frac{\vartheta_3(0,q_n)\vartheta_3(0,q_m)}{|\eta(\tau)|^2},
\qquad q_n=e^{-\kappa\Theta/R^2},\quad
q_m=e^{-\kappa\Theta R^2/4},
\end{aligned}
\label{eq:compactZ}
\ee
where $q=e^{2\pi i\tau}$ and
$\vartheta_3(z,q)=\sum_{j\in\mathbb Z}q^{j^2}e^{2ijz}$.
The first expression makes the contributions of the left- and right-moving charges explicit, while the second follows from the rectangular geometry $\tau=i\Theta/L$, for which the momentum and winding sums factorize~\cite{Ginsparg:1988,DiFrancesco:1997,ItzyksonZuber:1986}. When applying the torus Ward identity, however, the holomorphic modulus derivatives must be evaluated using the unfactorized charge-lattice expression, with the antiholomorphic modulus held fixed. The rectangular-torus condition is imposed only after differentiation.

The local fluctuations of the compact scalar are described by the same Gaussian theory as in the noncompact case. Within each winding sector, the periodic fluctuation field therefore has the torus Green function
\be
\begin{aligned}
G_\phi(v)
&=\langle\widetilde\phi(w_2)\widetilde\phi(w_1)\rangle_{\mathbb T^2}\\
&=-\log\left|\frac{\vartheta_1(\pi v/L|\tau)}
{\vartheta_1'(0|\tau)}\right|^2
+\frac{2\pi}{L\Theta}(\operatorname{Im}v)^2,
\qquad v=w_2-w_1.
\end{aligned}
\label{eq:compactgreen}
\ee
Here $\widetilde\phi$ denotes the periodic Gaussian fluctuation with its constant mode removed. The difference from the noncompact theory arises through the discrete charge sectors, whose contributions must be included when constructing correlation functions of the full compact field.

As before, we examine quenches generated by the holomorphic current $J=i\partial\phi_L$ and by the momentum vertex operator $V_a=:e^{ia\phi}:$. The compactification restricts the vertex charge to $a=N/R$, where $N\in\mathbb Z_{>0}$.

We first consider the current quench. The current two-point function receives contributions both from Gaussian contractions of the fluctuating field and from the zero-mode charges. The latter are now determined by discrete thermal averages over the momentum and winding sectors. Combining these contributions gives
\be
\begin{aligned}
G_J(w_1,w_2)=-P_\Theta(v)+\kappa^2\bigg[
&\frac{q_n}{R^2}\partial_{q_n}\log\vartheta_3(0,q_n)\\
&+\frac{R^2q_m}{4}\partial_{q_m}\log\vartheta_3(0,q_m)\bigg].
\end{aligned}
\label{eq:compactcurrent}
\ee
The first term is the oscillator contribution, already encountered in the noncompact theory, while the remaining terms describe the fluctuations of the quantized left-moving charge. These terms depend on the compactification radius and the Euclidean time circumference through the momentum and winding sums.

\begin{figure}[t]
\centering
\includegraphics[width=\textwidth]{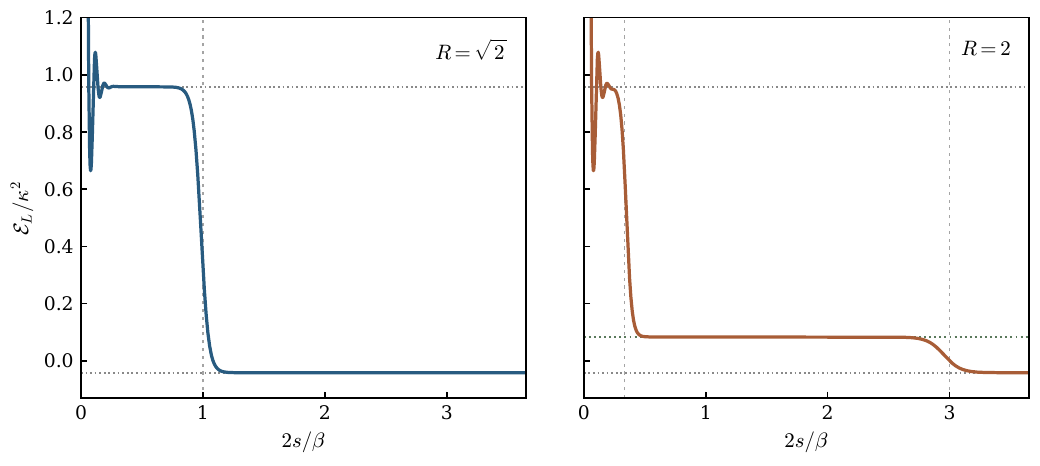}
\caption{Compact-current quench energy $\cE_L/\kappa^2$ at $R=\sqrt2$ (left) and $R=2$ (right), with $\beta=95$. Horizontal dotted lines mark the current, charged-primary (where distinct) and vacuum levels; vertical lines mark the cold-limit crossings. The calculation uses $|n|,|m|\le16$ and $320$ oscillator modes, with $L=10$, $X=5$, $x_0=0$, $\epsilon=0.5$, $\delta=0.35$.}
\label{fig:compactcurrent}
\end{figure}

Substitution of the correlator~(\ref{eq:compactcurrent}) and partition function~(\ref{eq:compactZ}) into the torus Ward identity~(\ref{eq:masterward}) gives the local left-moving energy. Figure~\ref{fig:compactcurrent} compares the current quench at the self-dual radius $R=\sqrt2$ with the generic-radius example $R=2$. The former approaches the vacuum through a single excited-state plateau, whereas the latter develops a distinct intermediate charge-primary plateau in the cold regime.

We next turn to the momentum-vertex quench. As in the noncompact theory, the oscillator contribution follows from the Gaussian contraction of exponential operators. Compactification introduces an additional dependence on the discrete charge sectors, which can be incorporated by summing over the momentum and winding lattice~\cite{DijkgraafVerlindeVerlinde:1988,MasonTuite:2003,CicconeDiPietroSerone:2024}. The resulting two-point function is
\be
\begin{aligned}
G_{V_a}(w_1,w_2)\propto{}&
\left|\frac{\vartheta_1'(0|\tau)}
{\vartheta_1(\pi v/L|\tau)}\right|^{2a^2}\\
&\times\frac{
\vartheta_3\!\left(\frac{\kappa a(v-\bar v)}{2R},q_n\right)
\vartheta_3\!\left(\frac{\kappa aR(v+\bar v)}{4},q_m\right)
}{\vartheta_3(0,q_n)\vartheta_3(0,q_m)}.
\end{aligned}
\label{eq:compactvertex}
\ee
The first factor describes the Gaussian fluctuations, whereas the theta functions in the second factor encode the contributions of the momentum and winding sectors. In particular, the dependence on the vertex charge $a$ reflects the shift of the momentum quantum number induced by the operator insertion.

For the quench geometry, the operators are separated along the Euclidean time direction, so that $v=2is$ and $\bar v=-2is$. At these insertion points, the winding-dependent theta function reduces to its value at zero argument. Nevertheless, when computing the energy density through~(\ref{eq:masterward}), the derivatives of the general two-point function must be taken before imposing the insertion coordinates. The resulting energy profiles for different vertex charges and initial temperatures will be discussed below.

We now examine the late-time behavior of the two quenches. An important distinction from the noncompact theory is that the momentum and winding spectra are discrete, with nonzero excitation gaps for every finite compactification radius. As a result, Euclidean filtering can exponentially suppress excited charge sectors and select the vacuum at sufficiently late times.

More precisely, we consider $t\to\infty$ at fixed $\beta>0$, $R>0$, and observation point $X$. Since $\Theta(t)=\beta+2s(t)$ grows linearly with time, the contributions of nonzero momentum and winding sectors become exponentially small. The leading corrections are determined by the lowest such excitations.

For convenience, we introduce the finite-temperature oscillator sum
\be
g_\beta=\sum_{m=1}^{\infty}m e^{-\kappa m\beta}
=\frac{e^{-\kappa\beta}}{(1-e^{-\kappa\beta})^2}.
\ee
Expanding the current and vertex correlators at large $\Theta$, and evaluating the corresponding torus Ward identities, we obtain
\be
\begin{aligned}
\frac{\cE_{L,J}(X,t)}{\kappa^2}
={}&-\frac1{24}
+\left(\frac1{R^2}+\frac1{g_\beta R^4}\right)
e^{-\kappa\Theta(t)/R^2}\\
&+\left(\frac{R^2}{4}+\frac{R^4}{16g_\beta}\right)
e^{-\kappa R^2\Theta(t)/4}
+O\left(e^{-2\kappa\gamma_R\Theta(t)}\right),\\[3pt]
\frac{\cE_{L,V}(X,t)}{\kappa^2}
={}&-\frac1{24}
+\frac1{R^2}\cosh\left(\frac{2\kappa\beta a}{R}\right)
e^{-\kappa\Theta(t)/R^2}\\
&+\frac{R^2}{4}e^{-\kappa R^2\Theta(t)/4}
+O\left(e^{-2\kappa\gamma_R\Theta(t)}\right),
\qquad t\to\infty,
\end{aligned}
\label{eq:compactlate}
\ee
where $\gamma_R=\min\{R^{-2},R^2/4\}$.

In both cases, the energy density approaches the vacuum Casimir value $\cE_L=-\kappa^2/24$. The two exponential corrections correspond to the lowest nonzero momentum and winding sectors, respectively. The momentum contribution has an exponent proportional to $R^{-2}$, while the winding contribution has an exponent proportional to $R^2/4$. The two exponents coincide at the self-dual radius $R=\sqrt2$.

Since $\Theta(t)\sim2\delta t$, these contributions decay exponentially with Lorentzian time. This behavior contrasts with the algebraic relaxation of the noncompact boson, where the continuous momentum spectrum extends down to the vacuum energy. Compactification replaces the continuous zero-mode spectrum by isolated charge levels, allowing Euclidean filtering to suppress their contributions exponentially. The asymptotic expansion is taken at fixed $\beta$ and is not uniform in the zero-temperature limit.

The strict late-time limit at fixed $\beta$ does not exclude long intermediate plateaux when the initial thermal ensemble is sufficiently cold. The current has nonzero diagonal matrix elements in momentum and winding sectors with $p_L\ne0$. The lightest such primary has total conformal dimension $\gamma_R$ and chiral weight $\gamma_R/2$. Together with the two vacuum--current endpoint channels, it determines the cold lower envelope
\be
I_{*,J}(x)=\min\big\{x,\gamma_R(1+x),1\big\},
\qquad x=\frac{2s}{\beta},\qquad
\gamma_R=\min\big\{R^{-2},R^2/4\big\}.
\label{eq:compactcurrentcost}
\ee
For $R\ne\sqrt2$, the charged-primary branch dominates between the two crossings
\be
x_- = \frac{\gamma_R}{1-\gamma_R},\qquad
x_+ = \frac{1-\gamma_R}{\gamma_R}.
\label{eq:compactcurrentcross}
\ee
The homogeneous left-moving plateaux therefore approach $1-1/24$, $\gamma_R/2-1/24$ and $-1/24$ in units of $\kappa^2$. At $R=2$, the intermediate primary has $h=1/8$, with crossings at $x=1/3$ and $x=3$. At the self-dual radius $R=\sqrt2$, one has $\gamma_R=1/2$ and both crossings coincide at $x=1$, so the intermediate charge sector does not occupy a separate cold interval. At finite temperature the current matrix elements and descendant contributions shift and smooth the transitions, while the strict asymptotic relaxation remains governed by~(\ref{eq:compactlate}).

Momentum-vertex quenches can exhibit longer sequences at either radius. The operator $V_{N/R}$ shifts the integer momentum by $N$, so several input--output charge pairs can compete as the filtering length increases. Their output primary energies decrease successively along the cold lower envelope. 

\begin{figure}[t]
\centering
\includegraphics[width=\textwidth]{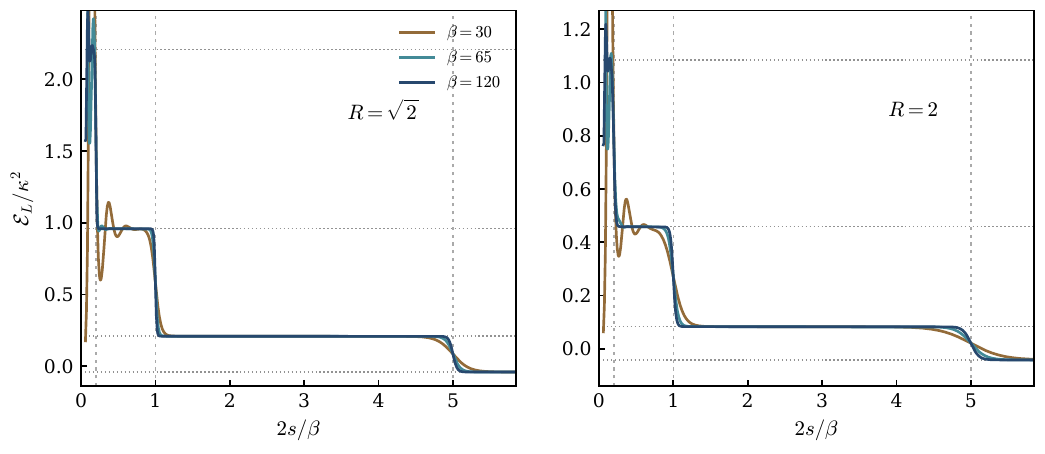}
\caption{Energy $\cE_L/\kappa^2$ after the vertex quench $V_{3/R}$ at $R=\sqrt2$ (left) and $R=2$ (right), for $\beta=30,65,120$. Horizontal dotted lines mark primary levels of output charges $3,2,1,0$; vertical lines mark cold-limit crossings. The sums use $|n|,|m|\le16$ and $320$ oscillator modes; $L=10$, $X=5$, $x_0=0$, $\epsilon=0.5$, $\delta=0.35$.}
\label{fig:compactvertex}
\end{figure}

This mechanism is illustrated in Fig.~\ref{fig:compactvertex}, where the vertex quench with $N=3$ develops a sequence of approximately homogeneous energy plateaux. Their heights approach the energies of primary states with successively smaller momentum charges. The corresponding transitions become more clearly resolved as the initial temperature decreases.

\begin{figure}[t]
\centering
\includegraphics[width=\textwidth]{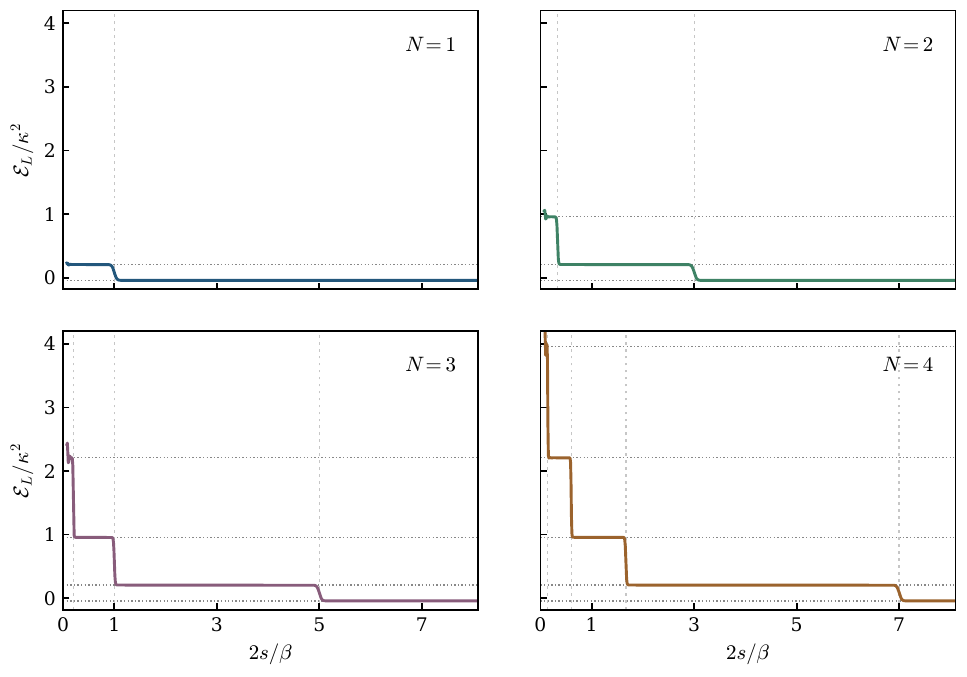}
\caption{Vertex-quench energy $\cE_L/\kappa^2$ for $N=1,2,3,4$ at $R=\sqrt2$, $\beta=140$. Horizontal dotted lines denote primary energy levels and vertical lines the predicted crossings. Other parameters are as in Fig.~\ref{fig:compactvertex}.}
\label{fig:compactcharges}
\end{figure}

The dependence on the vertex charge is shown more systematically in Fig.~\ref{fig:compactcharges}. Increasing $N$ allows a larger number of momentum sectors to participate in the evolution and produces correspondingly longer sequences of primary-energy plateaux. The number and positions of these plateaux are determined by the competition between the thermal weights of the initial states and the Euclidean suppression of the possible output states.

The intermediate plateaux arise on cold but finite filtering intervals, whereas the strict late-time limit is taken at fixed temperature. For every finite $\beta$ and $R$, the eventual relaxation is governed by~(\ref{eq:compactlate}), whereas the sequence of primary plateaux becomes well resolved in the low-temperature regime, where the competing spectral contributions remain separated over extended intervals. The spectral mechanism behind these transitions, including the selection of the dominant charge sectors and the determination of their crossing times, is developed in Sec.~\ref{subsec:effectivechannels} and App.~\ref{app:coldproof}.

\subsection{Free Majorana fermion}
\label{subsec:fermion}

For the free Majorana fermion, the spin structure determines the lowest-energy states accessible to Euclidean filtering. Different spatial boundary conditions consequently produce different limiting energy densities.

We work with the nonchiral free Majorana theory, with left- and right-moving central charges $c_L=c_R=1/2$, and choose the holomorphic fermion $\psi_L$, of conformal weights $(1/2,0)$, as the local quench operator. To define the theory on the torus, one must specify the fermionic boundary conditions along both its spatial and Euclidean time cycles. The ordinary thermal trace, without an insertion of $(-1)^F$, corresponds to antiperiodic boundary conditions along the Euclidean time direction of circumference $\Theta=\beta+2s$.

Along the spatial circle, we impose either antiperiodic boundary conditions in the Neveu--Schwarz (NS) sector or periodic boundary conditions in the Ramond (R) sector~\cite{Ginsparg:1988,DiFrancesco:1997,AlvarezGaume:1986},
\be
\psi_L(x+L)=\begin{cases}
-\psi_L(x),&\mathrm{NS},\\
+\psi_L(x),&\mathrm{R}.
\end{cases}
\label{eq:fermionBC}
\ee
Combined with the antiperiodic thermal boundary condition, these choices define the NS--NS and R--NS spin structures. Their partition functions form a modular-covariant family. The modular-invariant bosonic Ising theory requires a suitable combination of fermionic sectors and is considered below.

The two spatial boundary conditions lead to different fermionic mode expansions. In the NS sector, the allowed mode numbers are half-integers, $r=1/2,3/2,\ldots$, whereas the R sector contains integer modes, $n=1,2,\ldots$, together with a fermionic zero mode. This difference modifies both the ground-state energies and the thermal spectra.

Taking the thermal traces over the corresponding Fock spaces gives
\be
\begin{aligned}
Z_{\mathrm{NS,NS}}
&=\left|q^{-1/48}\prod_{r\in\mathbb Z_{\geq0}+1/2}
(1+q^r)\right|^2
=\left|\frac{\vartheta_3(0|\tau)}{\eta(\tau)}\right|,\\
Z_{\mathrm{R,NS}}
&=2\left|q^{1/24}\prod_{n=1}^\infty(1+q^n)\right|^2
=\left|\frac{\vartheta_2(0|\tau)}{\eta(\tau)}\right|,
\end{aligned}
\label{eq:fermionZ}
\ee
where $q=e^{-\kappa\Theta}$. The factors preceding the oscillator products account for the respective ground-state energies. In particular, the Ramond sector has a ground state of nonzero conformal weight, and the additional factor of $2$ reflects the fermionic zero-mode degeneracy of the nonchiral theory.

The two-point functions of the quench operator can be obtained by summing the thermal contractions of the fermionic modes in each spin structure. Expressing the resulting sums in terms of Jacobi theta functions, we find
\be
G_{\psi,\nu}(w_1,w_2)=\frac{\pi}{L}
\frac{\vartheta_1'(0|\tau)}{\vartheta_1(\pi v/L|\tau)}
\frac{\vartheta_\nu(\pi v/L|\tau)}{\vartheta_\nu(0|\tau)},
\qquad
\begin{cases}\nu=3,&\mathrm{NS},\\\nu=2,&\mathrm{R},\end{cases}
\quad v=w_2-w_1.
\label{eq:fermionG}
\ee
The two expressions have the same general structure, but differ in their dependence on the spin structure through the theta-function ratio. In the Ramond case, the correlator also incorporates the contribution of the fermionic zero mode.

\begin{figure}[t]
\centering
\includegraphics[width=\textwidth]{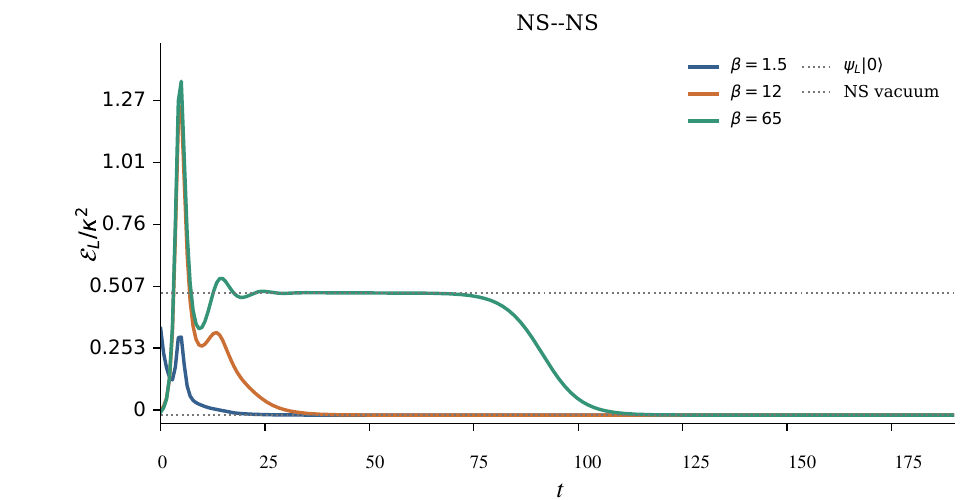}
\caption{Majorana-fermion quench energy $\cE_L/\kappa^2$ in the NS--NS sector at $\beta=1.5,12,65$. Dotted lines mark the excited state $\psi_L|0\rangle$ and NS vacuum. The calculation retains $140$ positive modes and $59$ Fourier terms; $L=10$, $X=5$, $x_0=0$, $\epsilon=0.5$, $\delta=0.35$.}
\label{fig:fermionNS}
\end{figure}

\begin{figure}[t]
\centering
\includegraphics[width=\textwidth]{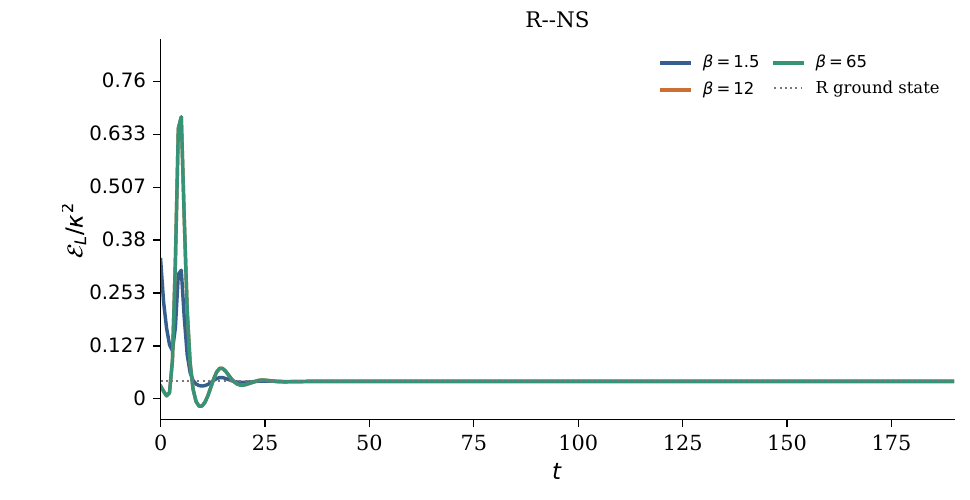}
\caption{Majorana-fermion quench energy $\cE_L/\kappa^2$ in the R--NS sector at $\beta=1.5,12,65$. The dotted line marks the Ramond ground-state level $1/24$. Numerical settings are as in Fig.~\ref{fig:fermionNS}.}
\label{fig:fermionR}
\end{figure}

Substituting these correlators and the corresponding partition functions into the torus Ward identity~(\ref{eq:masterward}), we obtain the left-moving energy density following the local fermion quench. The resulting profiles for different initial temperatures are shown in Figs.~\ref{fig:fermionNS} and~\ref{fig:fermionR}.

To understand the late-time behavior, we consider the limit $t\to\infty$ at fixed initial inverse temperature $\beta>0$ and observation point $X$. As the Euclidean filtering parameter $s(t)$ increases, contributions from higher-energy states are exponentially suppressed. The asymptotic energy density is therefore controlled by the lowest-energy states accessible within each spatial spin sector.

Expanding the corresponding torus correlators at large $s(t)$, we obtain
\be
\begin{aligned}
\frac{\cE_{L,\psi}^{\mathrm{NS}}(X,t)}{\kappa^2}
&=-\frac1{48}
+\frac12\left(e^{\kappa\beta/2}-e^{-\kappa\beta/2}
+e^{-3\kappa\beta/2}\right)e^{-\kappa s(t)}
+O(e^{-2\kappa\delta t}),\\
\frac{\cE_{L,\psi}^{\mathrm R}(X,t)}{\kappa^2}
&=\frac1{24}+O(e^{-\kappa s(t)}),\qquad t\to\infty.
\end{aligned}
\label{eq:fermasymp}
\ee
Both sectors exhibit exponential relaxation, reflecting the discreteness of their excitation spectra. However, their limiting energy densities are different because the two spin structures have different ground states.

In the NS sector, the lowest-energy state is the vacuum, with conformal weight $h=0$. Accordingly, the energy density eventually approaches the Casimir value $\cE_L=-\kappa^2/48$. At sufficiently low initial temperatures, the evolution can nevertheless display an extended intermediate regime associated with the state $\psi_L|0\rangle$. Since the fermion has conformal weight $h=1/2$, this state carries the chiral energy $\cE_L=\kappa^2(23/48)$. The Euclidean filtering initially preserves this excited-state contribution before progressively favoring the vacuum, producing a transition between the two energy levels.

The Ramond sector behaves differently. Its periodic spatial boundary conditions introduce a fermionic zero mode, and the lowest-energy states have conformal weight $h_{\mathrm R}=1/16$. Consequently, the filtered energy density approaches the nonzero level $\cE_L=\kappa^2/24$, rather than the NS vacuum value. The zero mode allows the fermion operator to act within the degenerate Ramond ground-state subspace, so the long-time filtered state can retain support on this lowest-energy sector.

The distinct limiting energies of the NS and R sectors are determined by their accessible lowest-energy states, despite identical local fermionic dynamics.

\subsection{Ising model}
\label{subsec:ising}

The critical Ising CFT provides an example of how distinct quench operators yield different energy-relaxation patterns in the same theory. Although the Ising CFT admits a description in terms of free Majorana fermions, here we consider its bosonic modular-invariant formulation, obtained through an appropriate sum over fermionic spin structures and a fermion-parity projection. Unlike the fixed-spin-structure theories discussed in the previous subsection, this formulation contains three local bulk primary fields, namely the identity $\mathbf1$, the spin field $\sigma$ with conformal weights $(1/16,1/16)$, and the energy field $\varepsilon$ with weights $(1/2,1/2)$. The chiral Majorana fermion $\psi_L$ is not an ordinary local bulk primary of the bosonic theory.

The Ising model has central charge $c=1/2$, and its modular-invariant torus partition function is expressed in terms of the three Virasoro characters as~\cite{DiFrancesco:1997,DiFrancescoSaleurZuber:1987}
\be
\begin{aligned}
Z_I(\tau,\bar\tau)
&=|\chi_{\mathbf1}|^2+|\chi_\sigma|^2
+|\chi_\varepsilon|^2\\
&=\frac{|\vartheta_2(0|\tau)|+|\vartheta_3(0|\tau)|
+|\vartheta_4(0|\tau)|}{2|\eta(\tau)|}.
\end{aligned}
\label{eq:isingZ}
\ee
We investigate local quenches generated by the spin and energy operators, $\cO_\alpha=\sigma,\varepsilon$. These two choices are particularly instructive because their fusion rules allow different transitions between primary states. As a result, Euclidean filtering can select different sequences of dominant conformal families, leading to qualitatively different energy profiles.

To calculate the energy density, we need the corresponding torus two-point functions. These can be constructed from the fermionic propagators and the spin-field correlation functions, with the appropriate combination of spin structures~\cite{DiFrancescoSaleurZuber:1987,DiFrancesco:1997}. Introducing the dimensionless separation $u=\pi(w_2-w_1)/L$, the correlators take the form
\be
\begin{aligned}
G_\sigma(u,\tau)&\propto
\left|\frac{\vartheta_1'(0|\tau)}{\vartheta_1(u|\tau)}\right|^{1/4}
\frac{\sum_{\nu=1}^4|\vartheta_\nu(u/2|\tau)|}
{\sum_{\nu=2}^4|\vartheta_\nu(0|\tau)|},\\
G_\varepsilon(u,\tau)&\propto
\left|\frac{\vartheta_1'(0|\tau)}{\vartheta_1(u|\tau)}\right|^2
\frac{\sum_{\nu=2}^4|\vartheta_\nu(u|\tau)|^2/
|\vartheta_\nu(0|\tau)|}
{\sum_{\nu=2}^4|\vartheta_\nu(0|\tau)|}.
\end{aligned}
\label{eq:isingcorr}
\ee
Here we have suppressed overall normalization factors independent of the insertion coordinates and the torus modulus, since they cancel in the logarithmic derivatives entering the Ward identity. The different powers of the theta-function prefactors reflect the conformal dimensions of the two quench operators, while the remaining factors encode the dependence on the torus geometry and the contributions of the fermionic sectors.

\begin{figure}[t]
\centering
\includegraphics[width=0.86\textwidth]{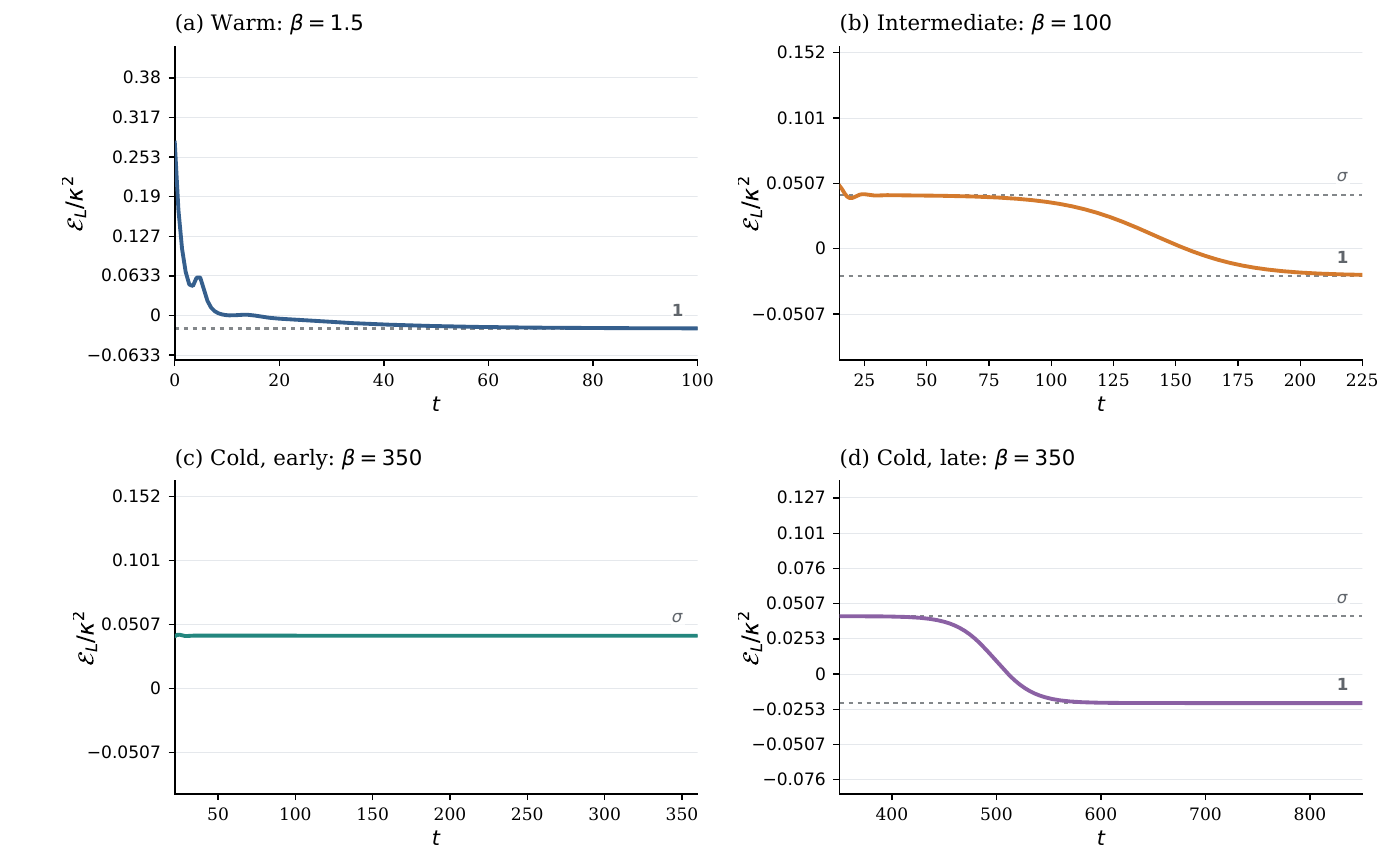}
\caption{Ising spin-field quench: $\cE_L/\kappa^2$ for $\beta=1.5,100,350$, with separate early- and late-time panels. Dotted lines mark the $\sigma$ and vacuum levels. Parameters: $L=10$, $X=5$, $x_0=0$, $\epsilon=0.5$, $\delta=0.35$; $55$-digit theta functions, $89$ Ward Fourier modes and derivative step $8\times10^{-5}$.}
\label{fig:isingsigma}
\end{figure}

\begin{figure}[t]
\centering
\includegraphics[width=\textwidth]{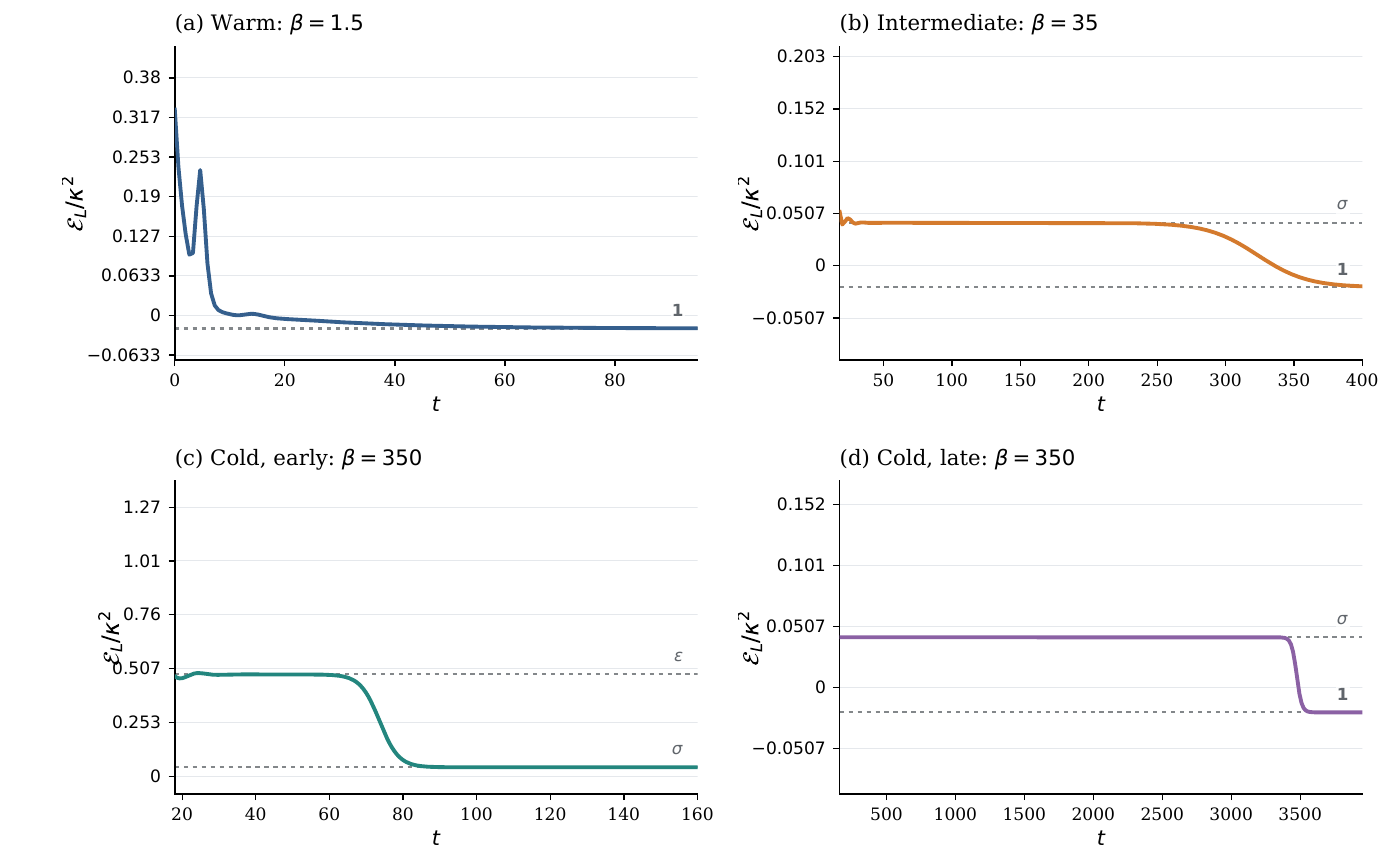}
\caption{Ising energy-field quench: $\cE_L/\kappa^2$ at $\beta=1.5,35,350$. Dotted lines mark the $\varepsilon$, $\sigma$ and vacuum levels, $23/48$, $1/24$ and $-1/48$, respectively. Numerical settings are as in Fig.~\ref{fig:isingsigma}.}
\label{fig:isingepsilon}
\end{figure}

Substituting the partition function~(\ref{eq:isingZ}) and the appropriate two-point function from~(\ref{eq:isingcorr}) into the torus Ward identity~(\ref{eq:masterward}), we obtain the corresponding local energy densities. Their time evolution for several initial temperatures is shown in Figs.~\ref{fig:isingsigma} and~\ref{fig:isingepsilon}.

We first consider the spin-field quench, $\cO=\sigma$. As shown in Fig.~\ref{fig:isingsigma}, the behavior depends strongly on the initial temperature. For a relatively warm thermal state, the energy generated by the local insertion relaxes toward the vacuum value without developing a clearly resolved intermediate plateau. As the initial temperature is lowered, however, the evolution exhibits an extended regime in which the local energy becomes approximately homogeneous and approaches the energy of the spin primary,
\be
\cE_L=\kappa^2\left(h_\sigma-\frac{c}{24}\right)
=\frac{\kappa^2}{24}.
\ee
This regime arises because, in a sufficiently cold initial ensemble, the dominant contribution initially comes from the excitation of the vacuum by the spin operator. As the Euclidean filtering increases, the relative importance of this contribution decreases, and the vacuum eventually becomes the dominant output state. The spin quench consequently yields a single nonvacuum primary plateau before relaxing to $\cE_L=-\kappa^2/48$, as expected from the restricted primary transitions allowed by its fusion rules.

The energy-field quench, $\cO=\varepsilon$, exhibits a richer structure. Figure~\ref{fig:isingepsilon} shows that, while a warm initial state again relaxes directly toward the vacuum after the local energy pulse, lowering the temperature allows intermediate primary levels to become visible. At moderately low temperatures, the evolution develops a plateau associated with the spin primary. For sufficiently cold initial states, an additional earlier plateau appears near the energy of the $\varepsilon$ primary,
\be
\cE_L=\kappa^2\left(h_\varepsilon-\frac{c}{24}\right)
=\frac{23\kappa^2}{48}.
\ee
The subsequent evolution passes through the spin-primary level $\cE_L=\kappa^2/24$ before eventually approaching the vacuum value $-\kappa^2/48$.

The appearance of the intermediate spin plateau can be understood from the Ising fusion rule $\varepsilon\times\sigma=\sigma$. Unlike the spin operator, the energy operator allows a diagonal transition within the spin conformal family. Consequently, thermally populated spin states can contribute to the filtered evolution while remaining in the same conformal family. At sufficiently low temperatures, this contribution becomes dominant over an extended time interval, separating the initial $\varepsilon$ regime from the final vacuum regime.

The two quenches therefore resolve different numbers of nonvacuum primary plateaux within the same Ising spectrum, reflecting the operator-dependent matrix elements that enter the filtered evolution.

We now examine the asymptotic relaxation at fixed initial inverse temperature $\beta>0$. In this limit, the Euclidean filtering parameter $s(t)$ grows without bound, and the contributions of excited output states become exponentially suppressed. For both quench operators, the vacuum is accessible through a nonvanishing matrix element from a thermally populated input state. The energy density therefore eventually approaches the identity level, independently of the intermediate plateau structure.

The leading correction to this limiting value is controlled by the lightest nonvacuum output state that couples to the quench operator. In the bosonic Ising model, this is the spin primary, with total conformal dimension $\Delta_\sigma=1/8$. To express its contribution, we introduce the thermal channel weights
\be
A_d^{(\alpha)}(\beta)=\sum_a e^{-\kappa\beta\Delta_a}
\big|\langle d|\cO_\alpha|a\rangle\big|^2,
\qquad d=0,\sigma,
\label{eq:isingthermalweights}
\ee
where $\alpha=\sigma,\varepsilon$, and the sum runs over the thermally populated input states, including their Virasoro descendants. These coefficients measure the total thermal weight of transitions into the vacuum and spin-primary output states, respectively.

At large $s(t)$, the relative contribution of the spin-primary output is suppressed by a factor $e^{-2\kappa s(t)\Delta_\sigma}=e^{-\kappa s(t)/4}$. The corresponding late-time expansion of the left-moving energy density is
\be
\frac{\cE_{L,\alpha}(X,t)}{\kappa^2}
=-\frac1{48}+\frac{A_\sigma^{(\alpha)}(\beta)}
{16A_0^{(\alpha)}(\beta)}
e^{-\kappa s(t)/4}
+o(e^{-\kappa\delta t/4}),
\qquad t\to\infty.
\label{eq:isinglate}
\ee
The spin-primary contribution fixes the same leading relaxation exponent for both quenches, while the amplitudes differ through the operator-dependent thermal matrix elements.

To make this dependence more explicit, we consider the low-temperature regime, where the leading primary transitions provide a useful approximation to the thermal channel weights. Keeping the dominant contributions gives
\be
\frac{A_\sigma^{(\sigma)}}{16A_0^{(\sigma)}}
\simeq\frac{e^{\kappa\beta/8}}{16}
\left(1+\frac14e^{-\kappa\beta}\right),
\qquad
\frac{A_\sigma^{(\varepsilon)}}{16A_0^{(\varepsilon)}}
\simeq\frac{e^{7\kappa\beta/8}}{64}.
\label{eq:isinglatecold}
\ee
The stronger temperature dependence of the energy-quench coefficient reflects the distinct thermal input states that enter its competing output channels.

The coefficients in~(\ref{eq:isingthermalweights}) retain descendant contributions at finite temperature, while the cold primary approximation isolates the dominant channels. Section~\ref{sec:effective} develops this selection in general and determines the ordering and crossover times of the Ising plateaux from the fusion rules.

\subsection{\texorpdfstring{$SU(2)_k$}{SU(2)k} WZW model}
\label{subsec:wzw}

We now consider the $SU(2)_k$ Wess--Zumino--Witten (WZW) model, which extends our analysis to a conformal field theory with a non-Abelian current algebra. In contrast to the free-field examples discussed above, the spectrum and correlation functions of this theory are constrained by an extended affine symmetry. This provides a useful setting for investigating how the representation structure of a conformal field theory influences the energy relaxation and the emergence of intermediate plateaux under Euclidean filtering.

The $SU(2)_k$ WZW model is characterized by a positive integer level $k$. Its three holomorphic currents $J^a$, $a=1,2,3$, generate the affine Kac--Moody algebra $\widehat{\mathfrak{su}}(2)_k$, with commutation relations
\be
[J_m^a,J_n^b]=i\epsilon^{abc}J_{m+n}^c+\frac{k}{2}m\delta^{ab}\delta_{m+n,0}.
\ee
The stress tensor is constructed from the currents through the Sugawara construction,
$T=(k+2)^{-1}\sum_a:J^aJ^a:$,
which gives the central charge $c_k=3k/(k+2)$.

The integrable highest-weight representations of the affine algebra are labeled by the spin
\be
j=0,\frac12,\ldots,\frac{k}{2},\qquad
h_j=\frac{j(j+1)}{k+2}.
\label{eq:wzwprimaries}
\ee
In the diagonal theory, each affine representation gives rise to a multiplet of bulk primary fields $\Phi^{(j)}_{m,\bar m}$ with conformal weights $(h_j,h_j)$, where $m,\bar m=-j,\ldots,j$. Together with their affine descendants, these multiplets span the Hilbert space. Thus, the theory contains $k+1$ integrable affine-primary families, although its decomposition into Virasoro representations involves infinitely many Virasoro primaries, since affine descendants can themselves be primary with respect to the Virasoro algebra~\cite{Witten:1984,KnizhnikZamolodchikov:1984,Ginsparg:1988,DiFrancesco:1997}.

We study a local quench generated by one of the holomorphic currents. Although the currents belong to the affine vacuum module as descendants, they are Virasoro primary fields with conformal weights $(1,0)$. The diagonal thermal ensemble is invariant under the global $SU(2)$ symmetry, implying that the two-point functions of $J^1$, $J^2$, and $J^3$ coincide up to their common normalization. We may therefore choose the Cartan current
$j^3=\sqrt{2/k}\,J^3$
without loss of generality.

This choice is particularly convenient because the Cartan zero mode acts diagonally on states within each spin multiplet. With the chosen normalization, the current modes satisfy
$[j^3_n,j^3_m]=n\delta_{n+m,0}$.
The current correlator can consequently be separated into a contribution from the nonzero modes, analogous to the oscillator contribution of a free boson, and a contribution from the zero mode, which carries information about the affine representations.

The torus two-point function can be obtained from the affine Ward identities and the thermal trace, following the current-algebra approach of~\cite{Bernard:1988,MathurMukhi:1988}. The nonzero current modes give a universal oscillator kernel, while the zero-mode contribution is determined by the thermal distribution of Cartan charges within the affine multiplets. These contributions can be expressed in terms of the affine characters.

Introducing $q=e^{-\kappa\Theta}$, the diagonal torus partition function and the normalized current two-point function take the form
\be
\begin{aligned}
Z_k(\tau,\bar\tau)&=\sum_{j=0}^{k/2}|\chi_j(q)|^2,\\
G_{j^3}(w_1,w_2)&=\kappa^2\left[g_\Theta(w_2-w_1)
+\frac1{Z_k}\sum_{j=0}^{k/2}|\chi_j(q)|^2M_j(q)\right].
\end{aligned}
\label{eq:wzwcorr}
\ee
Here $\chi_j(q)=\chi_j(q,1)$ is the character of the affine representation of spin $j$. The quantities $M_j(q)$ describe the corresponding normalized second moments of the Cartan charge and can be obtained by introducing a chemical-potential variable $y$ into the characters,
\be
M_j(q)=\frac{2}{k}\left.\frac{(y\partial_y)^2\chi_j(q,y)}{\chi_j(q,1)}\right|_{y=1},
\label{eq:wzwmoment}
\ee
where
$\chi_j(q,y)=\Tr_{\mathcal H_j}\big(q^{L_0-c_k/24}y^{J^3_0}\big)$.
Thus, the zero-mode contribution is determined by the charge fluctuations within each affine representation, weighted by its contribution to the thermal partition function.

The oscillator contribution is encoded in the kernel
\be
\begin{aligned}
g_\Theta(u+iy)=\sum_{n=1}^\infty \frac{n}{1-e^{-\kappa n\Theta}}
\Big[e^{-\kappa ny+i\kappa nu}
+e^{-\kappa n(\Theta-y)-i\kappa nu}\Big],\qquad 0<y<\Theta.
\end{aligned}
\label{eq:wzwgtheta}
\ee
This series follows from the commutation relations of the normalized Cartan current and the cyclicity of the thermal trace. The two exponential terms account for propagation along the complementary Euclidean segments of the thermal circle. The resulting expression is absolutely convergent for $0<y<\Theta$.

The decomposition in~(\ref{eq:wzwcorr}) separates the oscillator kernel of nonzero current modes from the character moments of the affine charge spectrum. Both contributions enter the energy density through the torus Ward identity~(\ref{eq:masterward}).

For numerical evaluation, the affine characters and charge moments can be expressed using the Weyl--Kac character formula. Introducing the sums
\be
\begin{aligned}
A_r^{(p,a)}(q)&=\sum_{n\in\mathbb Z}(2pn+a)^r q^{(2pn+a)^2/(4p)},\\
\chi_j(q)&=\frac{A_1^{(k+2,2j+1)}(q)}{A_1^{(2,1)}(q)},\\
M_j(q)&=\frac{1}{6k}\left[
\frac{A_3^{(k+2,2j+1)}}{A_1^{(k+2,2j+1)}}-
\frac{A_3^{(2,1)}}{A_1^{(2,1)}}\right],
\end{aligned}
\label{eq:wzwcharacter}
\ee
we obtain explicit expressions for the quantities entering the current two-point function. All $k+1$ integrable affine representations are included in the calculation. Substituting these expressions into~(\ref{eq:masterward}) then determines the local energy density following the Cartan-current quench.

An important feature of the WZW model is that the current operator can connect thermally populated states within the same affine representation. In particular, the lightest nontrivial affine-primary states belong to the spin-$1/2$ multiplet, whose total conformal dimension is
$\Delta_{1/2}=3/[2(k+2)]$.
The Cartan zero mode acts on these states without changing their affine representation, allowing a diagonal contribution from the spin-$1/2$ sector to the filtered dynamics.

At sufficiently low initial temperatures, this contribution can compete with those associated with the current-excited vacuum state $j^3|0\rangle$ and the vacuum itself. As the Euclidean filtering increases, the relative weights of these contributions change, potentially producing three successive regimes in which the energy density approaches the levels of these states.

The corresponding chiral energies are
$\cE_L=\kappa^2(1-c_k/24)$
for the current-excited vacuum,
\be
\frac{\cE_{L,1/2}}{\kappa^2}=\frac{3}{4(k+2)}-\frac{c_k}{24},
\label{eq:wzwspin}
\ee
for the spin-$1/2$ primary multiplet, and
$\cE_L=-\kappa^2c_k/24$
for the vacuum. These levels provide the characteristic values around which approximately homogeneous energy plateaux may develop.

The resulting behavior depends on both the initial temperature and the affine level $k$. For $k>1$, sufficiently cold preparations can resolve an intermediate spin-$1/2$ plateau between the current and vacuum regimes. At $k=1$, however, the spin-$1/2$ contribution does not dominate over an extended interval in the leading cold-channel approximation, and the corresponding intermediate plateau is absent. The origin of this distinction will be explained through the spectral selection criterion in Sec.~\ref{sec:effective}.

\begin{figure}[t]
\centering
\includegraphics[width=\textwidth]{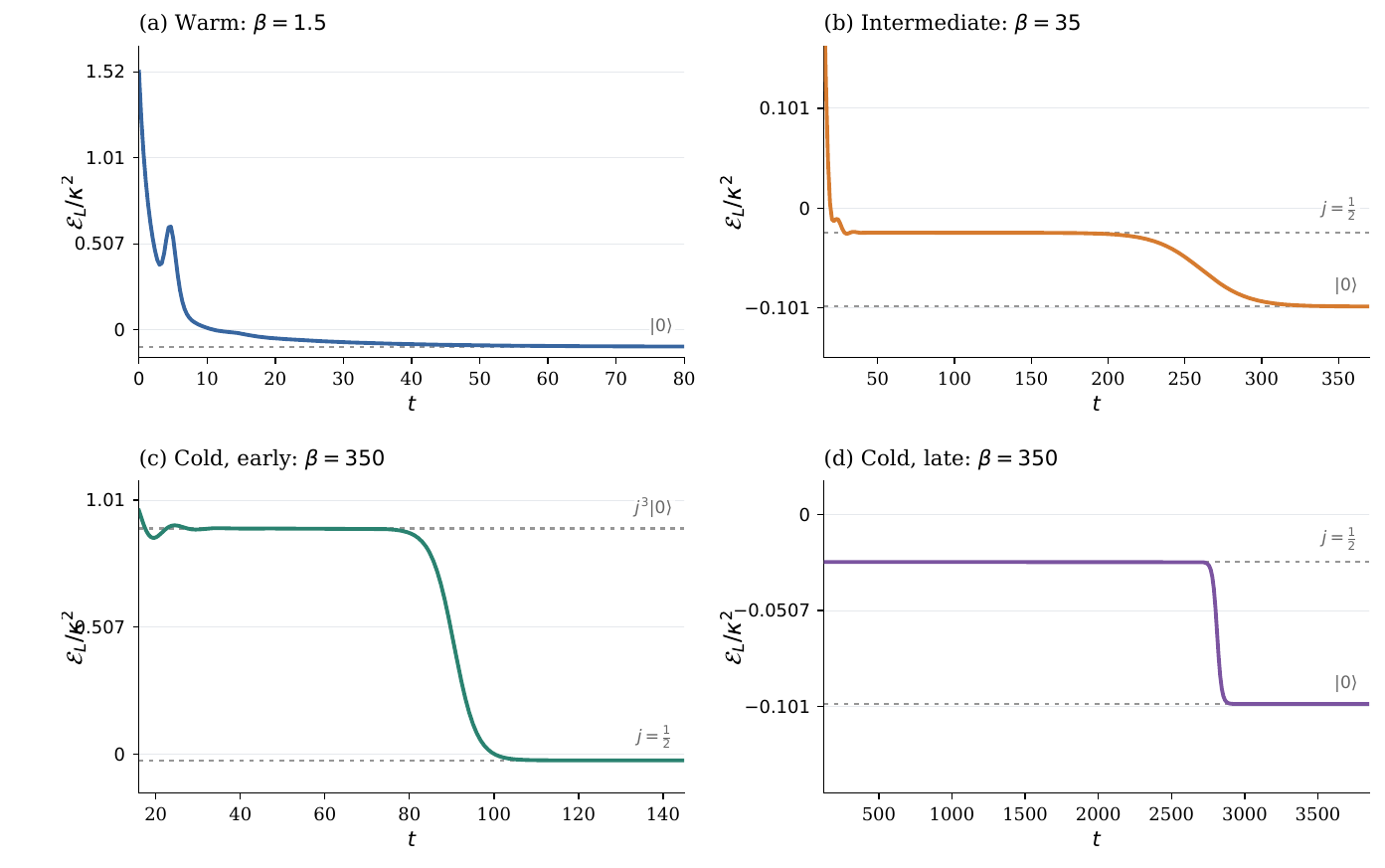}
\caption{Cartan-current quench energy $\cE_L/\kappa^2$ in the $SU(2)_8$ WZW model at $\beta=1.5,35,350$. Dotted lines mark the current, spin-$1/2$ and vacuum levels. The calculation retains $220$ oscillator modes, $59$ Ward Fourier terms and Weyl--Kac terms through $|n|\le12$; $L=10$, $X=5$, $x_0=0$, $\epsilon=0.5$, $\delta=0.35$.}
\label{fig:wzwtemp}
\end{figure}

Figure~\ref{fig:wzwtemp} illustrates the temperature dependence for the $SU(2)_8$ model. At relatively high initial temperature, the energy generated by the local current insertion relaxes toward the vacuum without a clearly resolved primary plateau. At lower temperature, an intermediate regime associated with the spin-$1/2$ multiplet becomes visible. For a sufficiently cold preparation, the evolution resolves both the initial current plateau and the intermediate spin-$1/2$ plateau before approaching the vacuum energy.

We next examine the strict late-time limit at fixed inverse temperature $\beta>0$. As the filtering parameter $s(t)$ grows, the contributions of excited output states are progressively suppressed. Since the vacuum is accessible through the current insertion acting on thermally populated states, the energy density eventually approaches the vacuum Casimir value $-\kappa^2c_k/24$. The leading correction is governed by the lightest accessible nonvacuum output, namely the spin-$1/2$ primary multiplet.

To describe this contribution, we introduce the thermal channel weights
\be
B_j(\beta)=\sum_{m_L,m_R=-j}^{j}\sum_a
 e^{-\kappa\beta\Delta_a}
\big|\langle j,m_L,m_R|j^3|a\rangle\big|^2.
\label{eq:wzwthermalcoeff}
\ee
Here the sum over $a$ includes the thermally populated input states, while the sum over $m_L,m_R$ accounts for the components of the output spin-$j$ multiplet. In the $j=0$ output sector, the selected state is the vacuum. The corresponding coefficient $B_0(\beta)$ receives contributions from the input states that can be connected to the vacuum by the current operator.

In terms of these coefficients, the leading late-time behavior is
\be
\begin{aligned}
\frac{\cE_{L,j^3}(X,t)}{\kappa^2}
&=-\frac{c_k}{24}+\frac{3B_{1/2}(\beta)}{4(k+2)B_0(\beta)}
 e^{-\frac{3\kappa}{k+2}(\epsilon+\delta t)}
+o\!\left(e^{-3\kappa\delta t/(k+2)}\right),\\
\frac{3B_{1/2}(\beta)}{4(k+2)B_0(\beta)}
&\simeq \frac{3}{2k(k+2)}
 e^{\kappa\beta(1-3/[2(k+2)])},
\qquad \kappa\beta\gg1.
\end{aligned}
\label{eq:wzwlate}
\ee
The first term is the vacuum Casimir energy. The exponentially suppressed correction describes the residual contribution of the spin-$1/2$ output sector, with a decay exponent determined by its total conformal dimension $\Delta_{1/2}$. The coefficient depends on the relative thermal weights of the spin-$1/2$ and vacuum output channels and therefore retains information about the initial ensemble and the matrix elements of the quench operator.

The first line of~(\ref{eq:wzwlate}) includes the thermal contributions through the coefficients $B_j(\beta)$, whereas the second gives their leading low-temperature approximation. In particular, the exponential dependence on $\beta$ reflects the different thermal costs of the input states contributing to the two competing output channels. The asymptotic expansion is understood at fixed $\beta$, while the cold approximation isolates the leading primary contributions to these coefficients.

\begin{figure}[t]
\centering
\includegraphics[width=\textwidth]{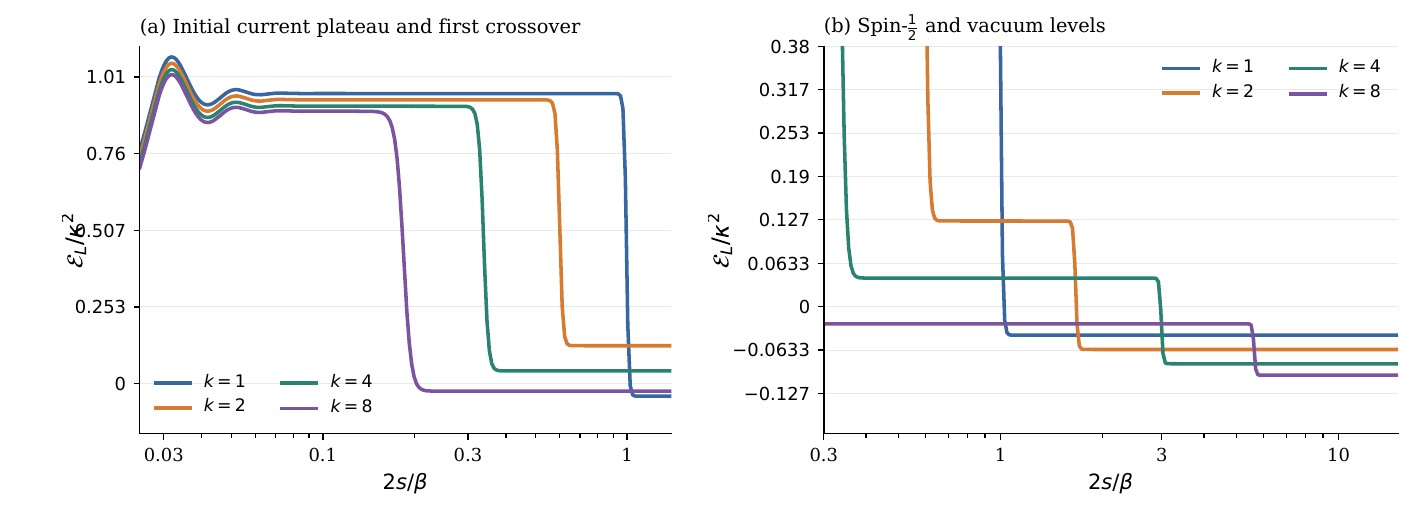}
\caption{Cartan-current quench energy $\cE_L/\kappa^2$ in $SU(2)_k$ at $k=1,2,4,8$ and $\beta=350$. The two panels resolve the current regime and the lower spin-$1/2$/vacuum levels on the $2s/\beta$ scale. All $k+1$ affine sectors are included; other parameters are as in Fig.~\ref{fig:wzwtemp}.}
\label{fig:wzwlevels}
\end{figure}

The dependence on the affine level is illustrated in Fig.~\ref{fig:wzwlevels}, which compares the energy evolution for $k=1,2,4,8$ at the same low initial temperature. The results show that the intermediate spin-$1/2$ regime is absent at $k=1$, while for larger levels the current, spin-$1/2$, and vacuum contributions can dominate over distinct time intervals. Increasing $k$ changes both the energy of the spin-$1/2$ primary and its competition with the other spectral channels, allowing the intermediate plateau to extend over a longer interval.

The Cartan zero mode permits diagonal transitions within affine representations. The cold-channel analysis in Sec.~\ref{sec:effective} determines when the resulting intermediate spin-$1/2$ regime dominates.

\section{Conformal data and the hierarchy of energy plateaux}
\label{sec:effective}

\subsection{Primary-channel selection and plateau transitions}
\label{subsec:effectivechannels}

The torus examples in Sec.~\ref{sec:torus} show that sufficiently cold local quenches can produce extended intervals of approximately homogeneous energy density. The Ising and $SU(2)_k$ current quenches exhibit intermediate primary levels, while compact-boson vertices with higher charge generate longer sequences. These observations suggest a common mechanism in which thermal preparation and Euclidean filtering select different accessible output states over successive intervals.

The spectra and operator matrix elements both affect this selection. The NS--NS Majorana quench relaxes from an excited regime to the vacuum, while the Ramond zero mode retains the lowest-energy Ramond sector. The continuous momentum spectrum of the noncompact boson leads to the algebraic corrections found in~(\ref{eq:ncasymp}). We therefore seek a description that identifies the allowed primary channels, bounds the number contributing to the leading cold sequence and determines the plateau heights and crossover scales from conformal data.

The starting point is the exact spectral decomposition of the torus two-point function. Inserting complete sets of energy eigenstates into the trace representation~(\ref{eq:Q}), we obtain
\be
G_{\mathbb T^2}(\beta,s)=\frac{1}{Z_{\mathbb T^2}(\beta+2s)}
\sum_{a,d}e^{-\beta E_a-2sE_d}
\big|\langle d|\cO(x_0)|a\rangle\big|^2.
\label{eq:spectral}
\ee
Here $a$ and $d$ label the initial and final energy eigenstates appearing in the spectral decomposition, including all Virasoro descendants. The factor $e^{-\beta E_a}$ determines the thermal weight of the input state, while $e^{-2sE_d}$ describes the suppression of the output state by Euclidean filtering. The matrix element $\langle d|\cO(x_0)|a\rangle$ determines whether the corresponding transition is allowed and with what strength it contributes.

Every term in~(\ref{eq:spectral}) is nonnegative, so the relative importance of the channels is determined directly by their exponential suppression and operator matrix elements. This representation allows us to examine both the strict late-time limit and the finite filtering intervals that support intermediate plateaux.

We first consider the strict late-time limit at fixed inverse temperature $\beta>0$. Suppose that the vacuum is an accessible output state, meaning that at least one thermally populated input state has a nonvanishing matrix element with the vacuum. Separating this contribution from the remaining terms in~(\ref{eq:spectral}), we can write
\be
G_{\mathbb T^2}(\beta,s)=\frac{e^{-2sE_{\rm vac}}}{Z_{\mathbb T^2}(\beta+2s)}
\left[W_0(\beta)+\sum_{d\ne0}W_d(\beta)e^{-2\kappa s\Delta_d}\right],
\qquad W_0(\beta)>0.
\label{eq:vacuumselection}
\ee
The coefficients $W_d(\beta)$ collect the thermally weighted matrix elements connecting the initial ensemble to the corresponding output states. Increasing $s$ suppresses the contributions of states with larger conformal dimensions.

For a discrete spectrum with a positive gap above the lowest accessible output, this suppression leads to exponential relaxation of the spatially averaged energy. The torus Ward identity also contains spatially dependent contributions, which decay as the Euclidean filtering becomes large. Consequently, when the vacuum is accessible, the local chiral energy approaches the vacuum Casimir level,
$\cE_L(X,t)\to-\kappa^2c/24$,
as $s(t)\to\infty$.

At fixed $\beta$, the leading corrections have the general structure
$O(e^{-2\kappa\delta t\Delta_{\rm gap}})
+O(e^{-\kappa\delta t})$,
where $\Delta_{\rm gap}$ is the smallest positive conformal dimension among the accessible output states. Their amplitudes depend on the thermal matrix elements of the quench operator. If the vacuum is not accessible, the lowest allowed output replaces it as the limiting state. This occurs, for example, in the Ramond sector of the Majorana theory. In contrast, a continuous spectrum extending to the lowest energy can produce power-law rather than exponential relaxation.

The Euclidean filtering also imposes a general constraint on the energy evolution. The expectation value of the output Hamiltonian can be obtained by differentiating the unnormalized spectral trace,
\be
\begin{gathered}
\langle H_{\rm out}\rangle=-\partial_{2s}\log
\left[Z_{\mathbb T^2}(\beta+2s)G_{\mathbb T^2}(\beta,s)\right],\\
\partial_{2s}\langle H_{\rm out}\rangle
=-\left(\langle H^2\rangle-\langle H\rangle^2\right)\le0.
\end{gathered}
\label{eq:outputvariance}
\ee
The second relation follows directly from the positivity of the energy variance and holds independently of any low-temperature approximation. It shows that the spatially averaged output energy is a nonincreasing function of the filtering parameter. The local energy density, however, need not evolve monotonically at every observation point, since the stress tensor can contain transient spatially dependent contributions. Thus, the monotonicity of the averaged energy is compatible with the local pulses and oscillations observed in the explicit examples.

Although the strict late-time limit selects the lowest accessible output, it does not determine whether intermediate states can dominate over extended intervals. To investigate these regimes, we must consider the competition between thermal preparation and Euclidean filtering on comparable scales.

We now restrict attention to a fixed compact unitary CFT with a discrete spectrum and finite multiplicities below every finite energy. We take the quench operator $\cO$ to be a Virasoro primary and assume that its Euclidean two-point trace converges at positive separations. Using the finite-volume scale defined in Sec.~\ref{sec:torus}, we consider the simultaneous low-temperature and long-filtering limit
\be
b=\kappa\beta\longrightarrow\infty,\qquad
Y=2\kappa s\longrightarrow\infty,\qquad
x=\frac{Y}{b}=\frac{2s}{\beta}>0\quad\text{fixed}.
\label{eq:scaled}
\ee
In this scaling limit, both Euclidean intervals become large while their ratio $x$ remains finite. The competition between spectral channels persists because the initial thermal suppression and the subsequent Euclidean filtering contribute at the same exponential order.

The positivity of~(\ref{eq:spectral}), together with the assumed spectral discreteness and convergence of the Euclidean trace, ensures that the leading exponential behavior is governed by a lower envelope of spectral costs. The convergence is uniform on closed intervals of $x>0$, even when the theory contains infinitely many primary representations.

Moreover, for a Virasoro-primary quench operator, the Ward identities relate matrix elements involving descendants to the corresponding primary three-point coefficients. A vanishing primary coupling therefore eliminates the associated descendant transitions, while nonzero descendant levels increase the exponential cost. As a result, the leading cold spectral envelope can be determined entirely from pairs of primary states connected by nonvanishing matrix elements of $\cO$.

After removing the common Casimir-energy factors, the primary contribution to the torus correlator takes the form
\be
\begin{aligned}
G_{\mathbb T^2}^{\rm prim}(\beta,s)
&=\frac{\kappa^{2\Delta_{\cO}}e^{\kappa c(\beta+2s)/12}}
{Z_{\mathbb T^2}(\beta+2s)}
\sum_{(a,d)\in\mathcal A_{\cO}}
A_{ad}\,e^{-b\Delta_a-Y\Delta_d},\\
A_{ad}&=\sum_\mu\big|C_{a\cO d}^{(\mu)}\big|^2>0,
\qquad \Delta_a=h_a+\bar h_a.
\end{aligned}
\label{eq:effectiveweights}
\ee
Here $\Delta_{\cO}=h+\bar h$ is the total conformal dimension of the quench operator. The set $\mathcal A_{\cO}$ consists of all pairs of primary states for which the corresponding operator matrix elements are nonzero, and $A_{ad}$ denotes the sum of their squared three-point coefficients, including the relevant multiplicities.

In the scaling limit~(\ref{eq:scaled}), each primary transition contributes an exponential weight proportional to $e^{-bI_{ad}(x)}$, where
\be
I_{ad}(x)=\Delta_a+x\Delta_d,
\qquad
I_*(x)=\min_{(a,d)\in\mathcal A_{\cO}}I_{ad}(x).
\label{eq:envelope}
\ee
The quantity $I_{ad}(x)$ represents the combined cost of preparing the initial primary state and filtering the corresponding output state. Since $b$ becomes large, the leading contribution comes from the channel, or set of channels, minimizing this cost.

Geometrically, each allowed primary transition defines a straight line in the $(x,I)$ plane, with intercept $\Delta_a$ and slope $\Delta_d$. The dominant channel at a given value of $x$ is determined by the lower envelope of these lines. As $x$ increases, the minimizing line can change, leading to a change in the dominant output primary and hence to a transition between energy plateaux.

A useful consequence of this representation is that only finitely many primary channels can contribute to the leading envelope in a fixed compact CFT. To see this, consider the two transitions connecting the vacuum and the state created by the quench operator. Their costs are $x\Delta_{\cO}$ and $\Delta_{\cO}$, so that
$I_*(x)\le\Delta_{\cO}\min(1,x)$.
On the other hand, every primary pair satisfies
$I_{ad}(x)\ge\min(1,x)(\Delta_a+\Delta_d)$.
Combining these inequalities, we find that a necessary condition for a primary pair to reach the lower envelope is
\be
\Delta_a+\Delta_d\le\Delta_{\cO}.
\label{eq:finiteprimary}
\ee
This bound is one of the central results of our analysis. For a fixed compact CFT, there are only finitely many primary states below any given conformal dimension. Therefore, although the full spectral decomposition may contain infinitely many primary families and their descendants, only finitely many primary pairs can determine the leading cold envelope.

The inequality~(\ref{eq:finiteprimary}) provides a necessary condition, rather than a guarantee that a given channel will become dominant. In particular, a nonvacuum channel can lie strictly below both vacuum endpoint branches over an open interval only if
$\Delta_a+\Delta_d<\Delta_{\cO}$.
Even when this condition is satisfied, the channel must still compete with all other allowed transitions. The detailed ordering of the exposed channels is therefore determined by the complete lower envelope. A proof of the finite-channel reduction, including the required convergence estimates, is presented in App.~\ref{app:coldproof}.

We next relate the dominant spectral channels to the actual filtered state. Consider a closed interval of $x$ on which a single input--output energy pair minimizes the cost and is separated from all competing pairs by a positive gap $g$. In this regime, the normalized density matrix converges exponentially to the state supported on the selected output eigenspace,
\be
\rho_*=
\frac{P_d\cO P_a\cO^\dagger P_d}
{\Tr(P_d\cO P_a\cO^\dagger P_d)},
\qquad
\|\rho(t)-\rho_*\|_1=O(e^{-bg/2}),
\label{eq:effective_rhostar}
\ee
where $P_a$ and $P_d$ project onto the selected input and output energy eigenspaces, respectively. Thus, away from spectral crossings, the Euclidean filtering exponentially suppresses contributions from competing energy sectors.

The limiting density matrix need not be pure. If the selected output energy is degenerate, the filtered state may remain a mixture within that eigenspace. For a unique scalar primary output, however, the limiting state is pure. The same exponential suppression controls energy-weighted contributions from higher states after a positive amount of Euclidean filtering.

To connect this spectral selection to the local energy density, we must also account for contributions involving Virasoro descendants. Input descendants are suppressed at order $e^{-b}$, while interference between primary and descendant output states can contribute at order $e^{-Y/2}$. Using the Virasoro energy bounds established in App.~\ref{app:coldproof}, we obtain
\be
\frac{\cE_L(X,t)}{\kappa^2}
=-\frac{c}{24}
+\frac{\displaystyle\sum_{(a,d)\in\mathcal A_{\cO}}
A_{ad}h_d e^{-b\Delta_a-Y\Delta_d}}
{\displaystyle\sum_{(a,d)\in\mathcal A_{\cO}}
A_{ad}e^{-b\Delta_a-Y\Delta_d}}
+O(e^{-b}+e^{-Y/2}).
\label{eq:effectivelocal}
\ee
This expression is uniform in the spatial coordinate and on compact positive intervals of $x$, including regions where different primary channels have comparable weights. It provides a general approximation to the local energy density in terms of the primary-state spectrum and the matrix elements of the quench operator.

The leading term is a weighted average of the output-primary chiral energies, with weights inherited from the cold spectral decomposition. Keeping only the finite set allowed by~(\ref{eq:finiteprimary}) introduces an exponentially small correction when the omitted channels have a positive gap.

When a single primary output $d$ dominates, the weighted expression reduces to
\be
\frac{\cE_L}{\kappa^2}\simeq h_d-\frac{c}{24},
\qquad
\frac{\cE_R}{\kappa^2}\simeq\bar h_d-\frac{c}{24}.
\label{eq:effectivestress}
\ee
The resulting energy density is approximately homogeneous at the level fixed by the selected conformal weights, while descendants and subleading channels produce the suppressed spatial dependence seen near crossovers.

The geometry of the lower envelope also constrains the ordering of the dominant states. Since $I_*(x)$ is the minimum of affine functions, it is concave, and its slope cannot increase as $x$ grows. Because the slope of each line is the total conformal dimension $\Delta_d$ of its output primary, the successive dominant output dimensions form a nonincreasing sequence. Different transitions may nevertheless lead to the same output energy, and a channel whose cost remains strictly above the lower envelope cannot generate a leading plateau.

We can now determine the characteristic times at which the dominant channels exchange their relative importance. Consider two successive exposed transitions $i=(a_i,d_i)$ and $j=(a_j,d_j)$. In the strict cold limit, their costs become equal at the crossing point $x_{ij}^{(0)}$. At large but finite $b$, the ratio of their matrix-element coefficients shifts the crossing to
\be
x_{ij}^{(0)}=\frac{\Delta_{a_j}-\Delta_{a_i}}{\Delta_{d_i}-\Delta_{d_j}},
\qquad
x_{ij}=x_{ij}^{(0)}+
\frac{\log(A_i/A_j)}{b(\Delta_{d_i}-\Delta_{d_j})}.
\label{eq:crossing}
\ee
The leading crossing depends on conformal dimensions, while the relative matrix-element weights produce the finite-temperature shift.

Under the linear filtering protocol $s(t)=\epsilon+\delta t$, the ratio $x=2s/\beta$ increases linearly with Lorentzian time. If two consecutive crossings occur at $x_i$ and $x_{i+1}$, the corresponding plateau has the nominal duration
\be
\Delta t_i\simeq\frac{\beta}{2\delta}(x_{i+1}-x_i),
\qquad
\Delta t_{\rm cross}\sim
\frac{1}{\kappa\delta\,|\Delta_{d_i}-\Delta_{d_j}|}.
\label{eq:duration}
\ee
The second estimate gives the characteristic width of the transition between two neighboring plateaux, up to a factor depending on the convention used to define the crossover region. It follows from the exponential competition between the two retained weights.

The plateau durations grow linearly with $\beta$, while isolated crossover widths remain of order $L/\delta$ when the differences of output dimensions are finite. Colder initial states can therefore resolve longer homogeneous regimes without comparably broadening their transitions.

The approximate local energy~(\ref{eq:effectivelocal}) also describes the interpolation through these crossings, as long as the omitted primary channels remain separated by a positive spectral gap. Near a crossing, two or more channels contribute with comparable weights, and the energy changes smoothly between their characteristic levels. Away from the crossing, a single channel dominates exponentially, producing the approximately constant energy associated with the corresponding primary state.

In this regime, this description relies on the simultaneous scaling limit~(\ref{eq:scaled}). The limits $\beta\to\infty$ and $t\to\infty$ probe different aspects of the spectral decomposition. Taking $\beta\to\infty$ at fixed time suppresses all thermally excited input states and selects the vacuum as the initial state. Conversely, taking $t\to\infty$ at fixed $\beta$ suppresses all but the lowest accessible output energy. Neither limit alone resolves the full competition responsible for intermediate plateaux.

The simultaneous cold scaling limit retains the competition between input and output energies. When several envelope crossings are closely spaced, a colder preparation is needed to resolve the predicted primary plateaux in the finite-temperature profile.

We apply the lower-envelope criterion to the Ising, $SU(2)_k$ current and compact-boson quenches studied in Sec.~\ref{sec:torus}.

\paragraph{The Ising quenches.}

We begin with the Ising model, where the small number of primary fields allows the spectral selection mechanism to be illustrated explicitly. The three primaries $\mathbf1$, $\sigma$, and $\varepsilon$ have total conformal dimensions
$\Delta_{\mathbf1}=0$, $\Delta_\sigma=1/8$, and
$\Delta_\varepsilon=1$. Although the spectrum is the same for both quenches considered in Sec.~\ref{subsec:ising}, the allowed primary transitions depend on the choice of quench operator. This leads to different lower envelopes and, consequently, different sequences of energy plateaux.

Consider first the spin-field quench, $\cO=\sigma$. The fusion rules
$\sigma\times\sigma=\mathbf1+\varepsilon$ and
$\sigma\times\varepsilon=\sigma$ allow four oriented primary transitions. Their spectral costs, defined in~(\ref{eq:envelope}), are
\[
\begin{aligned}
\mathbf1\to\sigma &: \frac{x}{8}, &
\sigma\to\mathbf1 &: \frac18,\\
\sigma\to\varepsilon &: \frac18+x, &
\varepsilon\to\sigma &: 1+\frac{x}{8}.
\end{aligned}
\]
However, not every allowed transition can become dominant. The costs associated with $\sigma\to\varepsilon$ and $\varepsilon\to\sigma$ remain strictly above those of $\sigma\to\mathbf1$ and $\mathbf1\to\sigma$, respectively, for all $x>0$. These channels therefore never reach the lower envelope and cannot produce additional leading primary plateaux.

The lower envelope is consequently determined by only two competing contributions,
$I_\sigma(x)=\min\{x/8,1/8\}$,
which exchange dominance at $x=1$. For $x<1$, the vacuum-to-spin transition dominates, and the filtered state approaches the $\sigma$ primary level. For $x>1$, the spin-to-vacuum channel becomes dominant, and the energy approaches the vacuum level. The corresponding homogeneous chiral energies are
$\cE_L/\kappa^2=1/24$ and $-1/48$, in agreement with Fig.~\ref{fig:isingsigma}.

The energy-field quench, $\cO=\varepsilon$, exhibits a different structure. In this case, the fusion rules
$\varepsilon\times\varepsilon=\mathbf1$ and
$\varepsilon\times\sigma=\sigma$
allow three oriented primary transitions. Their costs give the lower envelope
\be
\begin{aligned}
I_\varepsilon(x)&=\min\left\{\underbrace{x}_{\mathbf1\to\varepsilon},
\underbrace{\frac{1+x}{8}}_{\sigma\to\sigma},
\underbrace{1}_{\varepsilon\to\mathbf1}\right\},\\
\varepsilon&\ \xrightarrow{x=1/7}\ \sigma
\ \xrightarrow{x=7}\ \mathbf1.
\end{aligned}
\label{eq:isingeff}
\ee
All three channels reach the envelope, with successive homogeneous chiral energies
$\cE_L/\kappa^2=23/48,\,1/24,\,-1/48$.

The additional intermediate plateau has a simple origin in the $\mathbb Z_2$ symmetry of the Ising model. The energy operator is $\mathbb Z_2$-even and therefore admits a diagonal matrix element within the spin conformal family. By contrast, the spin operator is $\mathbb Z_2$-odd, so the corresponding $\sigma\to\sigma$ transition is forbidden. This difference in the allowed matrix elements explains why the energy quench can resolve two nonvacuum primary plateaux, whereas the spin quench has only one.

\begin{figure}[t]
\centering
\includegraphics[width=\textwidth]{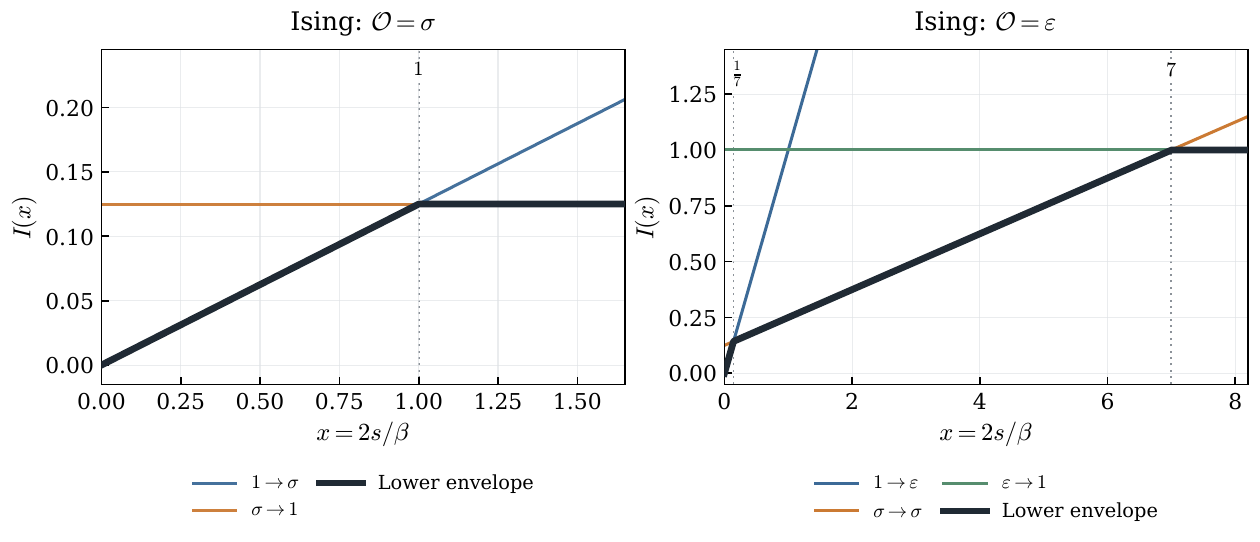}
\caption{Cold primary-channel costs $I_{a\to d}(x)$ for Ising spin (left) and energy (right) quenches, with $x=2s/\beta$. Colored lines denote the exposed channels and the thick dark line their lower envelope; dotted vertical lines mark channel crossings. The two nonexposed spin-quench channels are omitted.}
\label{fig:isingcost}
\end{figure}

The corresponding lower envelopes are shown in Fig.~\ref{fig:isingcost}, in agreement with the torus energy profiles in Figs.~\ref{fig:isingsigma} and~\ref{fig:isingepsilon}.

\paragraph{The Cartan-current quench in $SU(2)_k$.}

We next apply the same construction to the Cartan-current quench in the $SU(2)_k$ WZW model studied in Sec.~\ref{subsec:wzw}. An important distinction from the previous example is that the quench operator is a Virasoro primary of conformal weight $h=1$, but belongs to the affine vacuum module as a descendant. The spectral analysis must therefore include the current-excited state among the possible Virasoro-primary outputs, rather than restricting attention to the affine highest-weight states.

The affine symmetry strongly constrains the allowed transitions. In particular, the Cartan zero mode acts within each affine representation and can connect thermally populated states to output states in the same spin multiplet. The lightest nontrivial such multiplet has spin $j=1/2$ and total conformal dimension
$\Delta_{1/2}=3/[2(k+2)]$.

In the cold regime, the leading competition involves the current state created from the vacuum, the diagonal transition within the spin-$1/2$ multiplet, and the vacuum output obtained from a thermally populated current state. Their leading spectral weights are
\be
W_J=e^{-Y},\qquad
W_{1/2}=\frac2k e^{-\Delta_{1/2}(b+Y)},\qquad
W_0=e^{-b}.
\label{eq:wzwweights}
\ee
The corresponding costs and the resulting approximation to the homogeneous chiral energy are
\be
\begin{gathered}
I_J=x,\qquad I_{1/2}=\Delta_{1/2}(1+x),\qquad I_0=1,\\
\frac{\cE_{L,k}}{\kappa^2}\simeq-\frac{c_k}{24}
+\frac{W_J+h_{1/2}W_{1/2}}{W_J+W_{1/2}+W_0},\qquad
h_{1/2}=\frac{3}{4(k+2)}.
\end{gathered}
\label{eq:wzweff}
\ee
This expression describes how the energy interpolates between the current, spin-$1/2$, and vacuum levels as the filtering parameter increases. When one of the three weights dominates, the energy approaches the corresponding primary-state value, while comparable weights produce a crossover between neighboring levels.

The affine representation structure also explains why higher-spin primary multiplets do not generate additional leading plateaux. Since the Cartan zero mode preserves the affine representation, a diagonal transition within a spin-$j$ multiplet has the cost
$I_j(x)=\Delta_j(1+x)$.
The conformal dimensions increase with $j$, so for every $j>1/2$ and $x>0$, these costs lie strictly above the spin-$1/2$ branch. Excitations generated by nonzero current modes carry additional positive integer levels and likewise cannot lower the leading spectral cost. Consequently, the cold lower envelope is completely determined by the three channels in~(\ref{eq:wzwweights}).

Whether the spin-$1/2$ channel produces a distinct plateau depends on the level $k$. Its cost lies below both the current and vacuum branches only in the interval
\be
x_1=\frac{\Delta_{1/2}}{1-\Delta_{1/2}},\qquad
x_2=\frac{1-\Delta_{1/2}}{\Delta_{1/2}}.
\label{eq:wzwcurrentcross}
\ee
At $k=1$, we have $\Delta_{1/2}=1/2$, and the two crossing points coincide at $x_1=x_2=1$. The spin-$1/2$ channel therefore does not dominate over an open interval, and no extended intermediate plateau appears in the leading cold approximation.

For $k>1$, however, $\Delta_{1/2}<1/2$, so that $x_1<x_2$. The spin-$1/2$ branch is then exposed over a finite interval, producing an intermediate plateau between the current and vacuum regimes. Its leading duration is
$\Delta t\simeq\beta(x_2-x_1)/(2\delta)$.
As $k$ increases, the conformal dimension $\Delta_{1/2}$ decreases, separating the two crossing points and extending the interval over which the spin-$1/2$ multiplet dominates.

The affine-character calculations in Figs.~\ref{fig:wzwtemp} and~\ref{fig:wzwlevels} agree with this prediction and resolve the finite-temperature spatial dependence beyond the homogeneous primary approximation.

\begin{figure}[t]
\centering
\includegraphics[width=\textwidth]{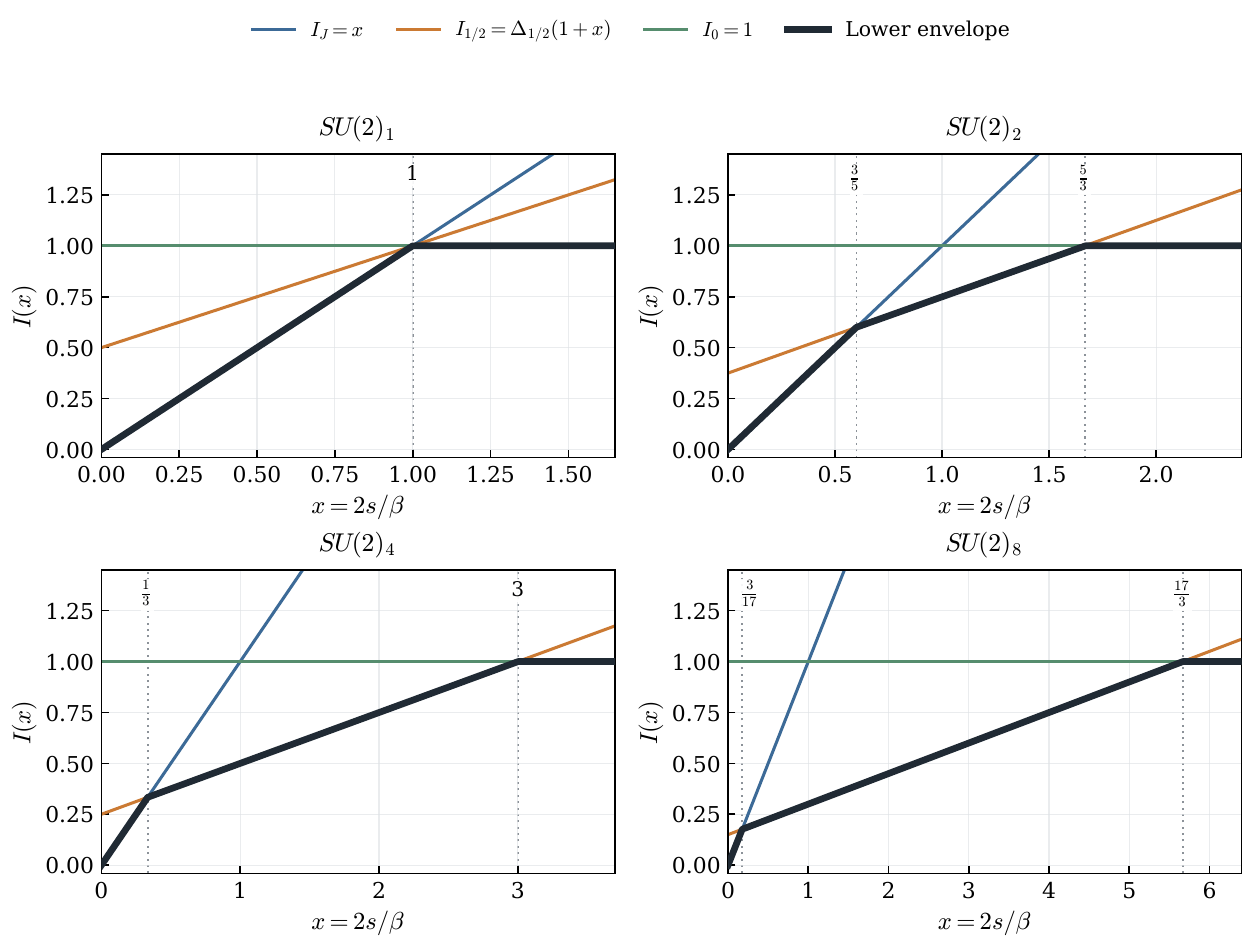}
\caption{Cold primary-channel costs for Cartan-current quenches in $SU(2)_k$, $k=1,2,4,8$. Colored lines represent the current, spin-$1/2$ and vacuum channels of Eq.~(\ref{eq:wzweff}); the thick dark line is their lower envelope. Dotted vertical lines mark the crossings of Eq.~(\ref{eq:wzwcurrentcross}).}
\label{fig:wzwcurrentcost}
\end{figure}

Figure~\ref{fig:wzwcurrentcost} compares the lower envelopes for $k=1,2,4,8$.

\paragraph{Compact-boson vertex quenches.}

Compact-boson momentum-vertex quenches allow the number of energy plateaux to be varied through the operator charge. Momentum quantization and the integer shift generated by the vertex determine the complete cold spectral envelope analytically.

We choose the quench operator $\cO=V_{N/R}$, where
$N\in\mathbb Z_{>0}$ and its total conformal dimension is
$\Delta_{\cO}=N^2/R^2$.
The vertex operator shifts the momentum quantum number according to $n\to n+N$, while preserving the winding number. Since nonzero winding increases the conformal dimensions of both the input and output states, the lowest-cost transitions occur in the zero-winding sector.

We can therefore label the relevant primary transitions by an integer $r$, corresponding to an input state of momentum $-r/R$ and an output state of momentum $(N-r)/R$. Their spectral costs are
\be
I_r(x)=\frac{r^2+x(N-r)^2}{R^2},\qquad r\in\mathbb Z.
\label{eq:compactcharges_cost}
\ee
The first term represents the thermal cost of the input momentum state, while the second describes the Euclidean filtering of the output state. Increasing $x$ changes the relative importance of these two contributions, progressively favoring output states with smaller momentum charge.

The minimization can be performed explicitly by completing the square,
\be
I_r(x)=\frac{1+x}{R^2}
\left(r-\frac{Nx}{1+x}\right)^2
+\frac{N^2x}{R^2(1+x)}.
\label{eq:compactcharges_square}
\ee
The second term is independent of $r$, so the dominant channel is the one for which $r$ is the integer nearest to $Nx/(1+x)$. As $x$ increases from zero to infinity, this quantity increases monotonically from $0$ to $N$. The minimizing integer therefore changes successively through
$r=0,1,\ldots,N$.

This shows that every one of these $N+1$ channels reaches the lower envelope over a finite interval of $x$. The crossings between neighboring channels are obtained by solving $I_r(x)=I_{r+1}(x)$, giving
\be
x_r=\frac{2r+1}{2(N-r)-1},\qquad r=0,\ldots,N-1.
\label{eq:compactcharges_cross}
\ee
The corresponding output states form the sequence
\be
V_{N/R}\longrightarrow V_{(N-1)/R}\longrightarrow\cdots
\longrightarrow V_{1/R}\longrightarrow\mathbf1,
\qquad
\frac{\cE_L}{\kappa^2}\simeq
\frac{(N-r)^2}{2R^2}-\frac1{24}
\label{eq:compactcharges_staircase}
\ee
where the energy expression applies on the interval in which the channel labeled by $r$ minimizes the spectral cost.

Thus, a vertex quench of charge $N/R$ produces $N$ distinct nonvacuum primary plateaux in the cold limit, followed by the vacuum. Each successive plateau corresponds to a smaller output momentum charge and therefore a lower conformal energy. This ordering is consistent with the general concavity property of the lower envelope discussed above.

The number of plateaux grows with the vertex charge $N$. This does not contradict the finite-channel bound~(\ref{eq:finiteprimary}), since increasing $N$ also increases the conformal dimension of the quench operator, $\Delta_{\cO}=N^2/R^2$, allowing a larger number of primary pairs to satisfy the necessary spectral condition.

For example, when $N=3$, the successive crossing points are $x=1/5,1,5$, corresponding to four dominant output sectors with charges $3,2,1,0$. The resulting energy staircase is shown in Fig.~\ref{fig:compactvertex}. Figure~\ref{fig:compactcharges} extends this comparison to $N=1,2,3,4$ and demonstrates how increasing the quench charge produces longer sequences of resolved primary-energy plateaux.

\begin{figure}[t]
\centering
\includegraphics[width=\textwidth]{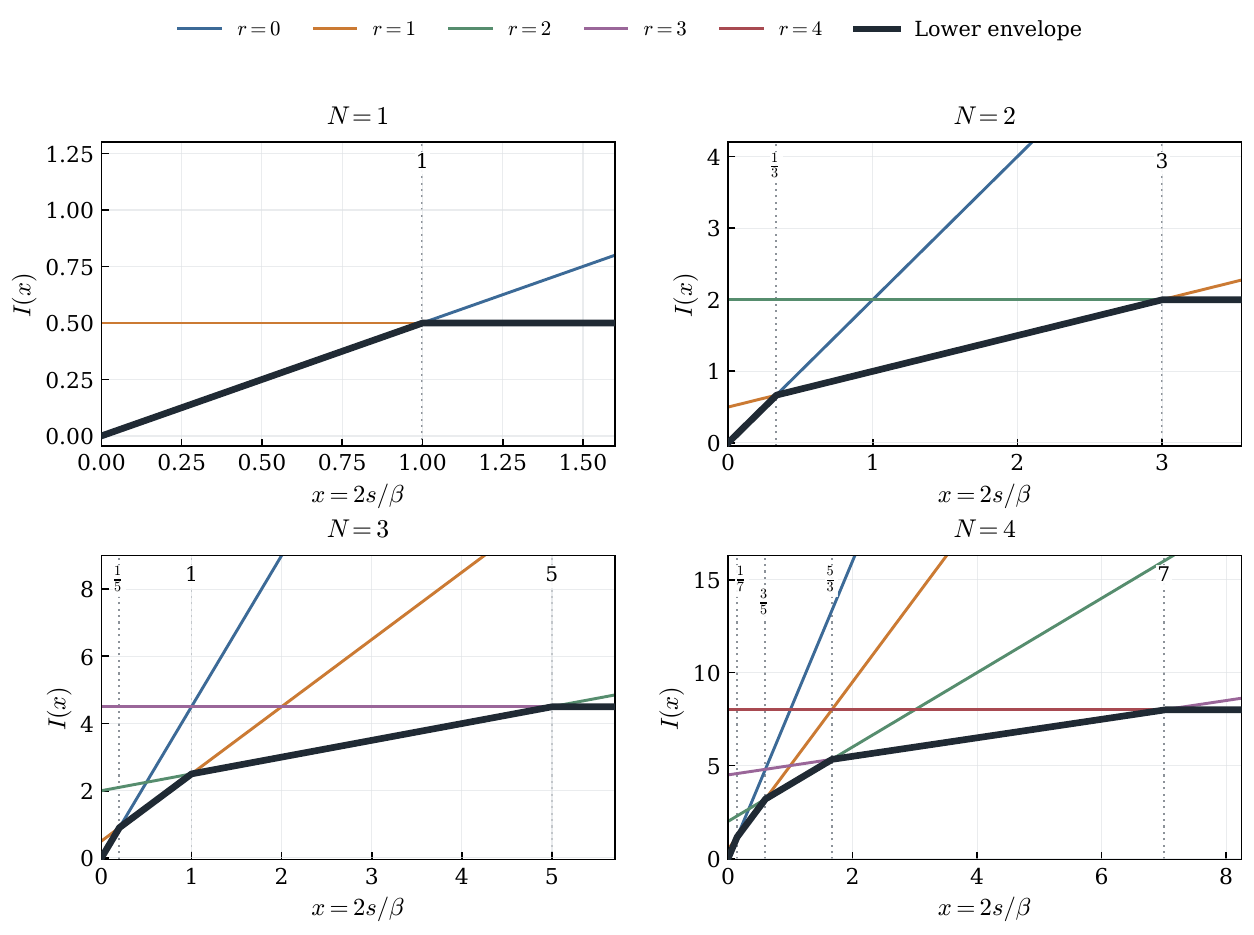}
\caption{Cold primary-channel costs for compact-boson vertex quenches $V_{N/R}$, $N=1,2,3,4$, at $R=\sqrt2$. Colored lines correspond to the channels $r=0,\ldots,N$ of Eq.~(\ref{eq:compactcharges_cost}), and the thick dark line is their lower envelope. Vertical dotted lines mark the crossings of Eq.~(\ref{eq:compactcharges_cross}).}
\label{fig:compactcost}
\end{figure}

Figure~\ref{fig:compactcost} shows the cold lower envelopes for $N=1,2,3,4$.

The oscillator excitations contribute only positive descendant levels to the spectral costs and therefore do not modify the leading cold envelope. Their effects appear instead in the subleading, spatially dependent corrections to the local stress tensor, which are controlled by~(\ref{eq:effectivelocal}).

\subsection{Diagonal unitary minimal models}
\label{subsec:effectiveminimal}

We now extend the spectral selection analysis to the diagonal unitary minimal models. These theories provide a useful setting for investigating the possible structure of energy plateaux, since their spectra contain finitely many Virasoro primary families and the allowed transitions between them are determined by known fusion rules. The lower-envelope construction can therefore be implemented directly from the conformal data, without requiring an explicit evaluation of the full torus correlation functions.

The unitary minimal models $\mathcal M(m,m+1)$, with integer $m\ge3$, contain $m(m-1)/2$ distinct Virasoro primary families. Their primary fields $\Phi_{r,s}$ are labeled by the Kac indices $(r,s)$, subject to the identification $(r,s)\sim(m-r,m+1-s)$. The corresponding central charge and conformal dimensions are~\cite{BPZ:1984,DiFrancesco:1997}
\be
\begin{gathered}
c_m=1-\frac{6}{m(m+1)},\qquad
h_{r,s}=\frac{[(m+1)r-ms]^2-1}{4m(m+1)},\qquad
\Delta_{r,s}=2h_{r,s},\\
1\le r\le m-1,\qquad 1\le s\le m.
\end{gathered}
\label{eq:minimalweights}
\ee

For a local quench generated by a primary operator $\cO=\Phi_{r_0,s_0}$, the possible spectral channels are determined by its nonvanishing matrix elements between the primary states of the theory. It is important to consider the fusion of $\cO$ with every thermally accessible primary, rather than restricting attention to the self-fusion $\cO\times\cO$. Indeed, thermally populated input states belonging to different conformal families can contribute to the filtered evolution and may become responsible for intermediate energy plateaux.

The minimal-model fusion rules provide a finite set of candidate transitions $a\to d$. More explicitly, for the fusion product $\Phi_{r_1,s_1}\times\Phi_{r_2,s_2}$, the first index of an allowed output primary takes the values
\be
r_3=|r_1-r_2|+1,\ |r_1-r_2|+3,\ldots,
\min(r_1+r_2-1,2m-r_1-r_2-1),
\label{eq:minimalfusionr}
\ee
with an analogous condition for the second index under $m\mapsto m+1$. The identifications of the Kac table are then imposed to obtain the distinct primary families.

Each transition with a nonvanishing bulk three-point coefficient contributes a spectral channel whose cost is
$I_{ad}(x)=\Delta_a+x\Delta_d$.
For a Hermitian scalar quench operator, the reversed transition is also allowed and has the same squared matrix element, although its spectral cost differs because the thermal and filtering factors act on the input and output states, respectively.

The allowed transitions determine the cost lines and their lower envelope. Its exposed slopes select the output dimensions, while its intersections fix the leading crossovers. Equations~(\ref{eq:effectivestress}) and~(\ref{eq:duration}) then determine the corresponding plateau heights and durations.

For each minimal model the construction involves finitely many primary transitions. The number of exposed channels depends on the quench operator and its allowed matrix elements, and the number of nonvacuum plateaux cannot exceed the number of accessible nonvacuum output families.

A particularly interesting family is obtained by choosing the quench operator $\cO=\Phi_{2,1}$, whose total conformal dimension is
$\Delta_{2,1}=(m+3)/(2m)$.
For this family, the fusion rules and conformal dimensions lead to a sequence of candidate dominant transitions of the form
\be
\mathbf1\to\Phi_{2,1},\qquad
\Phi_{j,j}\to\Phi_{j-1,j}\quad (2\le j\le m-1),
\qquad \Phi_{2,1}\to\mathbf1.
\label{eq:minimalpairs}
\ee
These channels describe the successive selection of primary outputs with decreasing conformal dimensions, ending with the vacuum. The resulting cold-envelope pattern predicts $m-1$ nonvacuum primary plateaux, followed by the identity level.

The corresponding crossing points can be determined by equating the costs of neighboring exposed channels. The first and last crossings occur at
$x_1=1/(2m+1)$ and $x_{\rm last}=2m+1$, respectively, while the intermediate crossings are
\be
x_j=\frac{2j+1}{2m+1-2j},\qquad 2\le j\le m-2.
\label{eq:minimalcross}
\ee
These crossings become more finely spaced as the minimal-model parameter $m$ increases. We have examined this pattern by enumerating the fusion-allowed transitions and constructing their lower envelopes using the rational Kac dimensions for $3\le m\le30$. The resulting sequences agree with~(\ref{eq:minimalpairs}) and~(\ref{eq:minimalcross}) throughout this range. The first examples are collected in Table~\ref{tab:minimalhull}, which lists the successive output dimensions and crossing ratios, including the final transition to the vacuum.

\begin{table}[ht]
\centering\small
\begin{tabular}{c|l|l}
$m$ & $\Delta_d$ in decreasing order, including vacuum & crossings $x$ \\ \hline
$3$ & $1,\,1/8,\,0$ & $1/7,\,7$ \\
$4$ & $7/8,\,1/5,\,3/40,\,0$ & $1/9,\,1,\,9$ \\
$5$ & $4/5,\,1/4,\,2/15,\,1/20,\,0$ & $1/11,\,5/7,\,7/5,\,11$ \\
$6$ & $3/4,\,2/7,\,5/28,\,2/21,\,1/28,\,0$ & $1/13,\,5/9,\,1,\,9/5,\,13$
\end{tabular}
\caption{Cold-limit $\Phi_{2,1}$ quenches in diagonal unitary minimal models: successive output dimensions $\Delta_d$ (including the vacuum) and crossing ratios $x=2s/\beta$.}
\label{tab:minimalhull}
\end{table}

An instructive example is provided by the tricritical Ising model $\mathcal M(4,5)$, which has central charge $c=7/10$. In this theory, the quench operator $\Phi_{2,1}$ is identified with the subleading spin field $\sigma'$, of total dimension $\Delta_{\sigma'}=7/8$. Its fusion rules allow eight oriented primary transitions, but only four of them reach the lower envelope.

Their envelope is
\be
\begin{gathered}
I_{\sigma'}(x)=\min\left\{\frac{7x}{8},\,\frac{3}{40}+\frac{x}{5},\,
\frac15+\frac{3x}{40},\,\frac78\right\},\\
\sigma'\ \xrightarrow{x=1/9}\ \varepsilon\ \xrightarrow{x=1}\
\ \sigma\ \xrightarrow{x=9}\ \mathbf1,\qquad
\frac{\cE_L}{\kappa^2}=\frac{49}{120},\ \frac{17}{240},\ \frac1{120},\ -\frac7{240}.
\end{gathered}
\label{eq:tricriticalprediction}
\ee
The exposed channels predict three nonvacuum plateaux, while the other four fusion-allowed transitions lie above the envelope and do not contribute at leading order.

\begin{figure}[t]
\centering
\includegraphics[width=\textwidth]{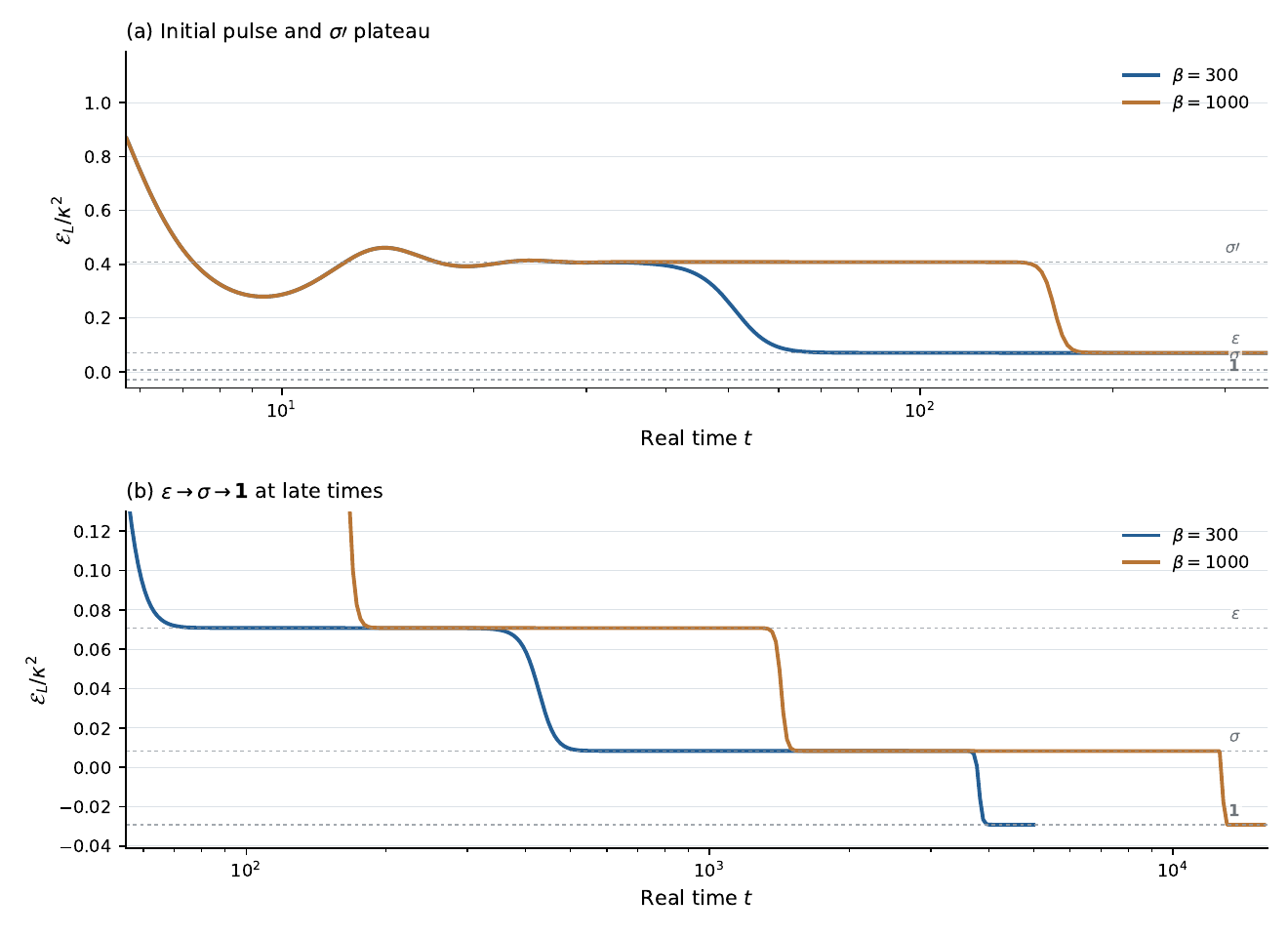}
\caption{Tricritical-Ising $\sigma'$ quench energy $\cE_L/\kappa^2$ from Virasoro sewing through chiral level $N=10$ at $\beta=300,1000$. Dotted lines mark the $\sigma'$, $\varepsilon$, $\sigma$ and vacuum levels. Parameters: $L=10$, $X=5$, $x_0=0$, $\epsilon=0.5$, $\delta=0.35$. Convergence and the cold-input approximation are discussed in App.~\ref{app:tricriticalsewing}.}
\label{fig:tricriticalsewing}
\end{figure}

To examine the descendant corrections beyond the leading plateau heights and crossover positions, we perform a numerical calculation based on Virasoro sewing of the two-arc Euclidean trace. The calculation includes physical Virasoro descendants through chiral level ten and reconstructs the local stress tensor using the genus-one Ward identity. We consider the low-temperature preparations $\beta=300$ and $\beta=1000$, for which contributions from excited thermal input descendants are strongly suppressed and can be neglected within the cold-input approximation.

The resulting energy profiles are shown in Fig.~\ref{fig:tricriticalsewing}. The calculation resolves the initial local energy pulse and the subsequent approximately homogeneous regimes, whose values approach the primary-state energies predicted by~(\ref{eq:tricriticalprediction}). The transitions between these regimes also occur near the corresponding cold-envelope crossing times.

Convergence and the domain of validity of the cold-input approximation are discussed in App.~\ref{app:tricriticalsewing}.

\subsection{Affine-primary quenches in $SU(2)_k$ WZW theory}
\label{subsec:effectiveaffine}

We next consider local quenches generated by bulk affine-primary operators in the diagonal $SU(2)_k$ WZW models. These theories provide another setting in which the allowed primary transitions can be determined from fusion rules. Unlike the Cartan-current quench discussed in Sec.~\ref{subsec:wzw}, where the quench operator belongs to the affine vacuum module, we now take the quench operator itself to be an affine primary. This allows us to investigate how the spin of the inserted operator controls the number and ordering of the resulting energy plateaux.

Let $\cO=\Phi_j$ be a bulk affine primary with conformal weights $(h_j,h_j)$, where $j$ is one of the integrable spins of the theory. The allowed transitions between affine-primary representations are governed by the $SU(2)_k$ fusion rule~\cite{Witten:1984,KnizhnikZamolodchikov:1984,DiFrancesco:1997}
\be
N_{aj}^{\phantom{aj}d}=1\ \Longleftrightarrow\
|a-j|\le d\le\min(a+j,k-a-j),\qquad a+j+d\in\mathbb Z.
\label{eq:wzwfusion}
\ee
Here $a$, $j$, and $d$ label integrable representations with spins in $\{0,1/2,\ldots,k/2\}$, and the allowed output spins increase in integer steps. As in the minimal-model analysis, the fusion rules determine the candidate spectral channels, while the existence of a contribution to the filtered correlator requires the corresponding bulk operator matrix element to be nonzero.

We assume that the relevant components of the bulk quench operator have nonvanishing couplings in the channels considered below. Under this assumption, the fusion constraints allow us to identify the least costly input representation for each possible output.

For an output spin $d\le j$, the smallest allowed input spin is $a=j-d$. Since the conformal dimensions increase with spin, this choice minimizes the thermal contribution to the spectral cost. On the other hand, channels with output spin $d>j$ have costs above the vacuum-input branch and cannot contribute to the leading lower envelope.

The cold-envelope problem can therefore be reduced to a finite set of channels connecting input and output spins whose sum is $j$. Writing the input spin as $r/2$, we obtain
\be
I_j(x)=\min_{r=0,\ldots,2j}\left[
\Delta_{r/2}+x\Delta_{j-r/2}\right],\qquad
\Delta_\ell=\frac{2\ell(\ell+1)}{k+2}.
\label{eq:wzwprimaryenvelope}
\ee
The two terms are the thermal input cost and the filtering cost of the output primary. As $x$ increases, the dominant output spin decreases.

A useful feature of this family is that the crossing points can be obtained analytically. Equating the costs of neighboring channels gives
\be
x_r=\frac{2r+3}{4j-2r+1},\qquad r=0,\ldots,2j-1.
\label{eq:wzwprimarycross}
\ee
These ratios are strictly increasing with $r$. Consequently, each channel in~(\ref{eq:wzwprimaryenvelope}) is exposed over a finite interval, and the dominant output spin decreases successively in steps of $1/2$.

The resulting cold primary-channel sequence is
\be
\Phi_j\longrightarrow\Phi_{j-1/2}\longrightarrow\cdots
\longrightarrow\Phi_{1/2}\longrightarrow\mathbf1,
\qquad N_{\rm nonvac}=2j.
\label{eq:wzwprimarysequence}
\ee
Thus, an affine-primary quench of spin $j$ is predicted to produce $2j$ nonvacuum primary plateaux, followed by the vacuum, provided the corresponding bulk couplings are nonzero.

The crossing ratios~(\ref{eq:wzwprimarycross}) are independent of the affine level $k$, because all primary dimensions contain the same overall factor $(k+2)^{-1}$, which cancels when neighboring spectral costs are equated. The plateau heights, however, depend on $k$ through the conformal weights and the central charge, with the chiral energy of an output primary given by
$\cE_L/\kappa^2=h_d-c_k/24$.

This construction gives a family of spectral predictions based solely on affine fusion rules and conformal dimensions. It applies to specified affine-primary quench operators with the assumed nonvanishing couplings and does not constitute a classification of all Virasoro-primary quenches in the WZW theory. For generic bulk affine-primary operators, obtaining the complete local energy density requires the corresponding torus conformal blocks and their derivatives in the Ward identity.

A particularly useful case is the maximal-spin primary, $\cO=\Phi_{k/2}$. This operator is a simple current of the $SU(2)_k$ fusion algebra, satisfying
$\Phi_{k/2}\times\Phi_a=\Phi_{k/2-a}$.
Thus, each affine representation is connected to a unique output representation, and the $k+1$ allowed input sectors generate the sequence
\be
\frac{k}{2}\longrightarrow\frac{k-1}{2}\longrightarrow\cdots
\longrightarrow\frac12\longrightarrow0,\qquad
x_r=\frac{2r+3}{2k-2r+1},\quad r=0,\ldots,k-1.
\label{eq:wzwmaxspinsequence}
\ee
The maximal-spin quench therefore provides a family in which the number of nonvacuum primary plateaux grows linearly with the affine level, reaching $k$ for the operator $\Phi_{k/2}$.

\begin{figure}[t]
\centering
\includegraphics[width=\textwidth]{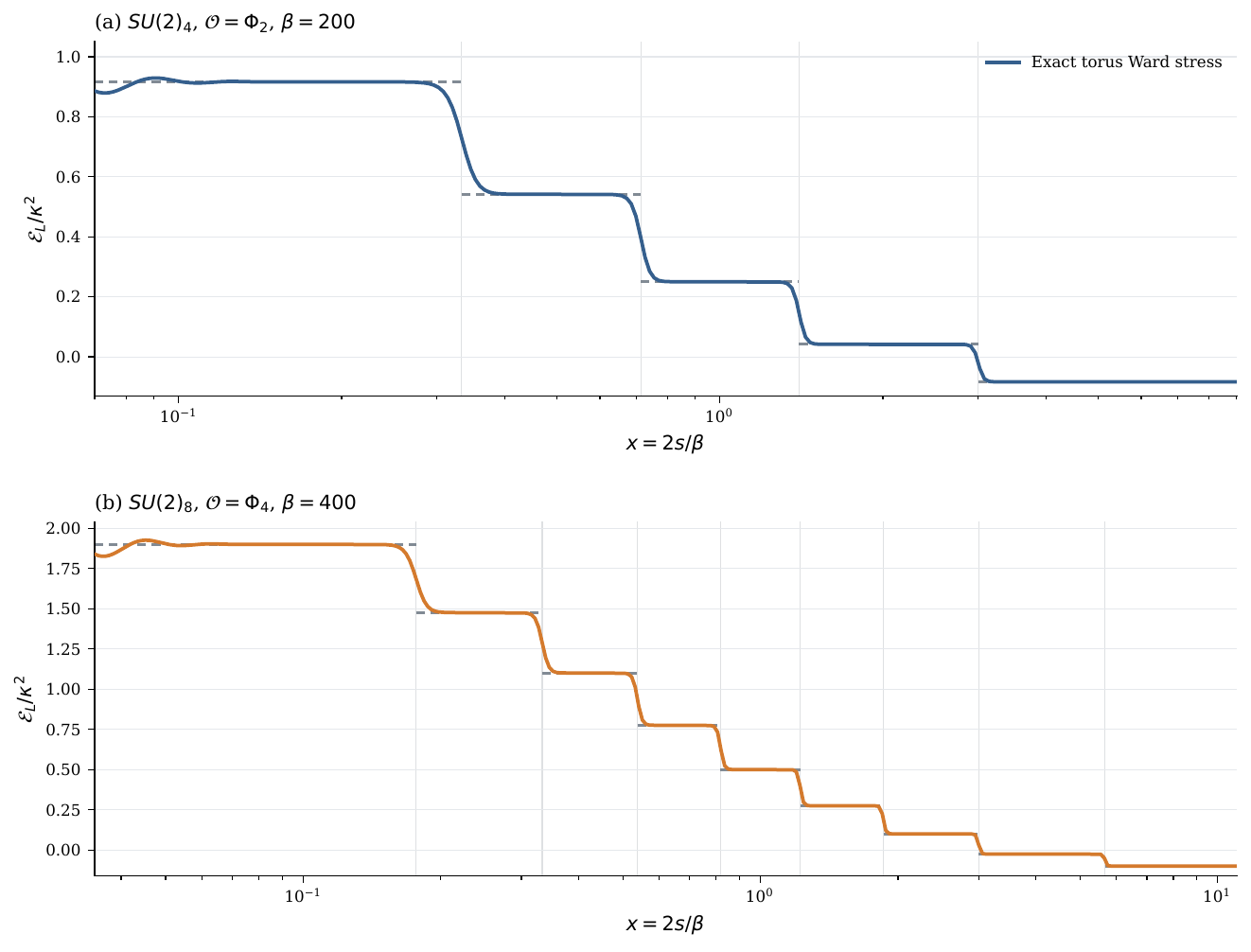}
\caption{Energy $\cE_L/\kappa^2$ after maximal-spin affine-primary quenches in $SU(2)_4$ ($\beta=200$) and $SU(2)_8$ ($\beta=400$). Solid curves are torus Ward results; horizontal dashed lines mark primary levels and vertical lines cold-limit crossings. All $k+1$ affine sectors are included, with Weyl--Kac sums through $|n|\le6$; $L=10$, $X=5$, $x_0=0$, $\epsilon=0.5$, $\delta=0.35$.}
\label{fig:wzwmaxspin}
\end{figure}

An additional advantage of this case is that the full torus two-point function is known in closed form. Using the result of Kirsch and Kucharski~\cite{KirschKucharski:2013}, the normalized correlator can be written as
\be
\begin{aligned}
G_{k/2}(z,\bar z;\tau,\bar\tau)
&=\frac{1}{Z_k(\tau,\bar\tau)}\sum_{a=0}^{k/2}\left|\frac{\chi_a^{(k)}(z|\tau)}{E(z|\tau)^{k/2}}\right|^2,\\
E(z|\tau)&=\frac{\vartheta_1(\pi z|\tau)}{\vartheta_1'(0|\tau)},\qquad
z=\frac{i\,2s}{L},\qquad \tau=\frac{i(\beta+2s)}{L}.
\end{aligned}
\label{eq:wzwmaxspincorr}
\ee
Here $\chi_a^{(k)}(z|\tau)$ denotes the generalized affine character with elliptic variable $z$, while $E(z|\tau)$ is the torus prime-form factor. An overall normalization independent of the insertion coordinates and modulus has been suppressed, since it cancels in the logarithmic derivatives entering the Ward identity.

The expression~(\ref{eq:wzwmaxspincorr}) makes it possible to evaluate the local energy density using the genus-one Ward identity and compare the resulting dynamics with the cold spectral predictions. Unlike the generic affine-primary case, where the plateau sequence follows from fusion and the lower-envelope construction alone, the maximal-spin correlator provides direct access to the finite-temperature energy profile, including the contributions beyond the leading primary approximation.

The comparison in Fig.~\ref{fig:wzwmaxspin} shows the local energies approaching the predicted primary-state levels in successive intervals for $SU(2)_4$ and $SU(2)_8$, with crossovers near the corresponding cold-envelope positions. The full correlator calculations retain all $k+1$ affine sectors and resolve four nonvacuum plateaux for $k=4$ and eight for $k=8$.

\FloatBarrier

\section{Energy plateaux in holographic CFTs}
\label{sec:holographic}

The preceding sections have established that Euclidean filtering can select successive primary-state contributions to the energy density. In the cold regime, the dimensions of the dominant output states determine the plateau heights and the competing input--output channels determine their leading crossover positions. We now examine how the same mechanism may appear in holographic CFTs at large central charge, where different energy sectors admit distinct bulk interpretations.

Nozaki, Numasawa and Takayanagi~\cite{NozakiNumasawaTakayanagi:2013} described a local operator quench in terms of a particle falling into AdS, providing a gravitational picture for the localized boundary excitation. The strength of its backreaction depends on how the dimension of the quench operator scales with the central charge. For a light operator with $\Delta_\cO=O(1)$, the bulk excitation carries finite energy in AdS units and its leading backreaction is suppressed by $1/c$. For a heavy operator with $\Delta_\cO=O(c)$, the excitation carries energy of order the inverse Newton constant and can alter the geometry at leading order. This light--heavy distinction will organize our analysis of the filtered state.

The thermal preparation introduces a separate issue concerning the bulk background. In large-$c$ CFTs with a sufficiently sparse light spectrum, the partition function on a spatial circle of circumference $L$ is dominated by thermal AdS when the Euclidean time period exceeds $L$, and by the BTZ saddle in the opposite regime~\cite{HartmanKellerStoica:2014,KrausSivaramakrishnanSnively:2017}. In our protocol this period is $\Theta=\beta+2s(t)$, so filtering can change the dominant thermal saddle when the initial temperature is high enough. The choice of background is consequently determined by the thermal parameters, while the backreaction produced by the quench depends on its conformal dimension. Even when the partition function is dominated by thermal AdS, a heavy operator insertion can have a nontrivial effect on the geometries contributing to the correlator.

The extension of the cold-channel picture also requires spectral control. Holographic CFTs have infinitely many conformal families, with rapidly growing degeneracies at high energy, and their relative contributions depend on OPE coefficients and conformal blocks as well as conformal dimensions. The thermal initial state populates several input sectors, while the two Euclidean arcs generated by the protocol weight the input and output energies differently. We therefore examine the full torus decomposition before considering its light and heavy limits.

Our starting point is the normalized torus two-point function $G_{\mathbb T^2}$, which determines the local energy density through the conformal Ward identity~(\ref{eq:masterward}). To analyze the contributions of different conformal families, it is convenient to use its decomposition into Virasoro conformal blocks. For a unitary CFT with equal left- and right-moving central charges and a unit-normalized Hermitian scalar primary $\cO$, the torus two-point function can be decomposed into primary representations propagating along the Euclidean arcs of lengths $\beta$ and $2s$. In the necklace channel of~\cite{AlkalaevBelavin:2017,AlkalaevMandryginPavlov:2022}, the resulting expansion takes the form
\be
\begin{aligned}
G_{\mathbb T^2}(w_1,w_2)
={}&\frac{\kappa^{2\Delta_\cO}e^{\kappa c(\beta+2s)/12}}{Z_{\mathbb T^2}(\beta+2s)}
\sum_{a,d}|C_{a\cO d}|^2
q_1^{h_a}q_2^{h_d}\bar q_1^{\bar h_a}\bar q_2^{\bar h_d}
\cF^{\rm Vir}_{h_a,h_d}(q_1,q_2)
\overline{\cF}^{\,\rm Vir}_{\bar h_a,\bar h_d}(\bar q_1,\bar q_2),\\
&q_1=e^{-\kappa\beta},\qquad q_2=e^{-2\kappa s}.
\end{aligned}
\label{eq:necklace}
\ee
The labels $a$ and $d$ run over the allowed pairs of holomorphic and antiholomorphic Virasoro representations, including primary multiplicities. Their left- and right-moving conformal weights need not coincide. Each block has unit leading coefficient and collects descendants in its own chiral sector. Although $q_i=\bar q_i$ on the rectangular torus, holomorphic and antiholomorphic modulus derivatives remain independent. In the formulas below, $\cF^{\rm Vir}_{ad}$ and $\cF^{\rm glob}_{ad}$ denote the holomorphic blocks for a fixed channel, with analogous barred-sector formulas. The normalization by $Z_{\mathbb T^2}(\beta+2s)$ is retained throughout the Ward calculation.

Each pair of primary families in~(\ref{eq:necklace}) carries the thermal suppression of the input state and the filtering weight of the output state, together with its OPE coefficient. The Virasoro blocks account for descendant contributions along both arcs. When the arcs are long, the primary terms give the cold-channel costs of Sec.~\ref{sec:effective}, provided that the remaining contributions are uniformly suppressed in the full spectral sum.

The large-$c$ form of this decomposition depends on the scaling of the external and intermediate dimensions. Light families may be described perturbatively on a fixed background, while heavy families can contribute through semiclassical blocks with different gravitational interpretations. In both regimes, we retain the partition-function normalization appropriate to the Ward identity and account for the dependence on $\beta+2s$. We first identify the assumptions needed for a controlled light-sector approximation and then examine the additional spectral competition generated by heavy insertions.

\subsection{Light quenches}
\label{subsec:holo-light}

We begin with a light quench operator, $\Delta_\cO=O(1)$ as $c\to\infty$, and work in the low-temperature regime $\beta>L$. Since $\Theta=\beta+2s$ increases during the evolution, this choice keeps the partition function without insertions in the thermal-AdS phase throughout the protocol. Under the standard large-$c$ sparseness assumptions~\cite{HartmanKellerStoica:2014,KrausSivaramakrishnanSnively:2017}, its leading contribution is
\be
Z_{\mathbb T^2}(\beta+2s)\simeq
\exp\!\left[\frac{\kappa c(\beta+2s)}{12}\right].
\label{eq:holovacZ}
\ee
We use this approximation in the deeper cold regime, with vacuum dominance at the accuracy required by the Ward calculation. To control the operator insertions, we additionally assume a unique vacuum, a sparse low-energy spectrum and large-$c$ factorization of light correlators, with finitely many light single-trace fields at fixed energy and multi-trace families generated perturbatively~\cite{Heemskerk:2009,KrausSivaramakrishnanSnively:2017}. These assumptions permit a bulk description in terms of perturbative matter excitations on the chosen thermal-AdS background, whose gravitational backreaction is suppressed at leading order.

The finite-primary bound~(\ref{eq:finiteprimary}) already restricts the channels that can determine cold plateaux. At each fixed $c$, a primary pair can reach the strict cold lower envelope only when $\Delta_a+\Delta_d\le\Delta_\cO$. Since $\Delta_\cO=O(1)$, every such pair has light conformal dimensions, and heavy intermediate families with $\Delta\sim c$ cannot generate additional leading cold plateaux. This spectral restriction does not by itself control the total heavy-state contribution uniformly in the simultaneous cold and large-$c$ limit.

To promote the finite-channel restriction to an approximation for the complete correlator, we use the light-state dominance results of Kraus, Sivaramakrishnan and Snively~\cite{KrausSivaramakrishnanSnively:2017}. Their analysis relies on bounds for light-operator matrix elements in medium-energy states and a controlled large-$c$ expansion of low-temperature correlators, with modular covariance constraining the heavy spectrum. We require the resulting suppression to remain uniform as the Euclidean arcs grow with $c$, including after the Ward derivatives. Under these conditions, the leading plateau analysis can be restricted to light intermediate families of fixed conformal weight.

For fixed $h_\cO,h_a,h_d=O(1)$, the separation of global and higher Virasoro descendants in the large-$c$ regime gives the factorization~\cite{AlkalaevBelavin:2017,ChoCollierYin:2019}
\be
\cF^{\rm Vir}_{ad}(q_1,q_2)
=\prod_{n=2}^{\infty}\frac{1}{1-(q_1q_2)^n}
\cF^{\rm glob}_{ad}(q_1,q_2)+O(c^{-1}).
\label{eq:holoblocklimit}
\ee
The global block contains the descendants generated by $L_{-1}$ on each arc, while the product accounts for the universal higher Virasoro modes. The necklace-channel global blocks and their descendant projectors have been developed in~\cite{AlkalaevMandryginPavlov:2022,Pavlov:2023,AlkalaevMandrygin:2023}. In terms of the normalized global descendant matrix elements $\mathcal M_{nm}^{ad}$, the expansion reads
\be
\begin{gathered}
\cF^{\rm glob}_{ad}(q_1,q_2)
=\sum_{n,m=0}^{\infty}
\frac{q_1^nq_2^m|\mathcal M_{nm}^{ad}|^2}
{n!m!(2h_a)_n(2h_d)_m},\\
\mathcal M_{nm}^{ad}
=\frac{\langle h_d|L_1^m\cO(1)L_{-1}^n|h_a\rangle}
{\langle h_d|\cO(1)|h_a\rangle}.
\end{gathered}
\label{eq:hologlobalseries}
\ee
Here $(u)_n$ denotes the Pochhammer symbol, and descendants generated by $L_{-1}$ are omitted on arcs carrying the vacuum representation. Expanding in the two Euclidean propagation parameters gives the leading terms~\cite{AlkalaevBelavin:2017,Pavlov:2023}
\be
\begin{aligned}
\cF^{\rm glob}_{ad}(q_1,q_2)
=1&+\frac{(h_a+h_\cO-h_d)^2}{2h_a}q_1
+\frac{(h_d+h_\cO-h_a)^2}{2h_d}q_2\\
&+O(q_1^2,q_1q_2,q_2^2),
\end{aligned}
\label{eq:hologlobal}
\ee
where the corresponding linear term is absent for a vacuum arc. All coefficients shown remain finite when the relevant nonvacuum weights are held fixed as $c\to\infty$.

We next take both Euclidean arcs long and choose the simultaneous limit
\be
\log c\ll\kappa\beta\ll c,
\qquad 2\kappa s=\frac{2s}{\beta}\kappa\beta,
\qquad \frac{2s}{\beta}>0\quad\text{fixed}.
\label{eq:holocoldlimit}
\ee
This regime implies $q_1,q_2\ll1$, so the global block and the universal Virasoro factor approach one for every retained light channel. The upper restriction on $\kappa\beta$ keeps the $O(c^{-1})$ anomalous dimensions of fixed light families from accumulating finite corrections in the propagation exponent, while the lower restriction resolves channels whose couplings begin at different powers of $1/c$. Combining the preceding estimates with~(\ref{eq:necklace}) gives
\be
\frac{G_{\mathbb T^2}(w_1,w_2)}{\kappa^{2\Delta_\cO}}
\simeq\sum_{(a,d)\in\mathcal K}A_{ad}(c)
\exp[-\kappa(\beta\Delta_a+2s\Delta_d)],
\label{eq:hololighttrace}
\ee
where $\mathcal K$ includes all light primary transitions capable of competing at leading order, and $A_{ad}(c)$ is the squared OPE coefficient including the multiplicities of~(\ref{eq:effectiveweights}). The approximation also assumes that the discarded channels and descendants remain negligible after the Ward derivatives.

For the light spectrum, large-$c$ factorization supplies a perturbative expansion of normalized correlators and their OPE data in powers of $1/\sqrt c$~\cite{Heemskerk:2009,KrausSivaramakrishnanSnively:2017}. We assume that the leading nonzero contribution to each relevant squared coupling has the form
\be
A_{ad}(c)=A_{ad}^{(0)}c^{-\gamma_{ad}}[1+o(1)],
\qquad A_{ad}^{(0)}>0,\qquad\gamma_{ad}\ge0.
\label{eq:holoOPEscale}
\ee
The exponent and leading coefficient depend on the operators in each channel, and an exact selection rule may set a coupling to zero. For the remaining channels, combining this scaling with their Euclidean propagation weights gives the effective cost
\be
I_{ad}\!\left(\frac{2s}{\beta}\right)
=\Delta_a+\frac{2s}{\beta}\Delta_d
+\frac{\gamma_{ad}\log c-\log A_{ad}^{(0)}}{\kappa\beta}
+o\!\left((\kappa\beta)^{-1}\right),
\label{eq:holocost}
\ee
where the constant coefficient controls the finite-temperature shift of a crossing. The leading cold envelope is obtained by minimizing $\Delta_a+(2s/\beta)\Delta_d$ over all accessible light channels. An isolated minimizing output primary $d$ determines the homogeneous chiral plateau after spatial Ward harmonics have decayed,
\be
\frac{\cE_L}{\kappa^2}+\frac c{24}\simeq h_d,
\qquad
\frac{\cE_R}{\kappa^2}+\frac c{24}\simeq\bar h_d.
\label{eq:holoplateau}
\ee
The two vacuum endpoint channels have costs $(2s/\beta)\Delta_\cO$ and $\Delta_\cO$, while additional exposed lines determine possible intermediate plateaux. Their crossings and durations follow from~(\ref{eq:crossing}) and~(\ref{eq:duration}) after comparing all allowed light channels.

We first take a generalized-free scalar $\phi$ with $\Delta_\phi=1$ and the normalized quench operator $\cO=[\phi\phi]_{1,0}$ of dimension $\Delta_\cO=4$. Generalized-free factorization gives $|C_{\phi\cO\phi}|^2=1/2$~\cite{Heemskerk:2009}, and we assume that $\phi$ is the lightest nonvacuum state. Every nonvacuum-to-nonvacuum transition then obeys $\Delta_a+(2s/\beta)\Delta_d\ge1+2s/\beta$, with equality for $\phi\to\phi$, while primary orthogonality fixes the two vacuum endpoints. The complete generalized-free lower envelope is therefore
\be
\begin{gathered}
I_*\!\left(\frac{2s}{\beta}\right)
=\min\!\left\{\frac{8s}{\beta},1+\frac{2s}{\beta},4\right\},\\
\cO\ \xrightarrow{\,2s/\beta=1/3\,}\ \phi
\ \xrightarrow{\,2s/\beta=3\,}\ \mathbf1.
\end{gathered}
\label{eq:holoidentical}
\ee
The corresponding chiral plateau levels above the Casimir energy are $2$, $1/2$ and $0$, with the intermediate plateau lasting $4\beta/(3\delta)$ at leading order. In the thermal-$\mathrm{AdS}_3$ dual, $\phi$ describes a scalar particle with $m_\phi^2\ell_{\rm AdS}^2=\Delta_\phi(\Delta_\phi-2)=-1$ and excitation energy $E_\phi=\kappa\Delta_\phi=\kappa$. The quench operator creates a two-particle state with one relative radial excitation and energy $4\kappa$, so the three plateaux select the excited two-particle state, a single scalar particle and the AdS vacuum in succession.

\begin{figure}[t]
\centering
\includegraphics[width=\textwidth]{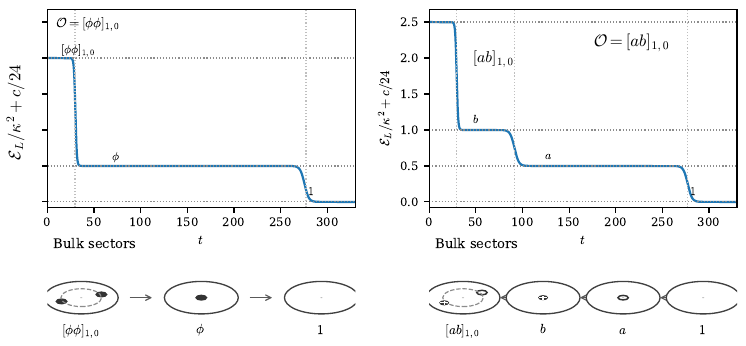}
\caption{Light-quench energy $\cE_L/\kappa^2+c/24$ (upper panels) and the corresponding thermal-$\mathrm{AdS}_3$ particle interpretation (lower panels) for the three- and four-channel examples~(\ref{eq:holoidentical}) and~(\ref{eq:threeexample}). Dotted lines mark primary levels and arrows indicate changes of the dominant sector. Parameters: $\beta=65$, $L=10$, $\epsilon=0.5$, $\delta=0.35$. The three-channel example uses a coupling weight $1/2$, while mixed-channel weights are set to unity in the four-channel illustration.}
\label{fig:hololight}
\end{figure}

We next take independent generalized-free scalar species $a$ and $b$ with $\Delta_a=1$ and $\Delta_b=2$, together with the quench operator $\cO=[ab]_{1,0}$ of dimension $\Delta_\cO=5$. The leading mixed coefficients satisfy $|C_{a\cO b}|^2=|C_{b\cO a}|^2=4/9$, and we impose an exact $\mathbb Z_2^{(a)}\times\mathbb Z_2^{(b)}$ symmetry under which $a$ and $b$ are odd under their respective factors. The quench operator is odd under both factors, which forbids the diagonal transitions $a\to a$ and $b\to b$ even when interactions are included. In the generalized-free spectrum the lowest dimensions in the neutral, singly odd and doubly odd sectors are $0$, $1$, $2$ and $3$, and every competing channel lies above one of the two mixed transitions or a vacuum endpoint. Consequently the full generalized-free envelope becomes
\be
\begin{gathered}
I_*\!\left(\frac{2s}{\beta}\right)
=\min\!\left\{\frac{10s}{\beta},1+\frac{4s}{\beta},
2+\frac{2s}{\beta},5\right\},\\
\cO\ \xrightarrow{\,2s/\beta=1/3\,}\ b
\ \xrightarrow{\,2s/\beta=1\,}\ a
\ \xrightarrow{\,2s/\beta=3\,}\ \mathbf1.
\end{gathered}
\label{eq:threeexample}
\ee
The chiral levels are $5/2$, $1$, $1/2$ and $0$, and the two intermediate durations are $\beta/(3\delta)$ and $\beta/\delta$. The dual contains scalar particles with $m_a^2\ell_{\rm AdS}^2=-1$ and $m_b^2\ell_{\rm AdS}^2=0$, whose one-particle energies are $\kappa$ and $2\kappa$, while the quench operator creates an excited two-particle state of energy $5\kappa$. Interactions such as $\Phi_a^2\Phi_b^2$ preserve the exact symmetry, which protects the two intermediate plateaux by excluding their lower-cost diagonal competitors. The sequence of dominant bulk sectors is the mixed two-particle state, the $b$-particle state, the $a$-particle state and the AdS vacuum, as illustrated in Fig.~\ref{fig:hololight}. The symmetry forbids a physical $b\to a$ decay, and all the selected energies remain $O(1)$ so that the leading thermal-AdS background is unchanged.

\FloatBarrier

\subsection{Heavy quenches}
\label{subsec:holo-heavy}

We next consider primary quench operators whose dimensions scale with the central charge. The associated excitation energy is of order $c$, so its gravitational backreaction enters at leading order and the filtered state may involve distinct semiclassical bulk geometries. The thermal-AdS choice for the unperturbed partition function does not determine which of these geometries contributes to the correlator with heavy insertions.

We consider a scalar quench operator $\cO_H$ with conformal weights
$h_H=\bar h_H=c\eta+o(c)$,
where $\eta>0$ remains fixed as $c\to\infty$. Its total conformal dimension is $\Delta_H=2h_H$, and the excitation energy $\kappa\Delta_H$ is of order $c$. In the semiclassical AdS$_3$ description, the nature of the geometry associated with such a heavy primary is characterized by the parameter
$\alpha_H^2=1-24h_H/c$.
The regime $0<h_H<c/24$ corresponds to a conical-defect geometry, while $h_H>c/24$ lies above the BTZ threshold~\cite{AsplundBernamontiGalliHartman:2015}.

Different heavy output states could therefore support different backreacted configurations, provided their channels dominate the filtered correlator over separated intervals. We examine this possibility through the necklace expansion~(\ref{eq:necklace}), whose denominator remains the unperturbed thermal partition function of~(\ref{eq:holovacZ}) in the cold thermal-AdS regime. Heavy insertions can modify the saddles contributing to the numerator~\cite{Abajian:2023}, making the competition between intermediate families essential to the bulk interpretation.

For intermediate primary weights that also scale with $c$, the semiclassical behavior of two-point torus blocks studied in~\cite{AlkalaevBelavin:2017} motivates an expansion of the form
\be
\log\cF^{\rm Vir}_{ad}(q_1,q_2)=-\frac c6 f_{ad}(q_1,q_2)+o(c).
\label{eq:heavyclassicalblock}
\ee
Here $f_{ad}$ is the corresponding classical conformal-block function, whose form depends on the external and intermediate conformal dimensions. The exponential dependence on $c$ reflects the fact that heavy conformal blocks can contribute at the same order as the classical gravitational action.

The necklace sum also contains families with dimensions that remain finite or grow more slowly than $c$, whose blocks need not follow the same semiclassical approximation. Their possible competition with heavy families must be retained when selecting dominant channels.

To isolate the primary contributions, we consider an additional cold limit in which both Euclidean intervals become parametrically large,
\be
\kappa\beta\gg\log c,\qquad
2\kappa s=x\kappa\beta\gg\log c,\qquad
x=\frac{2s}{\beta}>0\quad\text{fixed}.
\label{eq:heavycoldlimit}
\ee
For an individual channel whose descendant coefficients grow at most polynomially with $c$, the suppression by the Euclidean propagation factors is sufficient to make the descendant contributions subleading. The leading term is then determined by propagation of the corresponding primary states along the two Euclidean arcs.

For heavy insertions, the number of intermediate families and the large-$c$ behavior of their blocks and OPE coefficients can make this suppression nonuniform. We therefore require the descendant contributions to remain subleading after summing over the relevant families and taking the derivatives in the torus Ward identity. The light-state dominance results of~\cite{KrausSivaramakrishnanSnively:2017} apply to light external operators, so this uniformity is an additional assumption in the heavy sector.

Under this assumption, the contribution of an allowed primary transition $a\to d$ takes the leading form
\be
W_{ad}\simeq A_{ad}(c)e^{-\kappa\beta(\Delta_a+x\Delta_d)},\qquad
I_{ad}(x)=\Delta_a+x\Delta_d-\frac{\log A_{ad}(c)}{\kappa\beta},
\label{eq:heavycost}
\ee
where $A_{ad}(c)$ includes the squared OPE coefficients and the multiplicities of the relevant primary states. The effective cost $I_{ad}(x)$ combines the Euclidean propagation cost with the central-charge dependence of the operator matrix elements.

At each fixed $c$, the finite-primary selection criterion~(\ref{eq:finiteprimary}) continues to apply to the strict cold lower envelope. Every primary pair reaching that envelope must satisfy
$\Delta_a+\Delta_d\le\Delta_H$.
For a heavy quench, however, $\Delta_H=O(c)$, so this bound permits intermediate dimensions that themselves scale with the central charge. As $c$ increases, the number of potentially relevant primary families can also grow, and the bound no longer restricts the competition to a fixed set of light states.

The density of heavy states and their OPE coefficients can have exponential large-$c$ behavior~\cite{HartmanKellerStoica:2014,CollierMaloneyMaxfieldTsiares:2020}. As a result, many intermediate families can compete collectively with individual propagation channels. The spectral sum may then be controlled by a saddle whose characteristic output dimension changes continuously with $x$, leading to a smooth energy evolution. Isolated gravitational plateaux require further restrictions on this competition.

To illustrate the importance of the operator matrix elements, consider the two vacuum endpoint channels. Their leading propagation costs are $x\Delta_H$ and $\Delta_H$, corresponding to the creation of the heavy primary from the vacuum and the production of the vacuum from a thermally populated heavy state, respectively. These channels are always present for a normalized Hermitian quench operator, but they need not dominate the full spectral sum.

Suppose, for example, that another allowed transition connects two light states with $\Delta_a,\Delta_d=O(1)$, while its squared matrix element is exponentially suppressed as $A_{ad}(c)\sim e^{-\gamma c}$, with $\gamma>0$. Comparing its contribution with the larger of the two vacuum endpoint weights gives
\be
\log\frac{W_{ad}}{\max(W_{\mathbf1 H},W_{H\mathbf1})}
\simeq\kappa\beta\big[\Delta_H\min(x,1)-\Delta_a-x\Delta_d\big]-\gamma c.
\label{eq:heavylightcompetition}
\ee
Since $\Delta_H=O(c)$, the difference in propagation costs contributes at order $c\kappa\beta$. In the cold regime~(\ref{eq:heavycoldlimit}), this contribution can overcome the exponential suppression of the OPE coefficient. Thus, even a channel with a very small matrix element may dominate over the vacuum endpoint channels if its propagation cost is sufficiently lower.

The vacuum endpoint transitions can therefore be overtaken by other channels, including those with strongly suppressed matrix elements. This possibility is compatible with the monotonicity relation~(\ref{eq:outputvariance}), which constrains the averaged energy under filtering without requiring separate plateaux. We next consider how selection rules may isolate a small set of dominant channels.

One setting in which the spectral competition simplifies is provided by an exact discrete symmetry. Suppose the theory has a $\mathbb Z_2$ symmetry under which the heavy quench operator $\cO_H$ is odd, and assume that the heavy primary $H$ is the lightest state in the odd sector, while the even sector has a unique vacuum.

Because the quench operator changes parity, every allowed primary transition connects states of opposite parity. For an even-to-odd transition, the output dimension satisfies $\Delta_d\ge\Delta_H$, so its propagation cost obeys
$\Delta_a+x\Delta_d\ge x\Delta_H$.
Similarly, for an odd-to-even transition, the input dimension satisfies $\Delta_a\ge\Delta_H$, giving
$\Delta_a+x\Delta_d\ge\Delta_H$.

The vacuum endpoint channels saturate these inequalities. Consequently, the complete primary lower envelope is determined by
\be
I_*(x)=\Delta_H\min(x,1).
\label{eq:heavyprotected}
\ee
The resulting cold spectral structure contains only the heavy-primary and vacuum branches, which exchange dominance at $x=1$. No additional primary channel can lie below this envelope.

In a holographic theory, the exact discrete symmetry has a bulk interpretation as a discrete gauge symmetry~\cite{HarlowOoguri:2018}, which protects the primary lower envelope against competing transitions. Establishing dominance in the full correlator additionally requires bounding the collective contributions of states above the sector minima, including their derivatives in the torus Ward identity.

A different spectral structure can arise when the quench operator admits nonvanishing diagonal matrix elements within another primary family. Suppose there exists a primary state $a$ satisfying
$C_{aHa}\ne0$
and
$0<\Delta_a<\Delta_H/2$.
The corresponding diagonal transition has cost $(1+x)\Delta_a$, which can lie below both vacuum endpoint branches over a finite interval.

If all other allowed transitions have larger effective costs, the leading envelope is
\be
I_*(x)=\min\{x\Delta_H,(1+x)\Delta_a,\Delta_H\},\qquad
x_1=\frac{\Delta_a}{\Delta_H-\Delta_a},\quad
x_2=\frac{\Delta_H-\Delta_a}{\Delta_a}.
\label{eq:heavyintermediate}
\ee
The inequality $\Delta_a<\Delta_H/2$ ensures that $x_1<x_2$, so the diagonal channel is exposed over a nonzero interval. The resulting sequence of dominant output states is
$H\to a\to\mathbf1$,
with an intermediate plateau associated with the primary $a$.

If $\Delta_a=O(c)$, this intermediate state can itself carry leading gravitational backreaction and may correspond to a semiclassical geometry distinct from those associated with the initial heavy primary and the vacuum. The resulting plateau sequence could then describe the successive dominance of different backreacted bulk configurations under Euclidean filtering.

The diagonal channel in this construction is incompatible with the odd-operator selection rule of the preceding $\mathbb Z_2$ example, so the two cases correspond to different spectral assumptions.

Finally, consider the case in which the two vacuum endpoint channels are the only contributions surviving at leading order. Their cold weights and the corresponding homogeneous chiral energy reduce to
\be
\begin{gathered}
W_H\simeq e^{-2\kappa s\Delta_H},\qquad
W_0\simeq e^{-\kappa\beta\Delta_H},\qquad
p_H=\frac{W_H}{W_H+W_0},\\
\frac{\cE_L}{\kappa^2}+\frac c{24}\simeq h_Hp_H.
\end{gathered}
\label{eq:heavybranches}
\ee
Here $p_H$ is the relative weight of the heavy-primary output channel. For $2s<\beta$, the heavy contribution dominates in the cold limit, and the energy approaches the heavy-primary level. For $2s>\beta$, the vacuum contribution becomes dominant, and the energy approaches the vacuum Casimir value.

The two weights become equal at $2s=\beta$. Since $\Delta_H=O(c)$, their relative weight changes exponentially rapidly near this crossing, and the characteristic width in the dimensionless filtering variable is
$\Delta(2\kappa s)=O(c^{-1})$.
Thus, in the semiclassical limit, the transition between the two dominant energy levels becomes increasingly sharp.

Away from the crossover, the filtered state approaches either the heavy primary or the vacuum, admitting the corresponding backreacted or AdS vacuum interpretation. Near the crossing, however, both channels contribute with comparable weights. Because their energies differ by an amount of order $c$, the corresponding energy variance is of order $c^2$. The transition region therefore need not admit a description in terms of a single classical bulk geometry, even when each of the two limiting regimes does.

The light analysis admits a few-particle description in thermal AdS when the low-energy spectrum and the full correlator satisfy the stated bounds. For heavy quenches, the intermediate conformal families can grow in number with $c$, and their OPE coefficients and degeneracies may contribute at the same exponential order as the Euclidean propagation factors. The resulting competition must be controlled before a sequence of dominant heavy states can be interpreted geometrically.

When selection rules and bounds on the summed contributions isolate a small set of channels, their output dimensions can determine successive energy plateaux associated with distinct backreacted configurations. Near a crossover, several channels may contribute with comparable weights and the filtered state need not admit a single classical geometry. If the dominant spectral contribution instead varies continuously with the filtering ratio, the energy can evolve smoothly without well-separated gravitational plateaux.

\section{Conclusion and Discussion}

In this work, we have investigated the evolution of the energy density following local operator quenches in two-dimensional conformal field theories initially prepared in thermal states. The dynamics is governed by complex-time evolution, in which ordinary Lorentzian propagation is supplemented by Euclidean filtering. Using the conformal Ward identities on the cylinder and torus, we have expressed the local energy density in terms of correlation functions of the quench operator and examined its evolution in several conformal field theories.

Our results show that Euclidean filtering can give rise to sequences of approximately homogeneous energy plateaux, whose values approach the energies of different primary states. The underlying mechanism is the competition between conformal families accessible to the local operator insertion. At finite temperature, the initial ensemble contains contributions from excited states, allowing several input--output channels to participate in the evolution. As the filtering increases, their relative spectral weights change, and different output states can become dominant over successive time intervals. This mechanism is illustrated by the Ising model, compact boson, and WZW theories, while the algebraic relaxation of the noncompact boson demonstrates the qualitatively different behavior associated with a continuous spectrum.

To describe this phenomenon systematically, we developed a cold-limit spectral formulation in which the relevant primary transitions are assigned costs determined by the conformal dimensions of their input and output states. The dominant contributions are selected by the lower envelope of these costs. For a fixed compact unitary CFT, we showed that every primary pair contributing to the leading cold envelope must satisfy
$\Delta_a+\Delta_d\le\Delta_{\cO}$,
where $\Delta_{\cO}$ is the total conformal dimension of the quench operator. This bound restricts the leading spectral competition to finitely many primary channels, even when the theory contains infinitely many conformal families.

When a single channel dominates and is separated from the others by a positive spectral gap, Euclidean filtering selects the corresponding output eigenspace exponentially. The local energy density then approaches the homogeneous value determined by its conformal weights. When two or more channels have comparable weights, the energy interpolates between their characteristic levels. The leading crossover positions are fixed by the conformal dimensions, while the operator matrix elements determine the relative amplitudes and finite-temperature shifts. For sufficiently cold preparations, plateau durations grow linearly with the initial inverse temperature, allowing the different spectral regimes to become increasingly well resolved.

This framework accounts for the intermediate primary plateaux observed in the Ising and $SU(2)_k$ current quenches and explains the longer sequences generated by higher-charge compact-boson vertex operators. It also provides a systematic construction of candidate plateau sequences in diagonal unitary minimal models and affine-primary quenches in WZW theories. In particular, the $\Phi_{2,1}$ minimal-model family and the maximal-spin $SU(2)_k$ quenches illustrate how the number of leading primary plateaux can grow across a family of theories or quench operators. Comparisons with the corresponding torus calculations and Virasoro-sewing results support the spectral interpretation of these regimes.

The holographic analysis suggests a further interpretation of this mechanism in theories with large central charge. For light quenches, under suitable assumptions on the spectrum, operator matrix elements, and conformal-block expansion, the dominant primary channels can be associated with few-particle sectors in thermal $\mathrm{AdS}_3$. The resulting plateaux then describe the successive dominance of different bulk particle sectors under Euclidean filtering. Heavy quenches present a more complicated situation, since their excitation energies are of order $c$ and can produce leading gravitational backreaction. In this regime, the competition between the growing number of intermediate states, their OPE coefficients, and their spectral degeneracies prevents a universal prediction of isolated gravitational plateaux. When sufficiently separated dominant channels exist, they may correspond to distinct backreacted geometries, while regions of comparable spectral weights need not admit a description in terms of a single classical bulk configuration.

More generally, these results suggest that complex-time filtering can provide a way to probe the spectral structure of conformal field theories through local observables. The heights of resolved energy plateaux identify the conformal weights of the dominant output states, while their ordering and crossover positions constrain the allowed primary transitions. The finite-temperature shifts and interpolation between plateaux contain additional information about the corresponding operator matrix elements. Varying the initial temperature and the quench operator may therefore provide access to different parts of the conformal spectrum and its operator algebra. This possibility is particularly interesting in view of the interpretation of Euclidean filtering as an effective conditional evolution generated by repeated measurements and postselection.

Several directions remain open. An important next step is to construct a gravitational description of the finite-temperature complex-time quench that incorporates both the Euclidean filtering contour and the backreaction of heavy operator insertions. Such a construction could clarify how competing conformal families are represented by bulk configurations and whether the spectral crossovers correspond to transitions between dominant gravitational saddles.

Another natural extension is to study the evolution of entanglement entropy in the filtered thermal state, generalizing the zero-temperature analysis of~\cite{Bai:2026}. Comparing the entanglement dynamics with the changing dominant primary sectors may reveal whether the energy plateaux have accompanying signatures in the redistribution of quantum information. More broadly, understanding how Euclidean filtering selects different parts of the spectrum may provide further insight into relaxation and conditional dynamics in conformal field theories and their holographic duals.\footnote{The authors used AI tools during work on this manuscript. They reviewed and edited the content as needed and take full responsibility for the content of the publication.}

\appendix
\section{Cold spectral selection and local-stress bounds}
\label{app:coldproof}

For a fixed compact unitary CFT, remove the common Casimir factors from
$Z_{\mathbb T^2}G_{\mathbb T^2}$ and expand in exact energy eigenstates.
The resulting positive trace is
\be
\sum_{a,d}|\langle d|\cO|a\rangle|^2
 e^{-b\Delta_a-Y\Delta_d}.
\label{eq:app_positive_trace}
\ee
The sum converges at positive Euclidean separations by assumption, and all
energy eigenspaces below a finite dimension have finite multiplicity. On a
closed interval $x\in[x_-,x_+]\subset(0,\infty)$, choose one allowed
matrix element and denote the maximum of its cost by $M$. Every pair which
can reach the lower envelope satisfies
$\Delta_a+x_-\Delta_d\le M$, and hence
$\Delta_a\le M$ and $\Delta_d\le M/x_-$. Only finitely many eigenstate
pairs can satisfy these bounds, which proves attainment of the minimum and
its piecewise-linear form on the chosen interval.

Let $I_*(x)$ be this minimum. After factoring out $e^{-bI_*(x)}$,
the remainder is bounded from below by the smallest positive coefficient
among the finitely many minimizing pairs. It is bounded from above by
a reference trace at $b_0>0$ and $b_0x_-$, since for $b\ge b_0$
\be
\sum_{a,d}A_{ad}e^{-b[I_{ad}(x)-I_*(x)]}
\le e^{b_0\max_{x\in[x_-,x_+]}I_*(x)}
\sum_{a,d}A_{ad}e^{-b_0(\Delta_a+x_-\Delta_d)}.
\label{eq:app_reference}
\ee
The logarithm divided by $b$ therefore converges uniformly to the lower
envelope. Inside an interval with one minimizing input--output energy pair,
a gap $g>0$ separates its cost from every other state pair. Splitting the
exponential weights between $b_0$ and $b-b_0$ bounds the probability outside
the minimizing pair by $C e^{-(b-b_0)g}$. Multiplication by fixed powers of
the output energy can be controlled by evaluating the reference trace at
slightly shorter positive Euclidean lengths, which yields the analogous
energy-moment estimates.

For a Virasoro-primary quench operator, the commutators
\be
[L_n,\cO(z)]=(z^{n+1}\partial_z+(n+1)h_{\cO}z^n)\cO(z)
\label{eq:app_primaryward}
\ee
express descendant matrix elements through the corresponding primary
three-point coefficient~\cite{BPZ:1984,DiFrancesco:1997}. A vanishing primary coupling therefore removes all
its descendant matrix elements. Since the descendant levels add
$N_a+xN_d\ge0$ to the cost, the exact lower envelope is that of the
supported primary pairs. The endpoint channels give
$I_*(x)\le\Delta_{\cO}\min(1,x)$, and comparison with
$I_{ad}(x)\ge\min(1,x)(\Delta_a+\Delta_d)$ proves
(\ref{eq:finiteprimary}). The finitely many primary pairs obeying this
inequality determine the complete lower envelope for every $x>0$.

The density-matrix estimate can be obtained directly without enlarging the
Hilbert space. After removing the Casimir energy, write
\be
\rho=\frac{BB^\dagger}{\Tr(BB^\dagger)},\qquad
B=e^{-sH}\cO e^{-\beta H/2},\qquad
B_*=P_dBP_a.
\label{eq:app_B}
\ee
Orthogonality of the energy eigenspaces gives
$\|B-B_*\|_2^2/\|B\|_2^2=O(e^{-bg})$, where the norm is the
Hilbert--Schmidt norm. The inequality
$\|UV\|_1\le\|U\|_2\|V\|_2$ then gives
$\|\rho-\rho_*\|_1=O(e^{-bg/2})$ with $\rho_*$ in
(\ref{eq:effective_rhostar}). The finite-dimensional operator $\rho_*$
remains supported on the selected output energy eigenspace.

For the local chiral stress, write its dimensionless Fourier expansion as
\be
T_L(\theta)=L_0-\frac{c}{24}+\sum_{n\ne0}L_ne^{in\theta},
\qquad
\frac{\cE_L(X,t)}{\kappa^2}=\Tr[\rho_s T_L(\theta)],
\quad \theta=\kappa(X-x_0-t),
\label{eq:app_stressFourier}
\ee
where $\rho_s=BB^\dagger/\Tr(BB^\dagger)$ uses the filtered state before
Lorentzian evolution. Equivalently, one may use $\rho(t)$ with the unshifted
spatial argument $\theta=\kappa(X-x_0)$. This convention implements the
left-moving translation without counting the unitary evolution twice.
The Virasoro commutator and positivity of the energy imply an
energy-bound of the form
$\|L_n(L_0+1)^{-1}\|\le C_c(|n|+1)^{3/2}$, where
$C_c$ is finite for the fixed CFT. Since a mode $L_n$ changes the
conformal energy by $-n$, the heat-sandwiched operator
$e^{-r(L_0+\bar L_0)}T_L(\theta)e^{-r(L_0+\bar L_0)}$
is bounded uniformly in $\theta$ for every $r>0$. Its Fourier series
converges absolutely in operator norm. This permits the same reference-trace
estimate to be applied to the expectation value of the local stress.

Decompose $B$ into the part with primary input and output, the part with
primary input and descendant output, and the part with descendant input.
Their squared Hilbert--Schmidt norms relative to the full trace are
respectively of order one, $O(e^{-Y})$ and $O(e^{-b})$, uniformly on compact
$x$-intervals. The stress operator acts on the output space, so distinct
input levels do not interfere in its expectation value. Interference
between primary and descendant outputs is bounded by Cauchy--Schwarz and
gives $O(e^{-Y/2})$, while the descendant-input contribution remains
$O(e^{-b})$. Between primary output states, nonzero Virasoro modes have
vanishing matrix elements and the diagonal $L_0$ matrix elements give the
weighted primary expression. These estimates yield~(\ref{eq:effectivelocal})
uniformly in $X$ and through primary crossings, under the stated positive
shortened-arc trace condition.

To pass from the complete primary sum to the finite set in
(\ref{eq:finiteprimary}), let the smallest excess of
$\Delta_a+\Delta_d$ above $\Delta_{\cO}$ be positive, as ensured by the
discrete spectrum at fixed CFT. The omitted channels then exceed the lower
envelope by a uniform positive cost on every compact interval of $x>0$.
Their summed contribution and energy-weighted derivatives obey the same
exponential reference-trace bound, which supplies the additional finite-set
error in~(\ref{eq:effectivelocal}).

\section{Tricritical-Ising sewing and numerical checks}
\label{app:tricriticalsewing}

For the tricritical-Ising model $\mathcal M(4,5)$, we sew the six Virasoro families of dimensions $0,3/40,1/5,7/8,6/5,3$ at $c=7/10$. The scalar quench $\cO=\Phi_{2,1}=\sigma'$ has $h_{\cO}=\bar h_{\cO}=7/16$ and four Hermitian fusion pairs
\be
\mathbf1\leftrightarrow\sigma',\qquad
\sigma\leftrightarrow\varepsilon,\qquad
\sigma\leftrightarrow\varepsilon',\qquad
\sigma'\leftrightarrow\varepsilon''.
\label{eq:trisectionedges}
\ee
All eight orientations are retained. In the conventional unit-normalized primary basis, the real three-point couplings for the four unordered pairs are $C=1,1/2,3/4,7/8$, respectively, as determined by the tricritical-Ising operator algebra~\cite{Henriksson:2025}.

Virasoro descendants through chiral level $N=10$ are organized in a Poincar\'e--Birkhoff--Witt basis. The Gram matrices follow from
\be
[L_m,L_n]=(m-n)L_{m+n}+\frac{c}{12}m(m^2-1)\delta_{m+n,0},
\qquad [L_n,\cO(z)]=z^n(z\partial_z+(n+1)h_{\cO})\cO(z).
\label{eq:tricriticalcommutator}
\ee
We diagonalize each Gram matrix after normalizing its nonzero diagonal entries and retain only its positive eigenvalue subspace, thereby removing singular null descendants. For each output family $d$, input primary $a$ and output chiral level $n$, let $G_{d,n}$ be the Gram matrix and $v_{d\leftarrow a,n}$ the vector of three-point matrix elements generated by the second commutator in~(\ref{eq:tricriticalcommutator}), divided by the primary coefficient. The corresponding basis-independent nonnegative weight is
\be
w_{d\leftarrow a}(n)=v_{d\leftarrow a,n}^{\dagger}
G_{d,n}^{+}v_{d\leftarrow a,n},
\label{eq:tricriticalgramweight}
\ee
where $G^{+}$ is the Moore--Penrose inverse restricted to physical descendants. The left and right chiral factors coincide for the diagonal scalar quench operator.

At $\beta=300$ and $L=10$, incoming descendants are suppressed by $e^{-\kappa\beta}<10^{-81}$ before matrix-element factors. Retaining primary inputs gives the positive trace
\be
\begin{aligned}
Q_N(A,\beta)&=\kappa^{2\Delta_\cO}e^{\kappa c(A+\beta)/12}
\sum_{a\to d}|C_{a\cO d}|^2
 e^{-\kappa[\beta\Delta_a+A\Delta_d]}
\left[\sum_{n=0}^N w_{d\leftarrow a}(n)e^{-\kappa A n}\right]^2,
\end{aligned}
\label{eq:tricriticalsewingtrace}
\ee
where $A=2s=2\epsilon+2\delta t$. Both outgoing chiralities are included through $N=10$, with all six modules and eight oriented transitions. Omitted incoming descendants produce a relative $O(e^{-\kappa\beta})$ correction at fixed outgoing cutoff in this cold regime. The factor $\kappa^{2\Delta_\cO}$ restores the cylinder normalization and cancels in Ward logarithmic derivatives. Outgoing-level convergence is checked below. This remains a cold-input approximation to the torus correlator.

The torus Ward identity reconstructs local stress using the output-energy zero mode and Fourier coefficients determined by input--output energy-moment differences. We use $L=10$, $X=5$, $x_0=0$, $\epsilon=0.5$, $\delta=0.35$, $\beta=300,1000$ and $A\ge5$, with $t=(A-1)/0.7$. The cold crossings $A/\beta=1/9,1,9$ give
\be
\begin{array}{c|ccc}
\beta & \sigma'\to\varepsilon & \varepsilon\to\sigma & \sigma\to\mathbf1\\\hline
300 & 46.19 & 427.14 & 3855.71\\
1000 & 157.30 & 1427.14 & 12855.71
\end{array}
\label{eq:tricriticaltimeappendix}
\ee
for the leading transition times, with finite-$\beta$ corrections.

The calculation was checked in three ways. The physical Gram ranks through $N=10$ agree with the Rocha--Caridi characters of the six irreducible Virasoro modules. On a grid of $420$ values of $A$ for each $\beta=300,1000$, restricted to $A\ge5$, the maximal difference between the $N=8$ and $N=10$ left-moving stresses is $4.7071\times10^{-12}$ in units of $\kappa^2$. Repeating the sewing calculation with the diagonal Ising conformal data gives the exact Ising energy-quench Ward stress to within $3.1\times10^{-11}$ on the same cold grid. These comparisons assess convergence over the stated interval. At $A=3$, the difference between levels eight and ten increases to approximately $2.3\times10^{-7}$, which motivates the restriction $A\ge5$ in Fig.~\ref{fig:tricriticalsewing}.

\end{document}